\documentclass[a4paper,11pt]{article}
\pdfoutput=1 

\usepackage{jcappub} 

\usepackage[T1]{fontenc} 

\newcommand{\rnum}[1]{\lowercase\expandafter{\romannumeral #1}}

\title{\boldmath Revisiting van Citter-Zernike correlations in the presence of primordial gravitational waves}

\author[a,b,c]{F. S. Arani,\note{Corresponding author.}}
\author[a,d]{M. Bagheri Harouni}
\author[b,c,1]{Brahim Lamine}
\author[b,c]{Alain Blanchard}

\affiliation[a]{Department of Physics, University of Isfahan, Hezar Jerib Str., Isfahan 81746-73441, Iran}
\affiliation[b]{Université de Toulouse, UPS-OMP, IRAP, F-31400 Toulouse, France}
\affiliation[c]{CNRS, IRAP, 14 avenue Edouard Belin, F-31400 Toulouse, France}
\affiliation[d]{Quantum Optics Group, Department of Physics, University of Isfahan, Hezar Jerib Str., Isfahan 81746-73441, Iran}
\emailAdd{fshojaei@irap.omp.eu}
\emailAdd{m.bagheri@sci.ui.ac.ir}
\emailAdd{brahim.lamine@irap.omp.eu}

\abstract{In this paper, we develop a quantum field theory framework to describe the interaction between a gravitational wave (GW) background and an electromagnetic (EM) field emitted from a distant celestial source, such as a star. We demonstrate that a background of primordial gravitational waves (PGWs), as predicted by the inflationary scenario, induces a loss of spatial coherence in the EM field as it propagates over cosmological distances. This effect leads to the degradation of van Cittert-Zernike correlations, ultimately rendering them unobservable—a phenomenon referred to as blurring. Since spatial coherence is observed in very long baseline interferometry (VLBI) measurements of distant quasars, this places constraints on the amplitude of the PGW background. We quantitatively evaluate the blurring effect caused by PGWs in a two-mode squeezed state, which represents the standard quantum state predicted by the simplest inflationary models. However, due to the weak coupling between GWs and the EM field, we find that the induced incoherence is too small to be detected in current VLBI observations.}

\begin{document}
\maketitle
\flushbottom


\section{Introduction}\label{sec:1}

Interferometric methods have found vast applications in testing gravity either at the classical or quantum level. Starting from the famous Michelson-Morley interferometer to rule out the motion through ``ether,'' light interferometry is now routinely used as a crucial technique to detect the tiniest effects of gravitational waves (GWs) with incredible precision \cite{saulson1994fundamentals,barish1999ligo,ju2000detection,accadia2012virgo,blair2012advanced}. Whether these ripples in spacetime obey quantum mechanical rules at a fundamental scale, say the Planck scale, is an open question, and we lack any direct observational evidence in favor of that \cite{abdo2009limit, tamburini2011no}. 

It is believed that space and time, when viewed at Planck scales, form a foam-like structure due to quantum fluctuations of the vacuum metric tensor of spacetime \cite{garay1998spacetime,kempf1999three,rovelli2008loop}. Searching for Planck-scale quantum features of spacetime is an old quest, and many interferometric schemes based on the phase properties of electromagnetic (EM) radiation of distant objects have been proposed to inquire about it \cite{amelino2000gravity,lieu2003phase,ragazzoni2003lack,lammerzahl2005interferometry,lammerzahl2007search,khodadi2019probing}. For instance, the phase interferometric approach early introduced in \cite{lieu2003phase} inspects the Planck-scale physics through the spacetime foam-induced \textit{phase incoherence} of light emitted from distant objects. It has been perceived that the accumulation of tiny phase incoherence during the long journey of light through the quantum fluctuations of the vacuum spacetime leads to the loss of the phase of radiation at large distances. This approach was elaborated in \cite{ragazzoni2003lack} to rule out or put stringent limits on some Planck-scale phenomenological models by evidence of the diffraction pattern from the Hubble Space Telescope observations of SN 1994D at $z=5.34$.

However, further studies \cite{ng2003probing} declared that such effects ``are far below what is required in this approach to shed light on the foaminess of spacetime''.

Moreover, it turned out that ``propagation of spatial correlations is governed by the van Citter-Zernike theorem \cite{mandel1995optical}, and spatial correlations are immune to any underlying fuzzy Planck scale,'' hence the Planck scale remains inaccessible to interferometer detection with present technology \cite{coule2003planck}. Albeit, this result leaves room for quantum gravitational effects of length scales much larger than the Planck length.

One important example is the relic background of quantum tensor perturbations naturally generated by the pumping engine of inflation. At the heart of cosmology, the inflationary scenario predicts that quantum fluctuations of the gravitational field present in the very early Universe have evolved and been amplified during the successive expansion of the Universe and constitute a stochastic background of primordial gravitational waves (PGWs) \cite{grishchuk1977graviton,grishchuk1997implications,grishchuk2001relic,grishchuk2010discovering,rubakov1982graviton,fabbri1983effect,abbott1984constraints,allen1988stochastic,sahni1990energy,tashiro2004reheating,henriques2004stochastic,zhao2006relic,maggiore2007gravitational}. Due to its specific generating mechanism, PGW's spectrum spans a full range of frequencies, corresponding to wavelengths much larger than the Planck scale. Thereupon, the quest for PGWs has made one of the main targets of today's and upcoming GW detectors in different frequency bands: LIGO \cite{ligo}, Advanced LIGO \cite{advligo,waldman2011advanced}, VIRGO \cite{virgo,acernese2005status}, GEO \cite{geo,willke2002geo}, AIGO \cite{degallaix2005thermal,barriga2005optical}, LCGT \cite{kuroda2010status}, ET \cite{punturo2010einstein,hild2011sensitivity} aim at the frequency range $(10^2-10^3)$ Hz; the space interferometers, such as the future LISA \cite{larson2000sensitivity,larson2002unequal} which is sensitive in the frequency range $(10^{-4}-10^{-1})$ Hz, BBO \cite{crowder2005beyond,cornish2005lisa,cutler2006big} and DECIGO \cite{kawamura2006japanese,kudoh2006detecting} both sensitive in the frequency range $(0.1-10)$ Hz, and the pulsar timing arrays \cite{agazie2023nanograv, antoniadis2023second, zic2023parkes, reardon2023search, xu2023searching}, including PPTA \cite{hobbs2008gravitational, manchester2006parkes, jenet2006upper} and the planned SKA \cite{kramer2004strong} working in the frequency window $(10^{-9}-10^{-6})$ Hz. Besides, there are potential proposals able to detect the very-high-frequency part of PGWs, among which are the waveguide detector \cite{cruise2000electromagnetic,cruise2005correlation, cruise2006prototype,tong2008detecting}, the proposed Gaussian maser beam detector around GHz \cite{li2003electromagnetic,li2008perturbative,tong2008using}, and the $100$ MHz detector \cite{akutsu2008search}. Moreover, the very low-frequency portion of PGWs contributes to the anisotropy and polarization of cosmic microwave background (CMB) \cite{zaldarriaga1997all,kamionkowski1997statistics,keating1998large,pritchard2005cosmic,zhao2006analytic,xia2009approximate,zhao2009detecting}, yielding a magnetic-type polarization of CMB as a distinguished signal of PGWs, which would possibly be sensed using WMAP \cite{bevis2008fitting,komatsu2009five,dunkley2009five,jarosik2011seven,hinshaw2013nine}, Planck \cite{planck2006scientific,planck}, liteBIRD \cite{suzuki2018litebird}, the ground-based ACTPol \cite{niemack2010actpol} and the proposed CMBpol \cite{dunkley2009prospects}. 

Detection of PGWs would not only confirm the inflationary scenario but also validate the quantum description of gravity at scales much larger than the Planck length.

In the present study, we introduce a new schema to search for the unique imprints of the PGWs on the spatial correlations of the EM emission from the furthest objects and show how spatial correlations governed by the van Citter-Zernike theorem and measured by Very Long Baseline Interferometers (VLBI) are influenced by PGWs. The VLBI imaging technique takes advantage of the largest possible baselines (from several meters to thousands of kilometers) to precisely determine the location and fine structure of astronomical sources, including Active Galactic Nuclei (AGNs) \cite{cohen1973introduction,falcke2000radio,pushkarev2015milky,an2018capabilities,algaba2013high}. In a typical VLBI experiment, the incoming radiation of a distant source is collected by two or more spatially separated telescopes (see Fig.~\ref{fig3}). The existence of spatial correlations throughout the projected baseline of the interferometer leads to the appearance of an interference pattern (non-vanishing visibility), which is used to determine the geometrical properties of the object, such as its angular diameter.

Noticeably, VLBI imaging has found great attention in constraining cosmology. For instance, the Event Horizon Telescope (EHT) which is a VLBI array imaging supermassive black holes (SMBHs) on event horizon scales, allows tests of deviation from general relativity by performing high-precision measurements of the Kerr metric \cite{roelofs2021black,carballo2022toward}. Moreover, the Megamaser Cosmology Project (MCP) which is based on the VLBI sub-milliarcsecond angular mapping of compact objects, has provided an independent laboratory to constrain the Hubble constant, $H_0$, in parallel to other surveys such as CMB \cite{pesce2020megamaser}, likewise putting updated constraints on the mass of SMBHs \cite{kuo2013megamaser}. In particular, the capability of the VLBI method in accurate measurement of the angular size-redshift $\theta-z$ of intermediate-luminosity radio quasars has led to specifying AGNs as astrophysical standard rulers with intrinsic size $\ell_m \sim 11.03\,$pc that can be used to constrain cosmological parameters \cite{cao2017ultra,liu2023revisiting,liu2024model}. Recently, the capability of stellar interferometry as a potential tool to detect GWs in the lower frequency range $(10^{-6}-10^{-4})$ Hz is investigated \cite{park2021stellar}. However, to our best knowledge, VLBI measurements of the angular size-redshift have not yet been used to constrain the background of PGWs. Here, we promote the idea of employing stellar interferometric methods to characterize the underlying gravitational background of PGWs. In a more general sense, we promote the idea of using VLBI $\theta-z$ measurements of ever-distant sources to study the scalar or tensor perturbations of spacetime metric through the quantum-induced incoherence of phase correlations. In this way, we introduce VLBI setup as a new probe of the early universe perturbations, in parallel to other surveys such as CMB.

It is worth mentioning that, in the literature, there are numerous schema that focus on graviton-induced \textit{decoherence} (GID) of light \cite{lagouvardos2021gravitational,shojaei2023sensing} or particles \cite{lamine2006ultimate,kanno2021noise}. As a matter of fact, due to negligibly small coupling strength between gravity and radiation or matter fields, one usually needs a huge interaction time to observe the GID effect, provided that the system is completely isolated from other environmental effects that may lead to decoherence. In contrast, spatial correlations of a source are known to be immune to the typical environment-induced decoherence. In particular, the van Citter-Zernike theorem implies that the correlation length of initially spatial-incoherent sources grows with distance \cite{mandel1995optical}. However, we show that PGWs background acts as a competitive mechanism that tends to reduce spatial correlation length. Hence, sources with cosmological distances that have had enough time to interact with the underlying gravitational background seem proper candidates to explore the underlying quantum gravitational effects.

The paper is organized as follows. Sec.~\ref{sec:2} provides a brief introduction to the expansionary Universe, the generation of two-mode squeezed PGWs from vacuum fluctuations, and the definition of the related quantities of PGWs. In Sec.~\ref{sec:3}, a self-contained investigation of the EM-GWs interaction at the quantum level is included, followed by solving the Heisenberg equation of motion of the EM field. This will help us investigate the two-point correlation function of the EM radiation in Sec.~\ref{sec:4}, where the loss of correlations induced by two-mode squeezed PGWs is evaluated and characteristic length scales are defined. Sec.~\ref{sec:5} is devoted to results and discussion, where the main findings of the paper are presented, and further possible topics and caveats that can be addressed within the present schema are discussed. The summary and conclusion of the paper are provided in Sec.~\ref{sec:6}. A complete supplementary material containing the details of the derivation of the equations and their subtleties is provided in Apps.~\ref{app:A}-\ref{app:C}. Throughout the paper, we explicitly write $c$, $\hbar$, and $G$ in all equations. Table~\ref{tab:table1} contains the list of symbols and notations involved in the present study.

\begin{table*}[h]
\caption{\label{tab:table1} List of symbols and notations.}
\centering 
\begin{tabular}{|| c | c ||} 
\hline
 symbol & description \\ \hline\hline
 $\eta$ & conformal time \\
 $a(\eta)$ & scale factor in conformal time \\
 $\mathcal{H}(\eta)$ & comoving Hubble parameter in conformal time \\
 $(\Omega_K/c, \mathbf{K})$ & GWs four wave vector \\
 $\mathbf{K} = (K, \Theta_K, \Phi_K)$ & GWs wave vector in spherical coordinates \\
 $e_{ij}^{\lambda}[\hat{\mathbf{K}}]$ & polarization tensor \\
 $\hat{S}_{\mathbf{K}}(\zeta_K)$ & two-mode squeezing operator \\
 $\hat{R}_{\mathbf{K}}(\theta_K)$ & rotation operator \\
 $\zeta_K = r_K e^{2i\phi_K}$ & squeezing parameter \\
 $(r_K, \phi_K)$ & squeezing amplitude and phase \\
 $\theta_K$ & rotation parameter\\
 $\zeta_1$, $\zeta_s$, $\zeta_2$, $\zeta_E$ & increase parameters \\
 $A$ & initial amplitude of tensor perturbations at $K=K_H$ scale \\
 $(A_s,n_s)$ & scalar power spectrum and spectral index \\
 $(A_T, n_T)$ & tensor power spectrum and spectral index \\
 $T_{\text{reh}}$ & reheating temperature \\
 $k_0$ & pivot scale \\
 $\text{r}_{k_0}$ & tensor-to-scalar ratio at the pivot scale \\
 $K_E, K_H, K_2, K_s, K_1$ & characteristic wave numbers \\
 \hline
 $(\omega_k/c,\mathbf{k})$ & EM four wave vector \\
 $\mathbf{r}$ & observation point \\
 $(\mathbf{r}_1,\mathbf{r}_2)$ & location of two detectors on the Earth \\
 $(\bar{\bar{\varepsilon}},\bar{\bar{\mu}})$ & permittivity and permeability tensors \\
 $\nu_{k}(\mathbf{r},t)$ & temporal mode function of the EM field \\
 $\mathbf{f}_k(\mathbf{r},t)$ & spatial mode function of the EM field \\
 $\mathbf{A}(\mathbf{r},t)$ & vector potential of the EM field \\
 $\alpha_k$ & electromagnetic field expansion coefficient \\
 $\hat{\mathbf{u}}_{\mathbf{k}}$ & EM field polarization unit vector \\
 $\epsilon_{ijr}$ & Levi-Civita symbol \\
 $n_k(\mathbf{r},t)$ & refractive index corresponding to the GW's medium \\
 $\vartheta_s$ & angle between the baseline and the source's orientation \\
 $\theta$ & angular diameter of the source \\
 $\sigma$ & planar source \\
 $(\rho',\phi')$ & polar coordinates of the the surface element $d\sigma'$ on the planar source \\
 $\mathbf{r} = (r, \theta,\phi)$ & position vector in spherical coordinates \\
 $R$ & distance of the source to the Earth\\
 $z$ & redshift of the source \\
 \hline
\end{tabular}
\end{table*}


\section{PGWs in expansionary Universe}\label{sec:2}

\subsection{\label{subsec:2.1}Expansionary Universe}

The spatially flat expanding universe is often described by the Friedmann–Lemaître–Robertson– Walker (FLRW) metric assuming a power-law scale factor $a(\eta)\propto \eta^{\alpha}$, where $\eta$ is the conformal time. Each successive expansion stage is specified with a different index $\alpha$. 
The whole expansion history can be described by~\cite{tong2013using} 
\begin{eqnarray}\label{2.1}
a(\eta) =
 \begin{cases}
   \ell_0|\eta|^{1+\beta}\quad,\quad -\infty \leq\eta\leq \eta_1 \\
   a_z(\eta-\eta_p)^{1+\beta_s} \quad,\quad \eta_1 \leq\eta\leq \eta_s \\
   a_e (\eta-\eta_e) \quad,\quad \eta_s \leq\eta\leq \eta_2 \\
   a_m (\eta-\eta_m)^2 \quad,\quad \eta_2 \leq\eta \leq \eta_E \\
   \ell_H |\eta-\eta_a|^{-\gamma} \quad,\quad \eta_E \leq \eta \leq \eta_H 
 \end{cases}
\end{eqnarray}
where $\eta_H$ is the conformal time today and $a(\eta)$ has the dimensionality of length \cite{tong2012revisit, tong2013using}.

Usually the inflation index $\beta$ is related to the scalar spectral index of primordial perturbations according to $n_s=2\beta+5$ \cite{grishchuk2010discovering,zhang2010constraints,tong2009relic}. $n_s$ is a parameter introduced under the assumption that the scalar perturbation spectrum obeys a power-law behavior near the pivot scale $k_0$, with $n_s=1$ referred to as the Harrison-Zeldovich spectrum. The \textit{Planck2018} release favors $n_s \simeq 0.9649 \pm 0.0044$ \cite{akrami2020planck}, corresponding to $\beta \simeq -2.018 \pm 0.0022$. However, the exact relation between $\beta$ and $n_s$ depends crucially on the specific inflationary potential through the (first order) slow-roll parameter $\epsilon_V$. In this study, we consider $\beta$ as a free parameter of the model. However, for the sake of illustration, we mostly take $\beta= -2.0$.

The parameter $\beta_s$ describes the expansion during the reheating stage starting right after the end of inflation and may be related to the equation of state (EoS) parameter during the reheating stage, $w_{\text{reh}}$, and to the inflationary model parameters. In the literature, usually, $\beta_s=1$ is chosen, which may correspond to a quadratic inflation potential with EoS parameter $w_{\text{reh}}=0$ \cite{starobinsky1980new,tong2012revisit}. As discussed in App.~(\ref{app:A.3}), $\beta_s$ affects only the high-frequency part of the PGWs spectrum and leaves a negligible effect on the incoherence of the EM field that we study in this paper. Hence, we mostly adopt the value $\beta_s= 1$ in our investigations.

The parameter $\gamma$ determines the late expansion of the universe governed by dark energy $\Lambda$.

Throughout later numerical investigations, we take $\gamma=1$ which would correspond to a pure de Sitter acceleration phase \cite{miao2007analytic}.

On the other hand, the present scale factor is conveniently chosen as $a(\eta_H)=\ell_H$, so that the condition $|\eta_H-\eta_a|=1$ turns out \cite{miao2007analytic}.

The constant $\ell_H$ is determined by $\ell_H=\gamma/H_0$ as a result of Eq.~(\ref{2.1}). Here, $H_0=67.4\,$km s$^{-1}$ Mpc$^{-1}$ is the present Hubble constant \cite{collaboration2020planck}.

Considering $\beta$ and $\beta_s$ as free parameters, there remain $12$ constants in Eq.~(\ref{2.1}), $8$ of which are reduced by the continuity of $a(\eta)$ and its derivative at four jointing points $\eta_1$, $\eta_s$, $\eta_2$ and $\eta_E$, resulting in $4$ independent parameters. One usually expresses these $4$ parameters in terms of the increase of the scale factor during various stages, namely, $\zeta_1\equiv a(\eta_s)/a(\eta_1)$, $\zeta_s\equiv a(\eta_2)/a(\eta_s)$, $\zeta_2\equiv a(\eta_E)/a(\eta_2)$, and $\zeta_E\equiv a(\eta_H)/a(\eta_E)$. For the accelerating stage in the simple $\Lambda$CDM model, one has $\zeta_E= 1+ z_E \sim 1.33$, where $z_E$ is the redshift when the accelerating stage begins. For the matter-dominated stage, one has $\zeta_2 = (1+z_{\text{eq}}) \zeta_{E}^{-1} \sim 2547$ with $z_{\text{eq}} = 3387$ \cite{collaboration2020planck} being the redshift at matter-radiation equality (see App.~\ref{app:A.3}). 

The increase of scale factor during the reheating and radiation-dominated stages, namely $\zeta_1$ and $\zeta_s$, generally depends on the reheating temperature $T_{\text{reh}}$ at which the radiation stage begins (see Appendix~\ref{app:A} for details). Big Bang Nucleosynthesis (BBN) and the energy scale at the end of inflation put lower and upper bounds on $T_{\text{reh}}$ as $T_{\text{reh}} \geq 10\,$ MeV and $T_{\text{reh}} \leq 10^{16}\,$GeV, respectively \cite{martin2010first}. However, the CMB data modify the lower bound on the reheating temperature $T_{\text{reh}} \geq 6 \times 10^3\,$ GeV \cite{martin2010first}, and gravitino generation has given the upper bound $T_{\text{reh}} \leq 10^7$ GeV \cite{bailly2009re}. As discussed in App.~\ref{app:A.3}, $T_{\text{reh}}$ does not play a significant role in the incoherence of the EM field, since it only changes the high-frequency part of the PGWs spectrum (see Fig.~\ref{fig1}). However, the value of $T_{\text{reh}}$ could affect the range of other parameters, including $\beta, \beta_s, \zeta_1$, when considering specific inflationary scenarios (see \cite{tong2013relic} for example). In this paper, we mainly adopt $T_{\text{reh}}=10^8\,$GeV for the sake of illustration. The specific $T_{\text{reh}}$-dependence of $\zeta_s$ is determined by Eq.~(\ref{a.3}), though $\zeta_1$ is usually regarded as a free parameter. So far, four parameters $(\beta, \beta_s, T_{\text{reh}}, \zeta_1)$ determine the expansion of the universe. The introduction of primordial perturbations adds extra degrees of freedom, namely the scalar and tensor power spectrum, $A_{s,T}$, and the scalar and tensor spectral indices, $n_{s,T}$, at some pivot scale $k_0$, among which we only treat $A_T$ (or equivalently the tensor-to-scalar ratio $\text{r}_{k_0}$) as a free parameter and use the current constraints on $A_{s}$ and $n_s$ made by \textit{Planck2018} observations (see Sec.~\ref{subsec:2.3}). 


\subsection{\label{subsec:2.2} PGWs amplification in expanding Universe}

Inflationary-generated PGWs originate from tensor perturbations of the FLRW metric during the inflationary era. Starting from the vacuum state, they have been amplified during the expansion of the universe. Quantum description of cosmological perturbations has been extensively studied in the literature~\cite{grishchuk1993quantum,martin2016quantum}. The super-adiabatic amplification leads vacuum tensor perturbations to evolve into a two-mode squeezed (TS) state with an enormously large number of gravitons~\cite{grishchuk2001relic, grishchuk1993quantum, martin2016quantum}.

Starting from the perturbed Hilbert-Einstein action in FLRW universe, one ends up with the following equations for the evolution of the squeezing parameters ($r_K,\phi_K$) (see ~\cite{mukhanov2005physical,martin2016quantum} for example)
\begin{eqnarray}\label{2.2}
\frac{\mathrm{d}r_K}{\mathrm{d}\eta} &=& \frac{a'}{a} \cos(2\phi_K),  \\
\frac{\mathrm{d}\phi_K}{\mathrm{d}\eta} &=& -K -\frac{a'}{a} \coth(2r_K) \sin(2\phi_K) \, , \nonumber
\end{eqnarray}
with $(r_K, \phi_K)$ being the squeezing amplitude and angle, respectively (note that we shall use the symbol $r_K$ to represent the squeezing amplitude of the PGWs mode $K$, while the symbol $\text{r}_{k_0}$ is reserved for the tensor-to-scalar ratio at the pivot scale $k_0$). 
It is most favored in the literature to define two functions ($u_K(\eta), v_K(\eta)$ to represent the squeezing parameters $(r_K, \phi_K)$ as follows \cite{grishchuk1993quantum,martin2016quantum}
\begin{eqnarray}\label{2.3}
u_{K}(\eta) &\equiv& e^{i\theta_K(\eta)} \cosh r_K(\eta), \nonumber\\
v_{K}(\eta) &\equiv& e^{-i\theta_K(\eta) + 2i\phi_K(\eta)} \sinh r_K(\eta),
\end{eqnarray}
which satisfy the normalization $|u_{K}(\eta)|^2 - |v_{K}(\eta)|^2 = 1$. Here, the parameter $\theta_K(\eta)$ is called the rotation parameter and is non-relevant as long as amplification of vacuum perturbations matters. 

In the following, we consider the approximate solution to the system of equations~(\ref{2.2}) in the super-Hubble regime, $K \ll 2\pi \mathcal{H}$, where $\mathcal{H} = \frac{a'}{a}$ stands for the comoving Hubble parameter in conformal time. In the super-Hubble regime, the mode $K$ has been amplified sufficiently and the squeezing amplitude is high so that $r_K \gg 1$. 

For super-Hubble scales, the wavelength is larger than the (comoving) Hubble parameter $\mathcal{H}$, so that $K \ll 2\pi \mathcal{H}$ and the squeezing amplitude $r_K$ is very large. Hence $\coth(2r_K) \rightarrow 1$ and the equation for $\phi_K$ yields the following solution
\begin{eqnarray}\label{2.4}
\phi'_K(\eta) &=& -\frac{a'}{a} \sin(2\phi_K) \quad\Rightarrow\quad \tan(\phi_k) \propto \frac{1}{a^2(\eta)},
\end{eqnarray}
which implies $\tan(\phi_K)\rightarrow 0$ during the long-wavelength regime. The squeezing angle $\phi_K$ tends to either value $0$ or $\pi$, both resulting in $\cos(2\phi_K)\rightarrow 1$. Consequently, the equation of motion for $r_K(\eta)$ reduces to $r'_K(\eta) = a'/a$ with the simple solution $r_K(\eta) = \ln a(\eta)/a_{\ast}(K)$, where $a_{\ast}(K) \equiv a(\eta_{\ast}(K))$ denotes the value of the scale factor at the initial moment of horizon-crossing, $\eta_{\ast}(K)$, when the long-wavelength regime initiates for a given mode $K$. The initial value for $r_{K}$ is chosen as $r_{K}(\eta_{\ast})=0$. If one denotes by $\eta_{\ast\ast}(K)$ the time when the mode leaves the long-wavelength regime and re-enters the Hubble horizon and $K\geq 2\pi\mathcal{H}$, the final value of the squeezing amplitude is determined by 
\begin{eqnarray}\label{2.5}
e^{r_K} = \frac{a_{\ast\ast}(K)}{a_{\ast}(K)},
\end{eqnarray}
where $a_{\ast\ast}(K) = a(\eta_{\ast\ast}(K))$ denotes the value of the scale factor when the mode re-enters the horizon (see \cite{grishchuk2001relic} for more details). 

Considering the model of successive expansion of the universe introduced in Sec.~\ref{subsec:2.1}, one may show that the squeezing factor at present, $e^{r_{K}(\eta_H)}$, is related to the present spectral amplitude $h(K,\eta_H)$ according to $e^{r_K}(\eta_H) = \frac{1}{8\sqrt{\pi}} \frac{\ell_{H}}{\ell_{\text{Pl}}} \left(\frac{K_{H}}{K}\right) h(K,\eta_H)$, and is given by (see App.~\ref{app:A.2})
\begin{eqnarray}\label{2.6}
e^{r_{K}(\eta_H)} = \frac{1}{8\sqrt{\pi}} \frac{\ell_{H}}{\ell_{\text{Pl}}} \left( \frac{K_{H}}{K} \right)
\begin{cases}
A \left( \frac{K}{K_{H}} \right)^{2+\beta} \quad,\quad K \leq K_E \\
A \left( \frac{K}{K_{H}} \right)^{\beta-1} \frac{1}{(1+z_E)^3} \quad,\quad K_{E} \leq K \leq K_H \\
A \left( \frac{K}{K_{H}} \right)^{\beta} \frac{1}{(1+z_E)^3} \quad,\quad K_{H} \leq K \leq K_2 \\
A \left( \frac{K}{K_{H}} \right)^{1+\beta} \left( \frac{K_H}{K_2} \right) \frac{1}{(1+z_E)^3} \quad,\quad K_{2} \leq K \leq K_s \\
A \left( \frac{K}{K_{H}} \right)^{1+\beta-\beta_s} \left( \frac{K_s}{K_H} \right) \left( \frac{K_H}{K_2} \right) \frac{1}{(1+z_E)^3} \quad,\quad K_{s} \leq K \leq K_1 \\
\end{cases}
\end{eqnarray}
where the characteristic wave numbers $K_E, K_H, K_2, K_s$ and $K_1$ are determined once the increase parameters $\zeta_1, \zeta_s, \zeta_2$ and $\zeta_E$ are fixed (see App.~\ref{app:A.3}). The coefficient $A$ determines the initial perturbation amplitude that can be assessed by theoretical or observational normalization conditions, as discussed below. 


\subsection{\label{subsec:2.3} Quantum normalization condition (QNC)}

In the CMB literature, it is convenient to express the contribution of tensor perturbations around the pivot scale $k_{0}$ in terms of the so-called tensor-to-scalar ratio, $\text{r}_{k_0} \equiv \frac{A_{T}(k_0)}{A_s(k_0)}$ with $A_{T,s}(k_0)$ being the tensor and scalar power spectrum at the pivot scale $k_0$, respectively.

Remind that throughout the paper, the pivot scale is exceptionally denoted by $k_0$ and the symbol $\text{r}_{k_0}$ is reserved for the tensor-to-scalar ratio, not to be confused with the squeezing amplitude $r_K$. We shall take $k_0= 0.002\,$Mpc$^{-1}$ in our calculations \cite{collaboration2020planck}.

The perturbation field $h_{ij}$ at the initial moment (that is to say, before exiting the horizon during the inflationary era) can be treated as a quantum field in its vacuum state, possessing half energy quanta $\frac{1}{2}\hbar\Omega_K$ in each mode $K$. This assumption imposes a criterion on the initial amplitude $A$, called the quantum normalization condition (QNC) \cite{grishchuk2001relic, tong2013using, zhang2005relic}, which translates into a condition on the tensor-to-scalar ratio $\text{r}_{k_0}$. As a result of QNC, it can be shown that one ends up with the following theoretical constraint between the previously discussed parameters ($\beta, \beta_s, T_{\text{reh}}, \zeta_1, \text{r}_{k_0}$) \cite{tong2013using},
\begin{eqnarray}\label{2.7}
A = \sqrt{A_s \text{r}_{k_0}} \left( \frac{K_H}{k_0} \right)^{\beta} \zeta_E^{\frac{2+\gamma}{\gamma}} = 8\sqrt{\pi} \left( \frac{\ell_{\text{Pl}}H_0}{c} \right) \lambda^{-(2+\beta)} \zeta_1^{\frac{\beta_s-\beta}{1+\beta_s}} \zeta_s^{-\beta} \zeta_2^{\frac{1-\beta}{2}} \zeta_E^{1+\frac{1+\beta}{\gamma}}
\end{eqnarray}
\begin{figure*}[htb]
\centering
\includegraphics[
width=1\columnwidth]{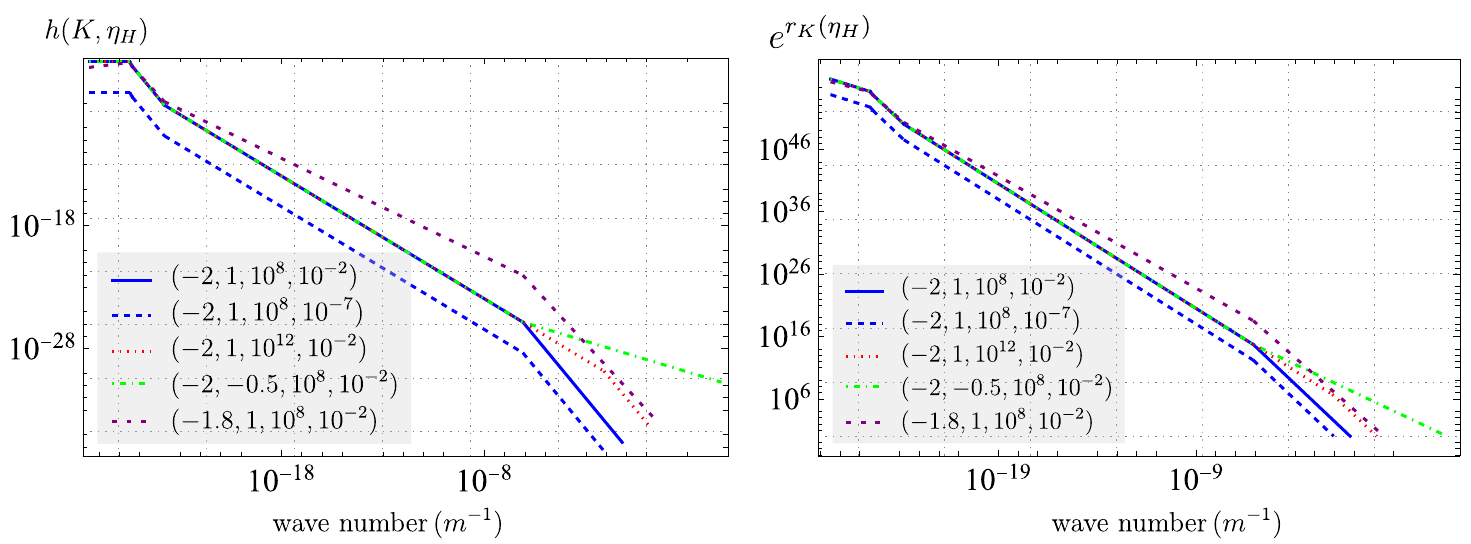}
\caption{\textit{Panel} (a): today PGWs spectral amplitude $h(K,\eta_H)$ and \textit{panel} (b): the corresponding squeezing factor $e^{r_{K}(\eta_H)}$, versus the physical wave number $0.01 K_{E} \leq K \leq K_{1}$ for different values of the parameters ($\beta, \beta_s, T_{\text{reh}}/\text{GeV}, \text{r}_{k_0}$).}
\label{fig1}
\end{figure*} 

Upper bounds $A_s \lesssim 2.1 \times 10^{-9}$ and $\text{r}_{k_0} < 0.056$ have been obtained by \textit{Planck2018} using TTTEEE + lowE + lensing + BAO~\cite{aghanim2020planck}.

Taking $\text{r}_{k_0}$ as a free parameter, one may convert Eq.~(\ref{2.7}) to express $\zeta_1$ in terms of the other parameters, namely

\begin{eqnarray}\label{2.8}
\zeta_1(\beta,\beta_s,\zeta_s,\text{r}_{k_0}) =\left[ \frac{c \big( A_{s}(k_0) \text{r}_{k_0} \big)^{1/2}}{8\sqrt{\pi}\ell_{\text{Pl}} H_0} \left( \frac{K_H}{k_0} \right)^{\beta} \gamma^{2+\beta} \zeta_s^{\beta} \zeta_2^{\frac{\beta-1}{2}} \zeta_E^{\frac{1-\beta}{\gamma}} \right]^{\frac{1+\beta_s}{\beta_s-\beta}},
\end{eqnarray}

and the expansionary model Eq.~(\ref{2.1}) is fully described by four independent parameters $(\beta, \beta_s, T_{\text{reh}}, \text{r}_{k_0})$. The spectral amplitude $h(K,\eta_H)$ and the corresponding squeezing factor $e^{r_K(\eta_H)}$ are shown in Fig~\ref{fig1} panel (a) and panel (b), respectively, for different values of the parameters $(\beta, \beta_s, T_{\text{reh}}, \text{r}_{k_0})$ (see App.~\ref{app:A.2}). It can be seen that changing the reheating parameters $\beta_s$ and $T_{\text{reh}}$ changes only the high-frequency part of the spectrum, i.e., $K_s < K < K_1$, which contains only a small amount of squeezing with respect to the ultralow-frequency part. On the other hand, changing $\beta$ and $\text{r}_{k_0}$ affects almost the whole frequency range of the spectrum.


\section{Quantum mechanical interaction between EM field and GWs}\label{sec:3}

\begin{figure*}[htb]
\centering
\includegraphics[
width=0.6\columnwidth]{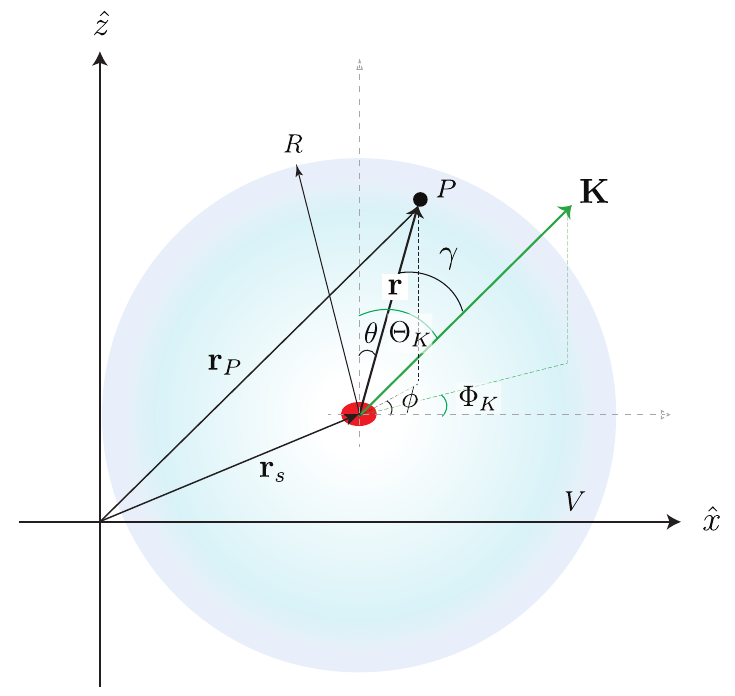}
\caption{Spherical wave emission from a point source and its interaction with GWs background. The interaction takes place in a sphere of radius $R$ and volume $V$. The vectors $\mathbf{r}=(r,\theta,\phi)$ and $\mathbf{K}=(K,\Theta_K,\Phi_K)$ are represented in spherical coordinates. The point $P$ shows a typical point within $V$ where the EM-GWs takes place. The vector $\mathbf{r}_s$ represents the location of the point source with respect to some coordinate system, and $\mathbf{r}$ shows the location of $P$ with respect to the point source.}
\label{fig2}
\end{figure*}

In this section, we present a self-contained investigation of the EM-GWs interaction at the quantum level. We consider the propagation of spherical waves in the presence of GW background, as schematically shown in Fig.~\ref{fig2}. The EM field is emitted from a point source located at $\mathbf{r}_s$ from the origin of the coordinate system and is going to be observed at a distance $R$ from it at a time instant $t$. We derive the Hamiltonian and solve the Heisenberg equation governing a single-mode EM field. This step helps us investigate Glauber's correlation functions of the EM field in the presence of such a quantum background in the next sections.

\subsection{Lagrangian of EM-GWs interaction}\label{subsec:3.1}

The imprint of GWs on light is comprehensively studied in the literature, either at the classical \cite{maggiore2007gravitational:book} or quantum level \cite{coradeschi2021can, anastopoulos2013master, lagouvardos2021gravitational, shojaei2023sensing}. 
Since we finally aim to investigate spatial correlations of distant emitters, it seems natural to consider a situation in which a single-mode spherical EM field, possessing wave number $k$ and frequency $\omega_k=ck$, propagates through the flat Minkowski spacetime and interacts with tensor perturbations. Basically, one could construct the EM-GWs Hamiltonian formalism in the FLRW setup. However, one already knows that the action of the EM field in the FLRW background is conformally invariant. Moreover, we proceed by the assumption that during the time interval that the EM field interacts with GWs, say the flight time $t$, the state of GWs is stationary. Especially, the spectrum of PGWs remains to be determined by $h(K,\eta_H)$ and the squeezing amplitude is going to be specified by Eq.~(\ref{2.7}). As a result, we may consider the interaction in the fabric of flat spacetime. The total Lagrangian of the EM-GWs system can be written as
\begin{eqnarray}\label{3.1}
\mathcal{L}_{\text{tot}} = \mathcal{L}^{(0)}_{\text{gw}} + \mathcal{L}^{(0)}_{\text{em}} + \mathcal{L}_{\text{int}},
\end{eqnarray}
The free Lagrangian of GWs is determined by
\begin{eqnarray}\label{3.2}
\mathcal{L}^{(0)}_{\text{gw}} &=& -\frac{c^4}{64\pi G} \Big( \frac{1}{c^2}(\dot{h}_{ij})^2 - (\partial_{\ell} h_{ij})^2 \Big)\, .
\end{eqnarray}
Quantization of free GWs is done by introduction of the quantum field $h_{ij}(\mathbf{r}_P,t)$ according to
\begin{eqnarray}\label{3.3}
\hat{h}_{ij}(\mathbf{r}_P,t) &=& \mathcal{C} \sum_{\lambda=+,\times} \int\frac{d^3\mathbf{K}}{(2\pi)^{3/2}} \frac{e^{\lambda}_{ij}[\hat{K}]} {\sqrt{2\Omega_{K}}} \Big( \hat{b}_{\mathbf{K},\lambda} e^{-i(\Omega_K t -\mathbf{K} \cdot \mathbf{r}_P )} + \hat{b}^{\dagger}_{\mathbf{K},\lambda} e^{i(\Omega_K t -\mathbf{K} \cdot \mathbf{r}_P)} \Big)\,.
\end{eqnarray}
Here, $\mathbf{r}_P$ represents a typical point in space wherein the EM-GWs interaction takes place (see Fig.~\ref{fig2}). In the above equation, $\mathcal{C}=\sqrt{16\pi c} \ell_{\text{Pl}} = \sqrt{16\pi c^3}(\hbar/E_{\text{Pl}})$, $(\hat{b}_{\mathbf{K},\lambda}, \hat{b}^{\dagger}_{\mathbf{K},\lambda})$ denote the bosonic operators of mode $(\mathbf{K}, \lambda)$ of tensor perturbations and $\mathbf{K} = K \hat{\mathbf{K}}$ denotes the conformal wave vector, $e_{ij}^{\lambda}[\hat{\mathbf{K}}]$ stands for the polarization tensor and $\lambda=+,\times$ represents the polarization state of tensor perturbations (remember that, throughout the paper, we shall use upper-case letters $K=(\Omega_K/c,\mathbf{K})$ to represent the four-momentum of GWs, while lower-case letters $k=(\omega/c,\mathbf{k})$ will be used to represent the four-momentum of the EM field). In the transverse-traceless (TT) gauge, the polarization tensor $e_{ij}^{\lambda}[\hat{\mathbf{K}}]$ fulfills the following conditions
\begin{equation}\label{3.4}
K^{i}\,  e_{ij}^{\lambda}[\hat{\mathbf{K}}] = 0\quad,\quad e^{\lambda}_{ii}[\hat{\mathbf{K}}]=0\quad\text{and}\quad
e_{ij}^{\lambda\ast}[\hat{\mathbf{K}}] \, e_{ij}^{\lambda'}[\hat{\mathbf{K}}] = 2\delta_{\lambda\lambda'}.
\end{equation}
Additionally, the two independent polarization states $\lambda=+,\times$ can be expressed in terms of two unit vectors $(\hat{\mathbf{n}}, \hat{\mathbf{m}})$ orthogonal to the propagation direction $\hat{\mathbf{K}}$ and to each other. In terms of the Euler's angles $(\Theta_{K}, \Phi_K)$, one may specify $(\hat{\mathbf{K}}, \hat{\mathbf{n}}, \hat{\mathbf{m}})$ by
\begin{eqnarray}\label{3.5}
    \hat{\mathbf{K}}= \begin{pmatrix} 
    \sin\Theta_K\cos\Phi_K\\
    \sin\Theta_K\sin\Phi_K \\ 
    \cos\Theta_K \\
	\end{pmatrix},\quad
    \hat{\mathbf{n}}=\begin{pmatrix} 
    \cos\Theta_K\cos\Phi_K \\
    \cos\Theta_K\sin\Phi_K\\ 
    -\sin\Theta_K \\
	\end{pmatrix},\quad
    \hat{\mathbf{m}}=\begin{pmatrix} 
    -\sin\Phi_K\\
    \cos\Phi_K \\ 
    0 \\
	\end{pmatrix}, 
\end{eqnarray}
and the polarization tensors are written as \cite{maggiore2007gravitational:book}
\begin{eqnarray}\label{3.6}
\begin{cases}
    \hat{e}_{ij}^{+}[\hat{\mathbf{K}}] =\hat{n}_{i}\hat{n}_{j}-\hat{m}_{i}\hat{m}_{j},\\
    \hat{e}_{ij}^{\times}[\hat{\mathbf{K}}] =\hat{m}_{i}\hat{n}_{j}+\hat{n}_{i}\hat{m}_{j}, 
  \end{cases} 
\end{eqnarray}
By defining the canonical conjugate of $h_{ij}$, the quantized Hamiltonian governing GWs is obtained as follows:
\begin{eqnarray}\label{3.7}
\hat{H}^{(0)}_{\text{gw}} &=& \hbar \sum_{\lambda=+,\times} \int d^3\mathbf{K} \, \Omega_K \, \hat{b}^{\dagger}_{\mathbf{K},\lambda}\hat{b}_{\mathbf{K},\lambda}\,.
\end{eqnarray}

On the other hand, free evolution of the EM field can be described by the Lagrangian
\begin{eqnarray}\label{3.8}
\mathcal{L}^{(0)}_{\text{em}} &=& -\frac{1}{4 \mu_0} F_{\mu\nu} F^{\mu\nu} = \frac{\epsilon_0}{2} |E|^2 - \frac{1}{2\mu_0} |B|^2 \, ,
\end{eqnarray}
where $F_{\mu\nu} = \partial_{\mu}A_{\nu} - \partial_{\nu}A_{\mu} $ is the electromagnetic tensor, $A_{\mu}$ is the vector potential, and the electric and magnetic fields are determined by $\mathbf{E} = - \partial_t \mathbf{A}$ and $\mathbf{B}= \nabla \times \mathbf{A}$.The interaction between EM and GWs is determined by the energy-momentum tensor of the EM field, $T^{\mu\nu}$ through the action $S_{\text{int}} = \frac{(32\pi G)^{3/2}}{2} \int d^4x \, h_{\mu\nu}T^{\mu\nu}$. In the TT gauge, one obtains
\begin{eqnarray}\label{3.9}
\mathcal{L}_{\text{int}} &=& -\frac{\varepsilon_0}{2}\, h_{ij} E^i E^j - \frac{1}{2 \mu_0} \, h_{ij} B^i B^j \,.
\end{eqnarray}
From Eq.~(\ref{3.8}), dynamics of the EM field in the presence of GWs can be followed. 


\subsection{Hamiltonian formalism for description of EM-GWs interaction}\label{subsec:3.2}

With the help of Eq.~(\ref{3.8}) and Eq.~(\ref{3.9}), the equation of motion governing the vector potential turns out as follows
\begin{eqnarray}\label{3.10}
\nabla\times\left[ \Bar{\Bar{\mu}}^{-1}(\mathbf{r}_P,t)\nabla\times \mathbf{A} (\mathbf{r}_P,t) \right] = - \Bar{\Bar{\varepsilon}}(\mathbf{r}_P,t) \, \partial^2_{t} \mathbf{A}(\mathbf{r}_P,t),
\end{eqnarray}
Here, the total contributions from the vacuum and GWs are combined and brought into the (spacetime-dependent) permittivity and permeability tensors $\bar{\bar{\varepsilon}}(\mathbf{r}_C,t)$ and $\bar{\bar{\mu}}(\mathbf{r}_C,t)$, defined by
\begin{eqnarray}\label{3.11}
\frac{\varepsilon_{ij}(\mathbf{r}_P,t)}{\varepsilon_0} = \frac{\mu_{ij}(\mathbf{r}_P,t)}{\mu_0} &=& \delta_{ij}-h_{ij}(\mathbf{r}_P,t) \\
&=& \delta_{ij}- \mathcal{C} \sum_{\lambda=+,\times} \int \frac{d^3\mathbf{K}}{(2\pi)^{3/2}} \frac{e^{\lambda} _{ij}[\hat{\mathbf{K}}]}{\sqrt{2\Omega_K}} \Big( h^{\lambda}_{\mathbf{K}} e^{-i (\Omega t-\mathbf{K}\cdot \mathbf{r}_P) } + h^{\lambda\ast}_{\mathbf{K}} e^{i \left( \Omega t-\mathbf{K}\cdot \mathbf{r}_P \right)} \Big) \nonumber,
\end{eqnarray}
In the second line, Eq.~(\ref{3.3}) is used for the expression of $h_{ij}(\mathbf{r}_P,t)$. For a general discussion, we have replaced the quantum ladder operators $\hat{b}_{\mathbf{K},\lambda}$ and $\hat{b}^{\dagger}_{\mathbf{K},\lambda}$ with classical field amplitudes $h^{\lambda}_{\mathbf{K}}$ and $h^{\lambda\ast}_{\mathbf{K}}$. However, the following arguments and results are unchanged, whether $h_{ij}(\mathbf{r}_P,t)$ is treated as a classical or quantum entity.

Note that the specification Eq.~(\ref{3.10}) is compatible with the optical medium analogy (OMA) framework, which is used to mimic the effect of GWs background with a magneto-dielectric medium~\cite{shojaei2023sensing}. As we shall see, this analogy helps us to find the mode solutions of $\mathbf{A}(\mathbf{r}_P,t)$ in a way similar to what is already established for the EM field in the presence of a magneto-dielectric media.

In order to solve Eq.~(\ref{3.10}), we proceed by considering a spherical wave expansion of the vector potential $\mathbf{A}(\mathbf{r}_P,t)$ as follows:
\begin{eqnarray} \label{3.12}
\mathbf{A}(\mathbf{r}_P,t)= \sum_{k}\sqrt{\frac{\hbar}{2}} \Big( \alpha_{k}\, \nu_{k}(\mathbf{r}_P,t) \, \mathbf{f}_{k}(\mathbf{r}_P,t) + \alpha_{k}^{\ast} \, \nu_{k}^{\ast}(\mathbf{r}_P,t) \, \mathbf{f}^{\ast}_{k}(\mathbf{r}_P,t) \Big).
\end{eqnarray}
Here, $k$ denotes different modes of the EM field possessing different wave numbers. Although the quantum pre-factor $\hbar$ is included intentionally, at the classical level one can substitute $\alpha_{k}\sqrt{\hbar}\rightarrow \mathcal{A}_{k}$. The time-harmonic behavior of $\mathbf{A}(\mathbf{r}_P,t)$ is included in the temporal mode function $\nu_{k}(\mathbf{r}_P,t)$ which recasts to $e^{\pm i\omega_{k}t}$ in the vacuum case, i.e., when there is no gravitational background. However, we let this function be $\mathbf{r}_P$-dependent as well, since the gravitational background is $\mathbf{r}_P$-dependent through $h_{ij}(\mathbf{r}_P,t)$. That means $\nu_k(\mathbf{r}_P,t)$ is a slowly varying function of $\mathbf{r}_P$. The spatial mode function $\mathbf{f}_{k}(\mathbf{r}_P,t)$ contains the spatial factor, which recasts to the spherical waves in the absence of GWs, and again we let it have a slowly varying envelope as well since the gravitational medium is time-dependent. Thus we may consider 
\begin{eqnarray} \label{3.13}
\mathbf{f}_k(\mathbf{r},t) \rightarrow \mathbf{f}_k(\mathbf{r}_P,t) \equiv \, \hat{\mathbf{u}}_{k}(\theta,\phi) f(t) \,\Big( \frac{e^{i k r}}{r} \Big),
\end{eqnarray}
with $f(t)$ being a slowly varying function of time and $\hat{\mathbf{u}}_{k}(\theta,\phi)$ a unit vector denoting the polarization of the electric field, which is orthogonal to $\mathbf{r}$ at each point $P$. Note that $r$ is the distance between point $P$ and the source element located at $\mathbf{r}$, as depicted in Fig.~\ref{fig2}. Here, the position vector $\mathbf{r}=(r,\theta,\phi)$ shows the location at which one monitors the EM field in spherical coordinates. In the following, we take a constant unit vector $\hat{\mathbf{u}}_k$ and neglect its variation in the vicinity of the observation point $\mathbf{r}_P$ (this assumption holds true for observation points much further than the source). The slowly varying envelope approximation (SVEA) implies that \cite{schleich1984general}
\begin{eqnarray} \label{3.14}
\left|\frac{\ddot{\mathbf{f}}_{k}(\mathbf{r},t)}{\dot{\mathbf{f}}_{k}(\mathbf{r},t)}\right|\ll \left|\frac{\dot{\mathbf{f}}_{k}(\mathbf{r},t)}{\mathbf{f}_{k}(\mathbf{r},t)}\right|\ll \omega_{k} \quad, \quad \left| \frac{\nabla^2 \nu_k(\mathbf{r}_P,t)}{\nabla \nu_k(\mathbf{r}_P,t)} \right| \ll \left| \frac{\nabla \nu_k(\mathbf{r}_P,t)}{\nu_k(\mathbf{r}_P,t)} \right| \ll k.
\end{eqnarray}
Note that since $\mathbf{r}_P=\mathbf{r}_s + \mathbf{r}$, derivatives with respect to $\mathbf{r}_P$ and $\mathbf{r}$ can be interchanged. Eq.~(\ref{3.14}) means that temporal change of the spatial mode function due to the existence of GWs happens on a time scale much larger than the characteristic oscillatory behavior of the EM wave. Similarly, spatial change of the temporal mode functions induced by the GW medium happens at scales much larger than the typical wavelength of the EM field. Inserting $\mathbf{f}_k(\mathbf{r},t)$ into the field expansion Eq.~(\ref{3.12}) and using the wave equation Eq.~(\ref{3.10}), one can show that
\begin{eqnarray} \label{3.15}
\hspace*{-1.2cm}
\nabla\times \Big( \Bar{\Bar{\mu}}^{-1}(\mathbf{r}_P,t) \nabla\times \mathbf{f}_k(\mathbf{r},t) \Big) &=& - k^2 \, f(t) \, \Big( \frac{e^{ikr}}{r} \Big) \hat{r} \, \times \Big[ \Bar{\Bar{\mu}}^{-1} (\mathbf{r}_P,t)(\hat{\mathbf{u}}_{k} \times \hat{r} ) \Big] + \mathcal{O}(\frac{1}{r^2}) \nonumber\\
&=& - \Big( \frac{\ddot{\nu}_k(\mathbf{r}_P,t)}{{\nu}_k(\mathbf{r}_P,t)} \Big) \, \Bar{\Bar{\varepsilon}}(\mathbf{r}_P,t) \, \hat{\mathbf{u}}_{k} \, f(t) \, \Big( \frac{e^{i k r}}{r} \Big) \,.
\end{eqnarray}
Here, $\hat{r}$ shows the unit vector from the source to the point $P$. In the derivation of Eq.~(\ref{3.15}), we have neglected the spatial derivative of $\Bar{\Bar{\mu}}(\mathbf{r}_P,t)$ and $\nu_k(\mathbf{r}_P,t)$, as well as the time derivative of $\mathbf{f}_k(\mathbf{r},t)$ under the adiabatic approximation $K \ll k$ and $\Omega_K \ll \omega_k$, implied by Eq.~(\ref{3.14}). Moreover, one can neglect higher-order terms $\propto \mathcal{O}(\frac{1}{r^2})$ that vanish fast at large distances. Thus, Eq.~(\ref{3.15}) contains the second-order spatial derivative of $\mathbf{f}_k(\mathbf{r},t)$ on the left-hand side and the second-order time derivative of $\nu_k(\mathbf{r}_P,t)$ on the right-hand side. To solve Eq.~(\ref{3.15}), we proceed by assuming the following time behavior for the temporal mode functions:
\begin{eqnarray}\label{3.16}
\ddot{\nu}_{k}(\mathbf{r}_P,t) \equiv -\omega_{k}^2(\mathbf{r}_P,t) \nu_{k}(\mathbf{r}_P,t).
\end{eqnarray}
Firstly, note that the $\mathbf{r}_P$-dependence of temporal mode function $\nu_{k}(\mathbf{r}_P,t)$ must satisfy the SVEA condition Eq.~(\ref{3.14}). In the following, we shall see that this is indeed the case. On the other hand, in the proposed form Eq.~(\ref{3.16}) a friction term is absent in the time evolution of the EM field, while its frequency may be affected by the underlying GWs background. This assumption intuitively supports the fact that the GW background does not produce or annihilate photons under the adiabatic approximation $\Omega_K \ll \omega_k$.
This means that the GW medium is not dispersive and absorptive in the optical range since the EM-GW interaction is off-resonant. In the corresponding quantum field theory, this assumption implies that EM-GW interaction does not create or annihilate photons, and the mean number of photons in the EM field is conserved. This point was first noted by Kip Thorne \cite{thorne2000gravitation} in the case of a slowly varying GW background.

Now, inserting Eq.~(\ref{3.16}) into Eq.~(\ref{3.15}) yields
\begin{eqnarray} \label{3.17}
\left[ \omega_k^2(\mathbf{r}_P,t) \varepsilon_{il}(\mathbf{r}_P,t) + k^2 \, \epsilon_{ijr} \, \epsilon_{mnl} \, \hat{r}_{j} \hat{r}_{n} \, \mu^{-1}_{rm}(\mathbf{r}_P,t) \right] \hat{\mathbf{u}}_{k,l} = 0,
\end{eqnarray}
with $\hat{\mathbf{u}}_{k,l}$ being the $l$-th component of the polarization vector $\hat{\mathbf{u}}_{k}$ and $\epsilon_{ijr}$ representing the Levi-Civita symbol. Now, if we define a new quantity 
\begin{eqnarray} \label{3.18}
n_k(\mathbf{r}_P,t) \equiv \frac{\omega_{k}}{\omega_k(\mathbf{r}_P,t)} \, ,
\end{eqnarray}
where $\omega_{k} = c \, k$ is the EM frequency in the vacuum (in the absence of GWs), combining Eq.~(\ref{3.17}) and Eq.~(\ref{3.18}) yields
\begin{eqnarray} \label{3.19}
\left[ \varepsilon_{il}(\mathbf{r}_P,t) + \frac{n_k^2(\mathbf{r}_P,t)}{c^2}\, \epsilon_{ijr}\, \epsilon_{mnl} \, \hat{r}_{j} \, \hat{r}_{n} \, \mu^{-1}_{rm}(\mathbf{r}_P,t) \right] \hat{\mathbf{u}}_{k,l} = 0 \, .
\end{eqnarray}
Eq.~(\ref{3.19}) is also obtained in \cite{bini2013light} for plane waves, where it turns out that the quantity $n_k(\mathbf{r}_P,t)$ is nothing but the refractive index of the GWs medium. The existence of eigen-solutions for $\hat{\mathbf{u}}_{\mathbf{k}}$ leads to the \textit{generalized Fresnel equation}:
\begin{eqnarray} \label{3.20}
n_k^2(\mathbf{r}_P,t)\, \varepsilon_{ij}(\mathbf{r}_P,t) \,\hat{r}_{i} \, \hat{r}_{j} - \text{det} \Big(\bar{\bar{\varepsilon}}(\mathbf{r}_P,t) \Big) = 0\,.
\end{eqnarray}
Eq.~(\ref{3.20}) identifies the refractive index of the effective medium according to
\begin{eqnarray} \label{3.21}
n_k(\mathbf{r}_P,t) =\sqrt{ \frac{\text{det} \Big( \bar{\bar{\varepsilon}}(\mathbf{r}_P,t) \Big)}{\varepsilon_{ij}(\mathbf{r}_P,t) \, \hat{r}_{i} \, \hat{r}_{j}}},
\end{eqnarray}
which is consistent with Eq.~(14) of \cite{bini2013light}. With the help of Eq.~(\ref{3.11}), one can check that $\text{det}\big(\bar{\bar{\varepsilon}}(\mathbf{r}_P,t)\big)=1$ in the linear order $\mathcal{O}(h)$, and Eq.~(\ref{3.19}) simplifies to
\begin{eqnarray} \label{3.22}
\hspace*{-1cm} n_k(\mathbf{r}_P,t) &=& \Big( \varepsilon_{ij}(\mathbf{r}_P,t) \, \hat{r}_{i} \, \hat{r}_{j} \Big)^{-1/2} = \Big( \delta_{ij} \,\hat{r}_{i}\, \hat{r}_{j} - h_{ij}(\mathbf{r}_P,t) \,\hat{r}_{i}\, \hat{r}_{j} \Big)^{-1/2} \nonumber\\
&=& 1 + \frac{1}{2} h_{ij}(\mathbf{r}_P,t) \, \hat{r}_{i} \,\hat{r}_{j} \,. 
\end{eqnarray}
Consequently, the EM frequency in the presence of GW background is determined by Eq.~(\ref{3.18}), namely
\begin{eqnarray}\label{3.23}
\omega_k (\mathbf{r}_P,t) = \omega_{k} \Big( \delta_{ij} - \frac{1}{2} h_{ij}(\mathbf{r}_P,t) \Big) \,\hat{r}_{i} \, \hat{r}_{j} \, ,
\end{eqnarray}
and the temporal mode functions $\nu_{k}(\mathbf{r}_P,t)$ are determined by Eq.~(\ref{3.16}).
Having identified the expression of $\mathbf{A}(\mathbf{r}_P,t)$ given by Eq.~(\ref{3.12}), in the following we construct the Hamiltonian of the EM-GWs system. Before doing so, it is useful to specify the orthogonality relation of the spatial mode functions $\mathbf{f}_{k}(\mathbf{r},t)$.


\subsection{Orthogonality relation of the mode functions $\mathbf{f}_{\mathbf{k}}(\mathbf{r},t)$} \label{subsec:3.3}

With the help of the optical medium analogy, it is convenient to define the orthogonality equation of the spatial mode functions $\mathbf{f}_{k}(\mathbf{r},t)$ in the presence of anisotropic media according to
\begin{eqnarray} \label{3.24}
\langle\, \mathbf{f}_k | \, \mathbf{f}_{k'} \, \rangle &\equiv& \int_{V} d^3 \mathbf{r} \, \varepsilon_{ij} (\mathbf{r}_P,t) f^{\ast}_{k,i}(\mathbf{r},t) \, f_{k',j}(\mathbf{r},t) \equiv \delta_{k,k'}.
\end{eqnarray}
(see \cite{shojaei2023sensing} and references therein). Here, $f_{k,i}(\mathbf{r},t)$ denotes the $i$-th component of the spatial mode function $\mathbf{f}_{k}(\mathbf{r},t)$. Note that since the medium is spacetime dependent, in general, the integration in Eq.~(\ref{3.24}) may not produce the Kronecker delta function $\delta_{k,k'}$. However, as we shall see, the orthonormality relation Eq.~(\ref{3.24}) holds in the adiabatic limit $K \ll k$. Firstly note that our EM field of interest could have a frequency range of radio to optical waves, which span from $\sim$ GHz to $10^{15}\,$Hz so that $10 \leq k \leq 10^7$ m$^{-1}$. On the other hand, PGWs span a wide range of frequencies $10^{-18}\leq \Omega_K \leq 10^{10}\,$Hz, so that $10^{-26} \leq K \leq 10^2 \,$ m$^{-1}$. Thus, the adiabatic condition $K \ll k$ is almost fulfilled for PGWs and radio waves. Moreover, the high-frequency part of PGWs has a negligible graviton content, and their effect on the EM observables is automatically vanishing due to small coupling strength (see Sec.~\ref{subsec:5.2} and Fig.~\ref{fig7}). Inserting the expression of $\mathbf{f}_{k}(\mathbf{r},t) = f(t) \, \hat{\mathbf{u}}_{k} \, \big( \frac{e^{i k r}}{r} \big) $ into Eq.~(\ref{3.24}), and with the use of Eq.~(\ref{3.11}), one obtains
\begin{eqnarray}\label{3.25}
\langle\, \mathbf{f}_k | \, \mathbf{f}_{k'} \, \rangle &=& \varepsilon_0 f^2(t) \int_V d^3 \mathbf{r} \, \big( \delta_{ij} - h_{ij}(\mathbf{r}_P,t) \big) \hat{\mathbf{u}}_{k,i} \hat{\mathbf{u}}_{k',j} \, \Big( \frac{e^{-i (k - k' ) r}}{r^2}\Big) \\
&=& \varepsilon_0 f^2(t) \int_V d^3 \mathbf{r}\Big( \frac{e^{-i (k - k' ) r}}{r^2} \Big) \delta_{ij} \, \hat{\mathbf{u}}_{k,i} \, \hat{\mathbf{u}}_{k',j} \nonumber\\
&-& \varepsilon_0 f^2(t) \, \mathcal{C} \sum_{\lambda=+,\times} \int \frac{d^3\mathbf{K}}{(2\pi)^{3/2}} \, \bigg\{ \frac{h_{\mathbf{K}}^{\lambda} e^{-i\big( \Omega t - \mathbf{K}\cdot \mathbf{r}_s \big)}}{\sqrt{2\Omega_K}} \int_V d^3 \mathbf{r} \, \Big( e_{ij}[\hat{\mathbf{K}}] \hat{\mathbf{u}}_{k,i} \, \hat{\mathbf{u}}_{k',j} \Big)\, \frac{e^{-i( k - k' - K \cos\gamma) r}}{r^2} \nonumber\\
&+& \frac{h_{\mathbf{K}}^{\lambda\ast} e^{i \big(\Omega t - \mathbf{K} \cdot \mathbf{r}_s \big)}}{\sqrt{2\Omega_K}} \int_V d^3 \mathbf{r} \Big( e_{ij}[\hat{\mathbf{K}}] \hat{\mathbf{u}}_{k,i} \, \hat{\mathbf{u}}_{k',j} \Big) \, \frac{e^{-i(k -k' + K \cos\gamma) r}}{r^2} \bigg\}. \nonumber
\end{eqnarray}
Here, we replaced $\mathbf{K} \cdot \mathbf{r} = K r \cos\gamma$, where $\gamma$ denotes the angle between $\mathbf{K}$ and $\mathbf{r}$. The radial integration can be performed using 
\begin{eqnarray}\label{3.26}
\text{lim}_{R\rightarrow \infty} \int_0^{R} dr \, r^2  \Big( \frac{e^{-i(k-k')r}}{r^2} \Big) = R \, \delta_{k-k',0} \,.
\end{eqnarray}
Thus, Eq.~(\ref{3.25}) becomes
\begin{eqnarray}\label{3.27}
\langle\, \mathbf{f}_k | \, \mathbf{f}_{k'} \, \rangle &=& \varepsilon_0 f^2(t) \int d^2 \Omega_{\hat{r}} \Big( \delta_{ij} \, \hat{\mathbf{u}}_{k,i} \, \hat{\mathbf{u}}_{k',j} \Big) \, R \, \delta_{k-k',0} \\
&-& \varepsilon_0 f^2(t) \, \mathcal{C} \sum_{\lambda=+,\times} \int \frac{d^3\mathbf{K}}{(2\pi)^{3/2}} \, \bigg\{  \frac{h_{\mathbf{K}}^{\lambda} e^{-i\big( \Omega t -\mathbf{K} \cdot \mathbf{r}_s \big)}  }{\sqrt{2\Omega_K}} \int d^2 \Omega_{\hat{r}} \, \Big( e_{ij}^{\lambda}[\hat{\mathbf{K}}] \hat{\mathbf{u}}_{k,i} \, \hat{\mathbf{u}}_{k',j} \Big) \, R \, \delta_{k-k',K\cos\gamma} \nonumber\\
&+& \frac{h_{\mathbf{K}}^{\lambda\ast} e^{i\big( \Omega t -\mathbf{K} \cdot \mathbf{r}_s \big)}}{\sqrt{2\Omega_K}} \int d^2 \Omega_{\hat{r}} \, \Big( e_{ij}^{\lambda}[\hat{\mathbf{K}}] \hat{\mathbf{u}}_{k,i} \, \hat{\mathbf{u}}_{k',j} \Big) \, R \, \delta_{k-k',-K\cos\gamma} \bigg\}. \nonumber
\end{eqnarray}
Here, the symbol $d^2 \Omega_{\hat{r}} = d(\cos\theta)\, d\phi$ stands for the spatial angular element. Thus, Eq.~(\ref{3.27}) explicitly shows the scattering of mode $k$ to $k'$ of the EM field due to the presence of GWs with momentum $K$. Thus, the momentum conservation implies either possibility $k' = k - K\cos\gamma$ or $k' = k + K\cos\gamma$ . Since $K\ll k$, each mode of the EM field scatters to itself, and we may set
\begin{eqnarray}\label{3.28}
\delta_{k-k', K\cos\gamma} &\simeq& \delta_{k-k'} \nonumber\\
\delta_{k-k', - K\cos\gamma} &\simeq& \delta_{k-k'}\,. 
\end{eqnarray}
\noindent Indeed, these conditions are fulfilled by almost all frequency parts of the PGWs spectrum, for which $K \ll k,k'$. For all GWs with $K\ll k$, the incoming EM field of mode $k$ scatters to itself so that $k \simeq k'$. So, we may proceed by retaining the contribution of all PGW modes in the integral Eq.~(\ref{3.27}), such that
\begin{eqnarray}\label{3.29}
\langle\, \mathbf{f}_k | \, \mathbf{f}_{k'} \, \rangle &=& \varepsilon_0 f^2(t) \, R \, \delta_{k,k'} \int d^2 \Omega_{\hat{r}} \Big( \delta_{ij} -h_{ij}(\mathbf{r}_s,t) \Big)  \, \hat{\mathbf{u}}_{k,i} \, \hat{\mathbf{u}}_{k',j} \\
&\equiv& \varepsilon_0 f^2(t) \, R \, \delta_{k,k'} \, m(t) \, , \nonumber
\end{eqnarray}
where the slowly varying function $m(t)$ is defined by
\begin{eqnarray}\label{3.30}
m(t) \equiv \int d^2 \Omega_{\hat{r}} \Big( \delta_{ij} -h_{ij}(\mathbf{r}_s, t) \Big) \hat{\mathbf{u}}_{k,i} \, \hat{\mathbf{u}}_{k,j} \, .
\end{eqnarray}
Comparing Eq.~(\ref{3.29}) with definition Eq.~(\ref{3.24}), the normalization factor $f(t)$ outcomes
\begin{eqnarray}\label{3.31}
f^2(t) = \frac{1}{R \, m(t)\varepsilon_0} \, ,
\end{eqnarray}
\noindent and spatial mode functions are determined by
\begin{eqnarray}\label{3.32}
\mathbf{f}_{k}(\mathbf{r},t) = \frac{\hat{\mathbf{u}}_{k}(\theta,\phi)}{\sqrt{R \, m(t) \varepsilon_0}} \, \Big( \frac{e^{i k r}}{r} \Big) \,.
\end{eqnarray}


\subsection{Derivation of the interaction Hamiltonian $\hat{H}_{\text{em-gw}}$ in the adiabatic limit} \label{subsec:3.4}

As we have discussed in Sec.~\ref{subsec:3.1}, the following Lagrangian density
\begin{eqnarray}\label{3.33}
\hspace*{-1.5cm}\mathcal{L}_{\text{em-gw}}(\mathbf{r}_P,t) &=& \mathcal{L}^{(0)}_{\text{em}} + \mathcal{L}_{\text{int}} \\
&=& \frac{1}{2}\Big( \varepsilon_{ij} (\mathbf{r}_P,t) E^{i} (\mathbf{r}_P,t) E^{j}(\mathbf{r}_P,t)
- \mu^{-1}_{ij}(\mathbf{r}_P,t) B^{i} (\mathbf{r}_P,t) B^{j}(\mathbf{r}_P,t)  \Big)\,. \nonumber
\end{eqnarray}
correctly describes the dynamics of the EM field in the presence of GWs, where the permittivity and permeability tensors are identified by Eq.~(\ref{3.11}). Following the standard second quantization procedure, one promotes the vector potential $\mathbf{A}(\mathbf{r}_P,t)$ to the Hermitian quantum field operator $\hat{\mathbf{A}}(\mathbf{r}_P,t)$. Thus, the expansion coefficients $(\alpha_k,\alpha^{\ast}_k)$ in the field expansion Eq.~(\ref{3.12}) become bosonic operators $(\hat{a}_k,\hat{a}^{\dagger}_k)$ satisfying the bosonic commutation relation $[\hat{a}_{k},\hat{a}^{\dagger}_k]=\delta_{k,k'}$. The canonical conjugate momentum of the canonical variable $\hat{A}_{i}(\mathbf{r}_P,t)$ is defined by
\begin{eqnarray}\label{3.34}
\hat{\Pi}_{i}(\mathbf{r}_P,t) &\equiv& \frac{\partial\mathcal{L}_{\text{em-gw}}(\mathbf{r}_P,t )}{\partial \dot{\hat{A}}_{i}(\mathbf{r}_P,t)} = \varepsilon_{ij}(\mathbf{r}_P,t) \dot{\hat{A}}_{j} (\mathbf{r}_P,t) \\
&=& \sum_{k} \sqrt{\frac{\hbar}{2}} \left( \hat{a}_{k} \, \dot{\nu}_{k}(\mathbf{r}_P,t) \, \varepsilon_{ij}(\mathbf{r}_P,t) \, f_{k,j}(\mathbf{r},t) + \hat{a}^{\dagger}_{k} \, \dot{\nu}^{\ast}_{k}(\mathbf{r}_P,t) \, \varepsilon _{ij}(\mathbf{r}_P,t) \, f^{\ast}_{k,j}(\mathbf{r},t) \right) \, . \nonumber
\end{eqnarray}
Note that $\nu_k(\mathbf{r}_P,t)$ and $\mathbf{f}_{k}(\mathbf{r},t)$ satisfy the SVEA condition Eq.~(\ref{3.14}). The Hamiltonian can be established by performing the Legendre transform on the Lagrangian $\mathcal{L}_{\text{em-gw}}(\mathbf{r}_P,t)$. The resulting Hamiltonian is
\begin{eqnarray}\label{3.35}
\hat{H}_{\text{em-gw}} &=& \frac{1}{2}\int_{V} d^3\mathbf{r} \, \Big[ \varepsilon_{ij}^{-1} (\mathbf{r}_P,t) \, \hat{\Pi}_{i}(\mathbf{r}_P,t) \, \hat{\Pi}_{j}(\mathbf{r}_P,t) \nonumber\\
&+& \mu_{ij}^{-1}(\mathbf{r}_P,t) \Big(\nabla\times \hat{\mathbf{A}}(\mathbf{r}_P,t)\Big)_{i} \Big(\nabla\times \hat{\mathbf{A}}(\mathbf{r}_P,t)\Big)_{j} \Big],
\end{eqnarray}
where the identity $\varepsilon_{ik}\varepsilon_{kl} ^{-1}=\delta_{il}$ is used. Here, spatial integration $\int_V d^3\mathbf{r}$ is performed over the volume $V$ wherein the EM-GWs interaction takes place. Plugging the expressions of $\hat{\mathbf{A}}(\mathbf{r}_P,t)$ and $\hat{\Pi}(\mathbf{r}_P,t)$ from Eqs.~(\ref{3.12}, \ref{3.34}) into the Hamiltonian Eq.~(\ref{3.35}) yields
\begin{eqnarray}\label{3.36}
\hat{H}_{\text{em-gw}} &=& \frac{1}{2} \int_{V} d^3\mathbf{r} \, \Big[ \varepsilon^{-1}_{ij}(\mathbf{r}_P,t) \, \hat{\Pi}_{i}(\mathbf{r}_P,t) \, \hat{\Pi}_{j}(\mathbf{r}_P,t) \\
&+& \mu^{-1}_{ij} (\mathbf{r}_P,t) \Big( \nabla \times \hat{A}(\mathbf{r}_P,t) \Big)_{i} \Big( \nabla \times \hat{A}(\mathbf{r}_P,t) \Big)_{j} \Big] \nonumber\\
&=& \frac{1}{2} \times \frac{\hbar}{2} \sum_{k,k'} \bigg\{ \int_V d^3 \mathbf{r} \, \varepsilon_{ij}(\mathbf{r}_P,t) \Big( \hat{a}_{k} \dot{\nu}_{k} \, f_{k,i} + \hat{a}^{\dagger}_{k} \dot{\nu}^{\ast}_{k} \, f^{\ast}_{k,i} \Big) \Big( \hat{a}_{k'} \dot{\nu}_{k'} f_{k',j} + \hat{a}^{\dagger}_{k'} \dot{\nu}^{\ast}_{k'} f^{\ast}_{k',j} \Big) \nonumber\\
&+& \int_V d^3\mathbf{r} \, \mu^{-1}_{ij}(\mathbf{r}_P,t) \Big( \hat{a}_{k} \nu_{k} (\nabla\times \mathbf{f}_{k})_i + \hat{a}^{\dagger}_{k} \nu^{\ast}_{k} (\nabla \times \mathbf{f}^{\ast}_{k})_i \Big) \nonumber\\
&\times& \Big( \hat{a}_{k'} \nu_{k'} (\nabla \times \mathbf{f}_{k'} )_j + \hat{a}^{\dagger}_{k'} \nu^{\ast}_{k'} (\nabla \times \mathbf{f}^{\ast}_{k'})_j \Big) \bigg\}. \nonumber
\end{eqnarray}
Re-arranging the terms, there would be four terms in the Hamiltonian, as follows
\begin{eqnarray}\label{3.37}
\hspace*{-0.5cm} \hat{H}_{em-gw}(t)&=& \frac{1}{2} \times \frac{\hbar}{2} \sum_{k,k'} \\
&\times& \bigg\{ \hat{a}_{k} \hat{a}_{k'} \int_V d^3 \mathbf{r} \Big( \varepsilon_{ij}(\mathbf{r}_P,t) f_{k,i} f_{k',j} \dot{\nu}_k \dot{\nu}_{k'} + \mu^{-1}_{ij}(\mathbf{r}_P,t) \nu_k \nu_{k'} (\nabla \times \mathbf{f}_{k})_i (\nabla \times \mathbf{f}_{k'})_j \Big) \nonumber\\
&+& \hat{a}_{k} \hat{a}^{\dagger}_{k'} \int_V d^3 \mathbf{r} \Big( \varepsilon_{ij}(\mathbf{r}_P,t) f_{k,i} f^{\ast}_{k',j} \dot{\nu}_k \dot{\nu}^{\ast}_{k'} + \mu^{-1}_{ij}(\mathbf{r}_P,t) \nu_k \nu^{\ast}_{k'} (\nabla \times \mathbf{f}_{k})_i (\nabla \times \mathbf{f}^{\ast}_{k'})_j \Big) \nonumber\\
&+& \hat{a}^{\dagger}_{k} \hat{a}_{k'} \int_V d^3 \mathbf{r} \Big( \varepsilon_{ij}(\mathbf{r}_P,t) f^{\ast}_{k,i} f_{k',j} \dot{\nu}^{\ast}_k \dot{\nu}_{k'} + \mu^{-1}_{ij}(\mathbf{r}_P,t) \nu^{\ast}_k \nu_{k'} (\nabla \times \mathbf{f}^{\ast}_{k})_i (\nabla \times \mathbf{f}_{k'})_j \Big) \nonumber\\
&+& \hat{a}^{\dagger}_{k} \hat{a}^{\dagger}_{k'} \int_V d^3 \mathbf{r} \Big( \varepsilon_{ij}(\mathbf{r}_P,t) f^{\ast}_{k,i} f^{\ast}_{k',j} \dot{\nu}^{\ast}_k \dot{\nu}^{\ast}_{k'} + \mu^{-1}_{ij}(\mathbf{r}_P,t) \nu^{\ast}_k \nu^{\ast}_{k'} (\nabla \times \mathbf{f}^{\ast}_{k})_i (\nabla \times \mathbf{f}^{\ast}_{k'})_j \Big)\, .\nonumber
\bigg\}.
\end{eqnarray}
Note that the spherical EM wave emitted from a point source located on the surface of a star propagates into a sphere with radius $R$ until it is received on the Earth, and $R$ is the distance of the source to the Earth. Then the interaction with GWs takes place within the volume encompassed in the half-sphere of radius $R$. The resulting Hamiltonian thus depends on the interaction time $t$ as well as on the distance $R$, which is indeed the characteristic length of the probe. In the following, we compute each term in Eq.~(\ref{3.27}) using the adiabatic approximation $K \ll k$ and derive the explicit form of the interaction Hamiltonian.

With the help of definition Eq.~(\ref{3.11}) for $\varepsilon_{ij}(\mathbf{r}_P,t)$, the first term in Eq.~(\ref{3.37}) contains a term as 
\begin{eqnarray}\label{3.38}
&& \int_V d^3 \mathbf{r} \, \varepsilon_{ij}(\mathbf{r}_P,t) \, f_{k,i}(\mathbf{r},t) \, f_{k',j}(\mathbf{r},t) \dot{\nu}_k(\mathbf{r}_P,t) \dot{\nu}_{k'}(\mathbf{r}_P,t) \\
&=& \frac{1}{R\, m(t)} \int_V d^3\mathbf{r} \, \Big(\delta_{ij} \hat{\mathbf{u}}_{k,i} \, \hat{\mathbf{u}}_{k',j} \Big) \, \Big( \frac{e^{i (k+k') r}}{r^2}\Big) \, \dot{\nu}_k(\mathbf{r}_P,t) \dot{\nu}_{k'}(\mathbf{r}_P,t) \nonumber\\
&-& \frac{1}{R \, m(t)} \sum_{\lambda=+,\times} \int \frac{d^3 \mathbf{K}}{(2\pi)^{3/2}} \bigg\{ \frac{h_{\mathbf{K}}^{\lambda} \, e^{-i\big( \Omega t -\mathbf{K} \cdot \mathbf{r}_s \big)} }{\sqrt{2\Omega_K}} \int_V d^3 \mathbf{r} \, \Big( \frac{ e^{i (k+k'+K \cos \gamma) r } }{r^2} \Big) \nonumber\\
&+& \frac{h_{\mathbf{K}}^{\lambda \ast} \, e^{i\big( \Omega t -\mathbf{K} \cdot \mathbf{r}_s \big)} }{\sqrt{2\Omega_K}} \int_V d^3 \mathbf{r} \, \Big( \frac{ e^{i(k+k'-K\cos\gamma) r}}{r^2} \Big) \bigg\} \Big( e_{ij}^{\lambda}[\hat{\mathbf{K}}] \, \hat{\mathbf{u}}_{k,i} \, \hat{\mathbf{u}}_{k',j} \Big)  \, \dot{\nu}_k(\mathbf{r}_P,t) \, \dot{\nu}_{k'}(\mathbf{r}_P,t) \, . \nonumber
\end{eqnarray}
In the second line, we used the expression of $\mathbf{f}_k(\mathbf{r},t)$ given by Eq.~(\ref{3.32}). Note that the function $e^{i(k+k'\pm K\cos\gamma) r} $ is highly oscillating for large values of its argument; hence the radial integral averages to zero unless $(k+k'\pm K\cos\gamma)r \rightarrow 0$, so that $e^{i(k+k'\pm K\cos\gamma)r} \rightarrow 1$. Moreover, taking into account that $K\ll k$, the result of the integral can be approximately written as
\begin{eqnarray}\label{3.39}
&& \int_V d^3 \mathbf{r} \, \varepsilon_{ij}(\mathbf{r}_P,t) \, f_{k,i}(\mathbf{r},t) \, f_{k',j}(\mathbf{r},t) \dot{\nu}_k(\mathbf{r}_P,t) \dot{\nu}_{k'}(\mathbf{r}_P,t) \\
&=& \frac{\delta_{k+k',0}}{R\, m(t)} \, \int_V d^3\mathbf{r} \, \Big( \delta_{ij} - h_{ij}(\mathbf{r}_s,t) \Big) \hat{\mathbf{u}}_{k,i} \, \hat{\mathbf{u}}_{k',j} \, \Big( \frac{\dot{\nu}_k^2 (\mathbf{r}_P,t) }{r^2} \Big)\, . \nonumber
\end{eqnarray}
Similarly, one may show that
\begin{eqnarray}\label{3.40}
\hspace*{-1.2cm}
\int_V d^3 \mathbf{r} \, \varepsilon_{ij}(\mathbf{r}_P,t) f_{k,i}(\mathbf{r},t) f^{\ast}_{k',j}(\mathbf{r},t) \dot{\nu}_k (\mathbf{r}_P,t) \dot{\nu}^{\ast}_{k'}(\mathbf{r}_P,t) &=& \frac{\delta_{k-k',0}}{R\, m(t)} \int_V d^3\mathbf{r} \, \Big( \delta_{ij} - h_{ij}(\mathbf{r}_s,t) \Big) \hat{\mathbf{u}}_{k,i} \, \hat{\mathbf{u}}_{k,j} \, \Big( \frac{|\dot{\nu}_k (\mathbf{r}_P,t)|^2 }{r^2} \Big)\, , \nonumber \\
\hspace*{-1.2cm} \int_V d^3 \mathbf{r} \, \varepsilon_{ij}(\mathbf{r}_P,t) f^{\ast}_{k,i}(\mathbf{r},t) f_{k',j}(\mathbf{r},t) \dot{\nu}^{\ast}_k (\mathbf{r}_P,t) \dot{\nu}_{k'}(\mathbf{r}_P,t) &=& \frac{\delta_{k-k',0}}{R\, m(t)} \int_V d^3\mathbf{r} \, \Big( \delta_{ij} - h_{ij}(\mathbf{r}_s,t) \Big) \hat{\mathbf{u}}_{k,i} \, \hat{\mathbf{u}}_{k,j} \, \Big( \frac{|\dot{\nu}_k (\mathbf{r}_P,t)|^2 }{r^2} \Big)\, , \nonumber \\
\hspace*{-1.2cm} \int_V d^3 \mathbf{r} \, \varepsilon_{ij}(\mathbf{r}_P,t) f^{\ast}_{k,i}(\mathbf{r},t) f^{\ast}_{k',j}(\mathbf{r},t) \dot{\nu}^{\ast}_k (\mathbf{r}_P,t) \dot{\nu}^{\ast}_{k'}(\mathbf{r}_P,t) &=& \frac{\delta_{k+k',0}}{R\, m(t)} \int_V d^3\mathbf{r} \, \Big( \delta_{ij} - h_{ij}(\mathbf{r}_s,t) \Big) \hat{\mathbf{u}}_{k,i} \, \hat{\mathbf{u}}_{k,j} \, \Big( \frac{\dot{\nu}^{\ast 2}_k (\mathbf{r}_P,t) }{r^2} \Big)\, . \nonumber\\
\end{eqnarray}

Next, we compute the second term in the first line of Eq.~(\ref{3.37}), which is of the form 
\begin{eqnarray}\label{3.41}
&& \int_V d^3 \mathbf{r} \, \mu^{-1}_{ij}(\mathbf{r}_P,t) \big(\nabla \times \mathbf{f}_{k}(\mathbf{r},t)\big)_i \big(\nabla \times \mathbf{f}_{k'}(\mathbf{r},t) \big)_j \, \nu_k(\mathbf{r}_P,t) \nu_{k'}(\mathbf{r}_P,t) \\
&=& \int_V d^3 \mathbf{r} \, \mu^{-1}_{ij}(\mathbf{r}_P,t) \epsilon_{ilm} \big(\partial_l f_{k,m}(\mathbf{r},t)\big) \big( \nabla \times \mathbf{f}_{k'}(\mathbf{r},t) \big)_j \nu_k(\mathbf{r}_P,t) \nu_{k'}(\mathbf{r}_P,t) \nonumber\\
&=& \int_V d^3 \mathbf{r} \, \epsilon_{ilm} \partial_l \Big[\mu^{-1}_{ij}(\mathbf{r}_P,t) \, f_{k,m} (\mathbf{r},t) \, \big( \nabla \times \mathbf{f}_{k'}(\mathbf{r},t) \big)_j \nu_k(\mathbf{r}_P,t) \nu_{k'}(\mathbf{r}_P,t) \Big] \nonumber\\
&-& \int_V d^3\mathbf{r} \, \epsilon_{ilm} \, f_{k,m}(\mathbf{r},t) \, \partial_l \Big[ \mu^{-1}_{ij}(\mathbf{r}_P,t) \big( \nabla \times \mathbf{f}_{k'}(\mathbf{r},t) \big)_j \nu_k(\mathbf{r}_P,t) \nu_{k'}(\mathbf{r}_P,t) \Big] \nonumber\\
&=& \int_V d^3 \mathbf{r} \, \epsilon_{mli} \, \partial_l \Big[ \mu^{-1}(\mathbf{r}_P,t) \nabla \times \mathbf{f}_{k'}(\mathbf{r},t) \Big]_{i} \nu_k(\mathbf{r}_P,t) \nu_{k'}(\mathbf{r}_P,t) f_{k,m}(\mathbf{r},t)\, \nonumber\\
&=& \int_V d^3 \mathbf{r} \Big( \nabla \times \Big[ \mu^{-1}(\mathbf{r}_P,t) \nabla \times \mathbf{f}_{k'}(\mathbf{r},t) \Big]\Big)_m \, f_{k,m}(\mathbf{r},t) \, \nu_k(\mathbf{r}_P,t) \nu_{k'}(\mathbf{r}_P,t) \nonumber\\
&=& \int_V d^3 \mathbf{r} \, \omega_{k'}^2(\mathbf{r}_P,t) \, \varepsilon_{il}(\mathbf{r}_P,t) \, f_{k',l}(\mathbf{r},t) \, f_{k,m}(\mathbf{r},t)\, \nu_k(\mathbf{r}_P,t) \nu_{k'}(\mathbf{r}_P,t) \, .\nonumber
\end{eqnarray}
Here, $\epsilon_{ilm}$ stands for the Levi-Civita symbol. Note that a surface integral vanishes assuming the EM field vanishes at infinity. Moreover, we have omitted derivatives of the slowly varying functions, such as $\partial_{l} \mu_{ij}^{-1}(\mathbf{r}_P,t)$ and $\partial_l \nu_k(\mathbf{r}_P,t)$, which are proportional to $K$ and are much smaller than $k$. In the last line of Eq.~(\ref{3.41}), we used the wave equation Eq.~(\ref{3.15}) and Eq.(\ref{3.16}). With the help of definition Eq.~(\ref{3.32}) for the spatial mode functions $\mathbf{f}_k(\mathbf{r},t)$, Eq.~(\ref{3.41}) can be written as follows
\begin{eqnarray}\label{3.42}
&& \int_V d^3 \mathbf{r} \, \mu^{-1}_{ij}(\mathbf{r}_P,t) \big(\nabla \times \mathbf{f}_{k}(\mathbf{r},t)\big)_i \big(\nabla \times \mathbf{f}_{k'}(\mathbf{r},t) \big)_j \, \nu_k(\mathbf{r}_P,t) \nu_{k'}(\mathbf{r}_P,t) \\
&=& \frac{1}{R \, m(t)} \int_V d^3 \mathbf{r} \Big( \delta_{ij} -h_{ij}(\mathbf{r}_s,t) \Big) \, \hat{\mathbf{u}}_{k,i} \, \hat{\mathbf{u}}_{k',j} \Big( \frac{e^{i(k+k')r}}{r^2} \Big) \, \nu_k(\mathbf{r}_P,t) \nu_{k'}(\mathbf{r}_P,t) \, \omega_{k'}^2(\mathbf{r}_P,t) \, .\nonumber
\end{eqnarray}
Again, the integral in Eq.~(\ref{3.42}) is non-vanishing whenever $e^{i(k+k')r} \rightarrow 1$, for which the integral can be approximated by
\begin{eqnarray}\label{3.43}
&& \int_V d^3 \mathbf{r} \, \mu^{-1}_{ij}(\mathbf{r}_P,t) \big(\nabla \times \mathbf{f}_{k}(\mathbf{r},t)\big)_i \big(\nabla \times \mathbf{f}_{k'}(\mathbf{r},t) \big)_j \, \nu_k(\mathbf{r}_P,t) \nu_{k'}(\mathbf{r}_P,t) \\
&=& \frac{\delta_{k+k'}}{R \, m(t)} \int_V d^3 \mathbf{r} \Big( \delta_{ij} -h_{ij}(\mathbf{r}_s,t) \Big) \, \hat{\mathbf{u}}_{k,i} \, \hat{\mathbf{u}}_{k',j} \Big( \omega_k^2(\mathbf{r}_P,t)\, \frac{\nu_k^2(\mathbf{r}_P,t)}{r^2} \Big)\, .\nonumber
\end{eqnarray}
A similar procedure can be done for other terms in Eq.~(\ref{3.37}). The result is
\begin{eqnarray}\label{3.44}
&& \int_V d^3 \mathbf{r} \, \mu^{-1}_{ij}(\mathbf{r}_P,t) \big(\nabla \times \mathbf{f}_{k}(\mathbf{r},t)\big)_i \big(\nabla \times \mathbf{f}^{\ast}_{k'}(\mathbf{r},t) \big)_j \, \nu_k(\mathbf{r}_P,t) \nu^{\ast}_{k'}(\mathbf{r}_P,t) \\
&=& \frac{\delta_{k-k'}}{R \, m(t)} \int_V d^3 \mathbf{r} \Big( \delta_{ij} -h_{ij}(\mathbf{r}_s,t) \Big) \, \hat{\mathbf{u}}_{k,i} \, \hat{\mathbf{u}}_{k',j} \Big( \omega_k^2(\mathbf{r}_P,t)\, \frac{|\nu_k(\mathbf{r}_P,t)|^2}{r^2} \Big)\, , \nonumber \\
&& \int_V d^3 \mathbf{r} \, \mu^{-1}_{ij}(\mathbf{r}_P,t) \big(\nabla \times \mathbf{f}^{\ast}_{k}(\mathbf{r},t)\big)_i \big(\nabla \times \mathbf{f}_{k'}(\mathbf{r},t) \big)_j \, \nu^{\ast}_k(\mathbf{r}_P,t) \nu_{k'}(\mathbf{r}_C,t) \nonumber \\
&=& \frac{\delta_{k-k'}}{R \, m(t)} \int_V d^3 \mathbf{r} \Big( \delta_{ij} -h_{ij}(\mathbf{r}_s,t) \Big) \, \hat{\mathbf{u}}_{k,i} \, \hat{\mathbf{u}}_{k',j} \Big( \omega_k^2(\mathbf{r}_P,t)\, \frac{|\nu_k(\mathbf{r}_P,t)|^2}{r^2} \Big)\, , \nonumber \\
&& \int_V d^3 \mathbf{r} \, \mu^{-1}_{ij}(\mathbf{r}_P,t) \big(\nabla \times \mathbf{f}^{\ast}_{k}(\mathbf{r},t)\big)_i \big(\nabla \times \mathbf{f}^{\ast}_{k'}(\mathbf{r},t) \big)_j \, \nu^{\ast}_k(\mathbf{r}_P,t) \nu^{\ast}_{k'}(\mathbf{r}_P,t) \nonumber \\
&=& \frac{\delta_{k-k'}}{R \, m(t)} \int_V d^3 \mathbf{r} \Big( \delta_{ij} -h_{ij}(\mathbf{r}_s,t) \Big) \, \hat{\mathbf{u}}_{k,i} \, \hat{\mathbf{u}}_{k',j} \Big( \omega_k^2(\mathbf{r}_P,t)\, \frac{\nu^{\ast 2}_k(\mathbf{r}_P,t)}{r^2} \Big)\, \nonumber.
\end{eqnarray}
\noindent Inserting Eqs.~(\ref{3.39}, \ref{3.40}) and Eqs.~(\ref{3.43}, \ref{3.3}) into the Hamiltonian Eq.~(\ref{3.37}) results in the following expression
\begin{eqnarray}\label{3.45}
\hspace*{-1.2cm} \hat{H}_{em-gw}(t) &=& \frac{\hbar}{4} \sum_{k,k'} \bigg\{ \hat{a}_{k} \hat{a}_{k'} \, \frac{\delta_{k+k',0}}{R\, m(t)} \, \int_V d^3 \mathbf{r} \, \Big( \delta_{ij} - h_{ij}(\mathbf{r}_s,t)) \Big) \hat{\mathbf{u}}_{k,i} \hat{\mathbf{u}}_{k',j} \, \Big( \frac{\dot{\nu}_k^2(\mathbf{r}_P,t) + \omega^2_k(\mathbf{r}_P,t) \nu_k^2(\mathbf{r}_P,t)}{r^2} \Big) \nonumber \\
&+& \hat{a}_{k} \hat{a}^{\dagger}_{k'} \, \frac{\delta_{k-k',0}}{R\, m(t)} \, \int_V d^3 \mathbf{r} \, \Big( \delta_{ij} - h_{ij}(\mathbf{r}_s,t)) \Big) \hat{\mathbf{u}}_{k,i} \hat{\mathbf{u}}_{k',j} \, \Big( \frac{|\dot{\nu}_k(\mathbf{r}_P,t)|^2 + \omega^2_k(\mathbf{r}_P,t) |\nu_k (\mathbf{r}_P,t)|^2 }{r^2} \Big) \nonumber \\
&+& \hat{a}_{k}^{\dagger} \hat{a}_{k'} \, \frac{\delta_{k-k',0}}{R\, m(t)} \, \int_V d^3 \mathbf{r} \, \Big( \delta_{ij} - h_{ij}(\mathbf{r}_s,t)) \Big) \hat{\mathbf{u}}_{k,i} \hat{\mathbf{u}}_{k',j} \, \Big( \frac{|\dot{\nu}_k(\mathbf{r}_P,t)|^2 + \omega^2_k(\mathbf{r}_P,t) |\nu_k (\mathbf{r}_P,t)|^2 }{r^2} \Big) \nonumber \\
&+& \hat{a}^{\dagger}_{k} \hat{a}^{\dagger}_{k'} \, \frac{\delta_{k+k',0}}{R\, m(t)} \, \int_V d^3 \mathbf{r} \, \Big( \delta_{ij} - h_{ij}(\mathbf{r}_s,t)) \Big) \hat{\mathbf{u}}_{k,i} \hat{\mathbf{u}}_{k',j} \, \Big( \frac{\dot{\nu}^{\ast 2}_k(\mathbf{r}_P,t) + \omega^2_k(\mathbf{r}_P,t) \nu_k^{\ast 2} (\mathbf{r}_P,t) }{r^2} \Big) \, .
\end{eqnarray}
Each term in the Hamiltonian Eq.~(\ref{3.45}) represents the scattering of photons mediated by long-wavelength GWs (since $K \ll k$ is assumed). Terms proportional to $\hat{a}_k \hat{a}_{k'}$ and $\hat{a}^{\dagger}_k \hat{a}^{\dagger}_{k'}$ correspond to photon creation by the GWs field, while other terms like $\hat{a}_k^{\dagger} \hat{a}_k$ represent the scattering of each mode to itself in the presence of GWs. The temporal mode functions $\nu_{k}(\mathbf{r}_C,t)$ are already defined by Eq.~(\ref{3.16}) in the adiabatic approximation $\Omega_K \ll \omega_k$. One can easily show that the following mode functions
\begin{eqnarray}\label{3.46}
\nu_k (\mathbf{r}_P,t) &=& \frac{1}{\sqrt{\omega_k(\mathbf{r}_P,t)}} e^{-i\int_{0}^{t} \omega_{k}(\mathbf{r}_P,t') dt' }, \\
\dot{\nu_k}(\mathbf{r}_P,t) &=& \frac{1}{\sqrt{\omega_k(\mathbf{r}_P,t)}} \, \frac{d}{dt} \big[ e^{-i\int_{0}^{t} \omega_{k}(\mathbf{r}_P,t') dt' } \big] \nonumber\\
&=& \frac{1}{\sqrt{\omega_k(\mathbf{r}_P,t)}} \Big( -i\frac{d}{dt} \big[ \int_{0}^{t} \omega_{k}(\mathbf{r}_P,t') dt' \big] \Big) e^{-i\int_{0}^{t} \omega_{k}(\mathbf{r}_P,t') dt' } \nonumber \\
&=& -i \omega_k(\mathbf{r}_P,t) \nu_k (\mathbf{r}_P,t). \nonumber
\end{eqnarray}
satisfy Eq.~(\ref{3.16}). Note that this solution is valid within the adiabatic approximation $\Omega_K \ll \omega_k$. Substitution of the mode functions Eq.~(\ref{3.46}) into the Hamiltonian Eq.~(\ref{3.45}) resets the contribution of $\hat{a}_{k} \hat{a}_{k}$ and $\hat{a}^{\dagger}_{k}\hat{a}^{\dagger}_{k} $ to zero, and Eq.~(\ref{3.44}) reduces to 
\begin{eqnarray}\label{3.47}
\hspace*{-1.2cm} \hat{H}_{em-gw}(t) &=& \hbar \sum_{k} \big( \hat{a}^{\dagger}_k \hat{a}_k + \frac{1}{2} \big) \, \frac{\delta_{k-k',0}}{R\, m(t)} \, \int_V d^3 \mathbf{r} \, \Big( \delta_{ij} - h_{ij}(\mathbf{r}_s,t) \Big) \hat{\mathbf{u}}_{k,i} \hat{\mathbf{u}}_{k,j} \, \frac{\omega_k(\mathbf{r}_P,t) }{r^2} \,.
\end{eqnarray}
Consequently, under the adiabatic condition $\Omega_K \ll \omega_k$ (and $K \ll k$), particle creation by GWs background is negligible, and the EM-GW coupling is intensity-dependent, i.e., proportional to $\hat{a}^{\dagger}_{k} \hat{a}_{k}$. One could alternatively derive the Hamiltonian Eq.~(\ref{3.47}) by energy consideration. Indeed, Eq.~(\ref{3.37}) implies that the Hamiltonian of the system is explicitly time-dependent, and, in the corresponding quantum field theory, the vacuum state of the field is not well-defined. Nevertheless, one may define \textit{instantaneous} vacuum by finding suitable mode functions $\nu_k(\mathbf{r}_P,t)$ such that the energy of the system, determined by the mean value of the Hamiltonian Eq.~(\ref{3.36}), is minimized at each moment. The minimization procedure yields the adiabatic mode functions Eq.~(\ref{3.46}) and the Hamiltonian Eq.~(\ref{3.47}) outcomes.

In order to obtain the explicit form of the interaction Hamiltonian $\hat{H}_{\text{int}}(t)$ given by Eq.~(\ref{3.47}), we insert from Eq.~(\ref{3.23}) the expression of $\omega_k(\mathbf{r}_P,t)$ into $\hat{H}_{\text{int}}(t)$. For simplicity, we consider only one mode of the EM field, specified by wave number $k$. By neglecting the vacuum energy $\frac{1}{2} \hbar \omega_k$, one has
\begin{eqnarray}\label{3.48}
\hspace*{-1.2cm} \hat{H}_{em-gw}(t) &=& \hbar \omega_k \hat{a}^{\dagger}_k \hat{a}_k  \,\frac{1}{R \, m(t)} \, \int_V d^3 \mathbf{r} \, \Big( \big[ \delta_{ij} - h_{ij}(\mathbf{r}_s,t) \big] \hat{\mathbf{u}}_{k,i} \hat{\mathbf{u}}_{k,j} \Big) \, \Big( \delta_{ab} - \frac{1}{2} h_{ab}(\mathbf{r}_P,t) \Big) \frac{\hat{r}_a \hat{r}_b }{r^2} \nonumber\\
&=& \hbar \omega_k \hat{a}^{\dagger}_k \hat{a}_k  \,\frac{1}{R \, m(t)} \, \int_V \frac{d^3 \mathbf{r}}{r^2} \, \Big( 1 - h_{ij}(\mathbf{r}_s,t)\, \hat{\mathbf{u}}_{k,i} \,\hat{\mathbf{u}}_{k,j} - \frac{1}{2} h_{ij}(\mathbf{r}_P,t) \, \hat{r}_{i} \, \hat{r}_{j} \Big) \, .
\end{eqnarray}
Note that $a,b$ are spatial indices and $\hat{r}_{a}$ represents the $a$-th Cartesian component of the unit vector $\hat{r}$. In the second line, we used the identity $\delta_{ij} \hat{\mathbf{u}}_{k,i} \hat{\mathbf{u}}_{k,j} =1$, and we retained terms up to $\mathcal{O}(h_{ij})$ in the linear approximation. Inserting from Eq.~(\ref{3.30}) for $m(t)$ and keeping terms up to $\mathcal{O}(h_{ij})$, it is easy to show that the Hamiltonian reads as follows
\begin{eqnarray}\label{3.49}
\hspace*{-1.2cm} \hat{H}_{em-gw}(t) &=& \hbar \omega_k \hat{a}^{\dagger}_k \hat{a}_k  \, \Big( 1- \frac{1}{4\pi R} \int_V \frac{d^3 \mathbf{r}}{r^2} h_{ij}(\mathbf{r}_P,t) \, \hat{r}_{i} \, \hat{r}_{j} \Big) \, .
\end{eqnarray}
The first term in the parentheses describes free evolution of the EM field. The second term shows the EM-GWs interaction Hamiltonian, which can be written as
\begin{eqnarray}\label{3.50}
\hspace*{-1.2cm} \hat{H}_{\text{int}}(R,t) &=& - \frac{1}{2}\,\hbar \omega_k \hat{a}^{\dagger}_k \hat{a}_k \, \Big( \frac{1}{2\pi R} \int_0^{R} dr \int d\Omega_{\hat{r}} \, h_{ij}(\mathbf{r}_P,t) \, \hat{r}_{i} \, \hat{r}_{j} \Big) \, .
\end{eqnarray}
The fully quantum version of the interaction Hamiltonian can be obtained by inserting from Eq.~(\ref{3.3}) the quantum field operator $\hat{h}_{ij}(\mathbf{r}_P,t)$ into the Hamiltonian Eq.~(\ref{3.50}), 
\begin{eqnarray}\label{3.51}
\hspace*{-1.2cm} \hat{H}_{\text{int}}(R,t) &=& - \frac{1}{2}\Big( \frac{\hbar \omega_k}{E_{\text{Pl}}} \Big) \frac{\sqrt{16 \pi c^3}}{(2\pi)^{3/2}} \sum_{\lambda=+,\times} \int \frac{d^3\mathbf{K}}{\sqrt{2\Omega_K}} \Big( \hat{b}_{\mathbf{K},\lambda} \, e^{-i \Omega_K t} \, \mathcal{G}^{\lambda}_{\mathbf{K}}(R) + \, \text{h.c.} \Big) \, \hat{a}^{\dagger}_k \hat{a}_k  \, ,
\end{eqnarray}
where the function $\mathcal{G}_{\mathbf{K}}(R)$ is defined by
\begin{eqnarray}\label{3.52}
\mathcal{G}^{\lambda}_{\mathbf{K}}(R) &\equiv& \frac{e^{i\mathbf{K} \cdot \mathbf{r}_s}}{2\pi R} \, \int d\Omega_{\hat{r}} \, e_{ij}^{\lambda}[\hat{\mathbf{K}}] \, \hat{r}_{i}\, \hat{r}_{j} \, \int_{0}^{R} dr \, e^{i\mathbf{K}\cdot\mathbf{r}} \, .
\end{eqnarray}
which bears the effect of the size of the probe (the distance of the source, in our case). Consequently, the Hamiltonian Eq.~(\ref{3.51}) describes the quantum interaction between a spherical EM field, propagated from a point source located at a distance $R$ from the Earth, with GWs background. 
The Hamiltonian shows an intensity-dependent coupling (proportional to $\hat{a}^{\dagger}_{k} \hat{a}_{k}$) between the EM field and GWs and is reminiscent of the optomechanical coupling in cavity optomechanics \cite{aspelmeyer2014cavity}. This analogy turns out to be useful in formulating quantum dynamics of the EM field. 

With the help of Hamiltonian Eq.~(\ref{3.51}), one can proceed to solve the Heisenberg equation of motion and find the dynamics of the ladder operators $\hat{a}_{k}(t)$ and $\hat{a}^{\dagger}_k(t)$. We left this step to App.~\ref{app:B}, and focused on the final result here. Consequently, it follows that the EM ladder operators in the presence of quantum GWs obey the following equation:
\begin{eqnarray}\label{3.53}
\hspace*{-1.2cm}\hat{a}_{k}(R,t) &=& \Big[ \prod_{\lambda,\mathbf{K}\in \Re^{3+}} \hat{D}_{\mathbf{K},\lambda}\big(-\kappa(K) \, \eta_{\mathbf{K}} (R,t) \big) \otimes \hat{D}_{-\mathbf{K},\lambda} \big(-\kappa(K) \, \eta_{-\mathbf{K}} (R,t) \big) \Big] \nonumber\\
&\times& e^{2i E(R,t) \left( \hat{a}_{k}^{\dagger}\hat{a}_{k} + \frac{1}{2} \right)} \, e^{-i\omega_{k} t} \hat{a}_{k}\, .
\end{eqnarray}
Here, we have used the half-Fourier space representation where $\mathbf{K}\in \Re^{3+}$, which turns out to facilitate the computations concerning two-mode squeezed PGWs. Different quantities, such as the displacement operator of GWs $\hat{D}_{\mathbf{K},\lambda}$, the coupling strength $\kappa(K)$, and the functions $E(R,t)$ and $\eta_{\mathbf{K}}(R,t)$ are defined in App.~\ref{app:B} by Eqs.~(\ref{b.5}, \ref{b.9}, \ref{b.17}). 
With the help of Eq.~(\ref{3.53}), we are equipped to investigate spatial correlations of the EM field emitted from distant objects in the presence of the two-mode squeezed PGWs background.


\section{Loss of spatial coherence induced by two-mode squeezed PGWs}\label{sec:4}

\subsection{Spatial correlations in the presence of GWs}\label{subsec:4.1}

To investigate the effect of PGWs background on the EM spatial correlations, we first describe the experimental setup depicted in Fig.~\ref{fig3}, which presents the physical concept of spatial correlations of the EM field based on which the angular size of extended objects is measured.


\subsubsection{Configuration of a VLBI-type system}\label{subsubsec:4.1.1} 

\begin{figure*}[htb]
\centering
\includegraphics[
width=0.75\columnwidth]{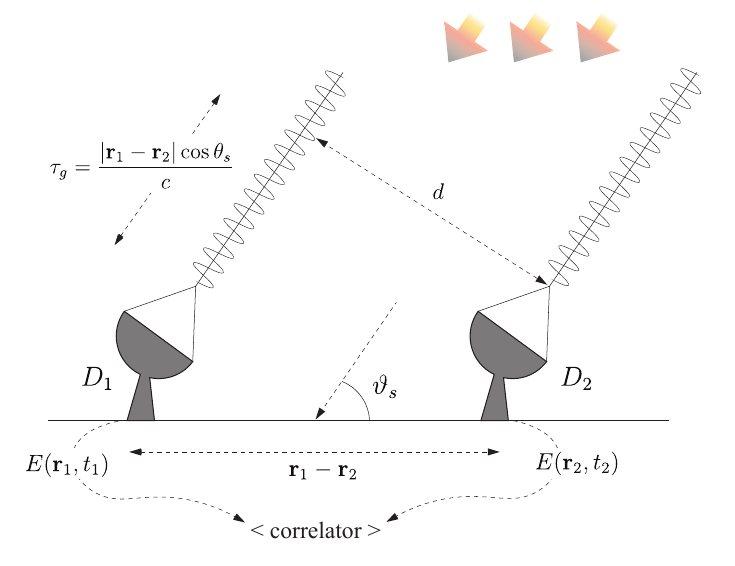}
\caption{Configuration of the VLBI system. The EM field is collected at two detectors, $D_1$ and $D_2$. The interference pattern is produced by beating the signals received by the two detectors, thanks to a correlator. The angle $\theta_s$ represents the orientation of the source with respect to the detectors baseline, and $\tau_g$ shows the geometrical delay.}
\label{fig3}
\end{figure*}

Fig.~\ref{fig3} shows a typical VLBI setup. The EM field radiated from a planar source $\sigma$ of radius $a$ (for example, a star) illuminates two detectors $D_{1}$ and $D_2$ located at $\mathbf{r}_1$ and $\mathbf{r}_2$, respectively. For simplicity, we take the origin of coordinates at the center of the planar source, and $\mathbf{r}_1$ and $\mathbf{r}_2$ denote the location of each telescope with respect to the center of the planar disc. The signals are then combined and correlated at point $P$ at zero difference-time (note that two paths $\overline{D_1 P}$ and $\overline{D_2 P}$ are equal). In the practical situation of a VLBI instrument, the orientation of the source with respect to the detector baseline makes an angle $\vartheta_s$ that causes a \textit{geometric delay time} $\tau_g=|\mathbf{r}_1-\mathbf{r}_2|\cos\theta_s/c$ that is often compensated for to achieve equal time correlations \cite{philip2016calibration}, so the configuration is practically symmetric.


\subsubsection{Equal-time first-order degree of coherence $g^{(1)}(\mathbf{r}_1,t;\mathbf{r}_2,t)$}\label{subsubsec:4.1.2}

For the VLBI setup shown in Fig.~\ref{fig3}, the electric field of the incoming light is collected by two detectors $D_1$ and $D_2$ located at $\mathbf{r}_1$ and $\mathbf{r}_2$, at time $t$. The total electric field received at the correlator at point $\mathbf{r}_C$ is thus determined by $\hat{E}(\mathbf{r}_C,t) = \hat{E}^{(+)}(\mathbf{r}_C,t) + \hat{E}^{(-)}(\mathbf{r}_C,t)$, where the positive and negative frequency parts of the electric field are defined by
\begin{eqnarray}\label{4.1}
\hat{E}^{(+)}(\mathbf{r}_C,t) = \hat{E}^{(+)}(\mathbf{r}_1,t-\tau_1) + \hat{E}^{(+)}(\mathbf{r}_2,t-\tau_2),
\end{eqnarray}
\noindent and its Hermitian conjugate. Here, $\tau_i$ accounts for the time delay of the EM signal through the arm $\overline{D_iP}$, with $i=1,2$, and $\hat{E}^{(+)}(\mathbf{r}_i,t-\tau_i)$ stands for the electric field at the location of the $i$-th detector at the delay time $t-\tau_i$. In the symmetric setup one has $\tau_1 = \tau_2$, which is also true in VLBI measurement thanks to the geometric delay line that ensures equal time correlation. Since we neglect the effect of the GWs background during the time delays $\tau_i$ (practically $\tau_i \lll t$), we can proceed by setting $t-\tau_1 = t-\tau_2 \equiv t$. Therefore, $t$ will refer to the time at which the EM signal arrives at the detectors, which can also be called the time of flight or the interaction time. The expression for the intensity at point $P$ is now given by 
\begin{eqnarray}\label{4.2}
I(\mathbf{r}_C,t) &=& \langle \hat{E}^{(-)}(\mathbf{r}_C,t) \hat{E}^{(+)}(\mathbf{r}_C,t) \rangle,
\end{eqnarray}

\begin{figure*}[htb]
\centering
\includegraphics[
width=0.8\columnwidth]{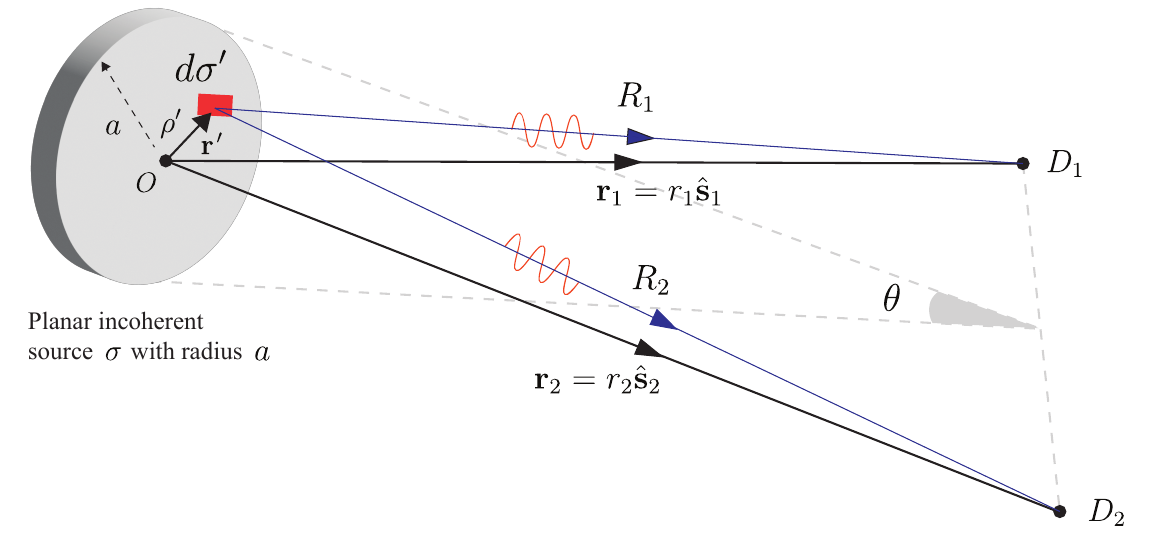}
\caption{An infinitesimal surface element $d\sigma'$ located at $\mathbf{r}'=(\rho'\cos\phi',\rho'\sin\phi')$ sends spherical EM radiation that is going to be detected at detectors $D_1$ and $D_2$. Here, $\mathbf{r}_i = r_i \hat{\mathbf{s}}_i $ with $i=1,2$ shows the location of detectors with respect to the origin $O$, and $R_{i}$ stands for the distance of the $i$-th detector with respect to the source element $d\sigma'$. The angular diameter of the source is shown by $\theta$.}
\label{fig4} 
\end{figure*}
The electric field $\hat{E}(\mathbf{r}_C,t)$ is a superposition of spherical waves emitted from different parts of the planar disc. To proceed, we assume that the surface of the planar disc with area $A=\pi a^2$ is divided into $N$ equal sub-surfaces, each of them radiating spherical waves independently. From the expression of $\hat{\mathbf{A}}(\mathbf{r}_C,t)$ given in Eq.~(\ref{3.12}), the electric field induced by spherical waves possessing wave number $k$, emitted from all point sources, each of which is located at $\mathbf{r}'$, at the observation point $\mathbf{r}_i$ is determined by
\begin{eqnarray}\label{4.3}
\hat{E}_k(\mathbf{r}_i,t) = \sum_{j=1}^{N} \mathcal{E}_j \Big( \hat{a}_j(R_{ij},t) \, \frac{e^{ik R_{ij}}}{R_{ij}} + \text{h.c.} \Big) \, ,
\end{eqnarray}
where $\mathcal{E}_j$ is the electric field amplitude and is irrelevant in subsequent calculations. We can proceed by a constant amplitude for all independent point sources, $\mathcal{E}_j \equiv \mathcal{E}$, assuming a uniform radiation intensity from the source. As can be seen from Fig.~\ref{fig4}, $R_{ij}$ stands for the distance between the $j$-th source element from the $i$-th detector. By combining Eq.(\ref{4.1}) and Eq.~(\ref{4.3}), one can identify the positive (and negative) frequency part of the total electric field at $\mathbf{r}_C$, as follows:
\begin{eqnarray}\label{4.4}
\hat{E}^{(+)}(\mathbf{r}_C,t) = \mathcal{E} \sum_{j=1}^{N} \Big( \hat{a}_{j} (R_{1j},t) \frac{e^{ikR_{1j}}}{R_{1j}} + \hat{a}_{j}(R_{2j},t) \frac{e^{i k R_{2j}}}{R_{2j}} \Big) \, ,
\end{eqnarray}
and its Hermitian conjugate. Here, $R_{1j}$ and $R_{2j}$ show the distance of the first and second detectors with respect to the $j$-th source point located at $\mathbf{r}'_j$. The evolution of the EM field through GW background is determined by Eq.~(\ref{3.52}). Plugging Eq.~(\ref{4.4}) into Eq.~(\ref{4.2}), the intensity $I(\mathbf{r}_C,t)$ is determined by 
\begin{eqnarray}\label{4.5}
\hspace*{-.3cm}I(\mathbf{r}_C,t) &=& |\mathcal{E}|^2 \sum_{j=1}^{N} \bigg\{ \frac{\big\langle \hat{n}_j (R_{1j},t) \big\rangle}{R_{1j}^2} + \frac{\big\langle \hat{n}_{j} (R_{2j},t) \big\rangle}{R_{2j}^2} + 2 \Re\Big[ \big\langle \hat{a}^{\dagger}_{j} (R_{1j},t) \, \hat{a}_j (R_{2j},t) \big\rangle \, \frac{e^{-ik \big(R_{1j} - R_{2j} \big)}}{R_{1j} R_{2j}} \Big]  \bigg\} \, . \nonumber\\
\end{eqnarray}
Here, cross-correlations of type $ \propto \big\langle \hat{a}^{\dagger}_{j} \hat{a}_{j'} \big\rangle = \big\langle \hat{a}^{\dagger}_{j} \big\rangle \big\langle \hat{a}_{j'} \big\rangle$ between independent modes $j$ and $j'$ (emitted from different point sources) vanish, assuming that each mode of the EM field is in a thermal state for which $\big\langle \hat{a}_{j} \big\rangle=0$. Since the EM-GWs Hamiltonian Eq.~(\ref{3.48}) preserves the number of photons in the adiabatic approximation, the mean number of photons is conserved and is equal to its initial value at the moment of emission. Hence, we can set
\begin{eqnarray}\label{4.6}
\big\langle \hat{n}_{k} (R_{ij},t) \big\rangle \equiv \big\langle \hat{n}_k \big\rangle \quad , \quad i=1,2 \,. 
\end{eqnarray}
where $\big\langle \hat{n}_k \big\rangle$ shows the mean number of photons with wave number $k$. The first two terms in Eq.~(\ref{4.5}) represent the intensity at the detectors, while the last term shows the role of correlations in mode $j$ in the interference pattern. In the absence of GWs, the annihilation operator simply reduces to the free-field case, $\hat{a}_{j}(R_{ij},t) \rightarrow e^{-i\omega_k t} \hat{a}_{j}$ (see Eq.~(\ref{3.52})) and Eq.~(\ref{4.5}) recasts to the free-field expression. The quantity inside the brackets is closely related to the equal-time first-order degree of coherence, which is usually defined by \cite{scully1999quantum}
\begin{eqnarray}\label{4.7}
g^{(1)}_{j}(R_{1j},t ; R_{2j}, t) &\equiv& \frac{ \big\langle \hat{a}_{j}^{\dagger}(R_{1j},t) \, \hat{a}_{j}(R_{2j},t) \big\rangle }{\big\langle \hat{n}_{k} \big\rangle} \, e^{-ik \big(R_{1j} - R_{2j} \big)} \,,
\end{eqnarray}
The expectation value in Eq.~(\ref{4.7}) is generally taken over the total state $\hat{\rho} = \hat{\rho}_{\text{gw}} \otimes \hat{\rho}_{j}$, where $\hat{\rho}_{\text{gw}}$ describes the state of GWs, and $\hat{\rho}_{j}$ stands for the density matrix of the EM field of mode $k$ emitted from $\mathbf{r}'_j$. The emission of distant objects can be assumed to be dominated by a thermal emission with a mean number of photons $\langle \hat{n}_{k} \rangle = [\exp(\hbar\omega_k/k_{\text{B}}T) -1 ]^{-1}$. However, as can be seen from Eqs.~(\ref{4.5} - \ref{4.7}), the state of the EM field does not play an important role, and only the mean number of photons appears. This is one of the main features that makes spatial correlations favorable with respect to temporal correlations that drastically depend on the EM state. Thus, the degree of coherence depends \textit{only} on the state of GWs. Combining Eqs.~(\ref{4.5} - \ref{4.7}) yields
\begin{eqnarray}\label{4.8}
\hspace*{-1.2cm}I(\mathbf{r}_C,t) &=&  \frac{2\, |\mathcal{E}|^2 \big\langle \hat{n}_k \big\rangle}{R ^2} \sum_{j=1}^N \Big\{ 1 + \Re\big[ g^{(1)}_{j}(R_{1j},t;R_{2j};t) \big] \Big\}\,. 
\end{eqnarray}
In deriving Eq.~(\ref{4.8}) we have replaced in the denominators $R_{1j} \simeq R_{2j} \equiv R$, while the path difference $R_{1j} - R_{2j}$ must be kept in the phase factors as it is the origin of spatial correlations. One may check that, in the absence of GW background, one has $g^{(1)}_j(R_{1j},t; R_{2j},t) \rightarrow e^{-ik \big(R_{1j} - R_{2j} \big)}$ and Eq.~(\ref{4.8}) reduces to the standard interference pattern in free space (see Eq.~(4.1.24) of \cite{scully1999quantum}). The summation over discrete index $j$ can be replaced by integration over the surface of the planar disc according to $\sum_{j=1}^N a_j(\mathbf{r}'_j) \rightarrow \int_{\sigma} \frac{d\sigma'}{A} \, a(\mathbf{r}')$ for arbitrary sequence $a_{j}(\mathbf{r}'_j)$, where $d\sigma'$ stands for the surface element and $A=\pi a^2$ is the area of the planar source. Thus, the expression of the intensity $I(\mathbf{r}_C,t)$ given in Eq.~(\ref{4.8}) is written as 
\begin{eqnarray}\label{4.9}
I(\mathbf{r}_C,t) &=& \frac{2\, |\mathcal{E}|^2 \big\langle \hat{n}_k \big\rangle}{R^2} \bigg\{ 1 +  \frac{1}{\pi a^2} \, \Re\, \Big[ \int_{\sigma} d \sigma' \, g^{(1)} (R_{1},t ; R_{2},t) \Big] \bigg\}. \qquad
\end{eqnarray}
Here, $\mathrm{d}\sigma' = \rho' \, d\rho' \, d\phi'$ is the surface element with $0 \leq \rho' \leq a$ and $0 \leq \phi' \leq 2\pi$, and the integration is performed over the area of the disc. The two-point correlation function $g^{(1)}(R_{1},t ;R_{2},t)$ should be evaluated from Eq.~(\ref{4.7}) where $R_1$ and $R_2$ are shown in Fig.~(\ref{fig4}). In the absence of GWs, Eq.~(\ref{4.9}) yields the van Citter-Zernike correlations
\begin{eqnarray}\label{4.10}
I(\mathbf{r}_C,t) &=& \frac{2\, |\mathcal{E}|^2 \big\langle \hat{n}_k \big\rangle}{R^2}\,\Big( 1 + \, j(\mathbf{r}_1,\mathbf{r}_2) \Big). \qquad
\end{eqnarray}
where the two-point function
\begin{eqnarray}\label{4.11}
j(\mathbf{r}_1,\mathbf{r}_2) &\equiv& \frac{ \int_{\sigma} d \sigma' \, e^{-ik(R_{1}-R_{2})} }{\pi a^2} \, ,
\end{eqnarray}
is called the \textit{mutual intensity} and bears spatial correlations in the absence of GWs background \cite{mandel1995optical}. In the next section, we obtain equal-time correlations $g^{(1)}(R_{1},t; R_{2},t)$ defined by Eq.~(\ref{4.7}) in the presence of GWs in a two-mode squeezed state, as predicted by the inflationary scenario.


\subsection{First-order degree of coherence in the presence of PGWs in two-mode squeezed state } \label{subsec:4.2}

In this section, we consider correlation function Eq.~(\ref{4.7}) when GWs are placed in the two-mode squeezed state, briefly denoted by $\big\vert\text{TS}\big\rangle$, for which the squeezing amplitude is governed by Eq.~(\ref{2.7}). Inserting $\hat{a}(R_{i},t)$ from Eq.~(\ref{3.53}) into Eq.~(\ref{4.7}), one obtains
\begin{eqnarray}\label{4.12}
\hspace*{-1.5cm} g_{\text{ts}}^{(1)}(1; 2) &=& \prod_{\lambda,\mathbf{K}\in \Re^{3+}} \big\langle \text{TS}\big\vert \hat{D}^{\dagger}_{\mathbf{K},\lambda} \big(-\kappa(K) \eta_{\mathbf{K}}(1)\big) \hat{D}_{\mathbf{K},\lambda} \big(-\kappa(K) \eta_{-\mathbf{K}}(2) \big) \\
&\times& \hat{D}^{\dagger}_{-\mathbf{K},\lambda} \big(-\kappa(K) \eta_{-\mathbf{K}}(1)\big) \hat{D}_{-\mathbf{K},\lambda} \big(-\kappa(K) \eta_{-\mathbf{K}}(2) \big) \big\vert \text{TS} \big\rangle \,. \nonumber
\end{eqnarray}
In the above equation, $(1)$ and $(2)$ are abbreviations for two points $(R_{1},t_1)$ and $(R_{2},t_2)$. Calculation of $g^{(1)}_{\text{ts}}(R_{1},t; R_{2},t)$ is left to App.~\ref{app:C}. Here, we focus on the final result and its implications. The final expression for the spatial correlation function reads as follows:
\begin{eqnarray}\label{4.13}
g^{(1)}_{\text{ts}} (R_{1},t; R_{2},t) = e^{-i \mathcal{C}^{\text{vac}}(R_{1},t; R_{2},t)} \, e^{-\mathcal{D}^{\text{vac}}(R_{1},t; R_{2},t)} \, e^{-\mathcal{D}^{\text{ts}}(R_{1},t; R_{2},t)} \, ,
\end{eqnarray}
where the two-point kernels $ \mathcal{C}^{\text{vac}}(R_{1},t; R_{2},t)$, $\mathcal{D}^{\text{vac}}(R_{1},t; R_{2},t)$ and $\mathcal{D}^{\text{ts}}(R_{1},t; R_{2},t)$ are defined accordingly
\begin{eqnarray}\label{4.14}
\mathcal{C}^{\text{vac}}(R_{1},t; R_{2},t) &\equiv& \sum_{\lambda} \int_{\mathbf{K}\in \Re^{3}} d^3\mathbf{K} \, \kappa^2(K) \, \mathcal{C}^{\text{vac}}_{\mathbf{K}}(R_{1},t; R_{2},t) \big]\, , \\
\mathcal{D}^{\text{vac}}(R_{1},t; R_{2},t) &\equiv& \frac{1}{2} \sum_{\lambda} \int_{\mathbf{k}\in \Re^3} d^3 \mathbf{K} \, \kappa^2(K) \,\mathcal{D}^{\text{vac}}_{\mathbf{K}}(R_{1},t; R_{2},t) \, ,\nonumber \\
\mathcal{D}^{\text{ts}} (R_{1},t; R_{2},t) &\equiv& \sum_{\lambda} \int_{\mathbf{K}\in \Re^{3}} d^3\mathbf{K} \, \kappa^{2}(K) \bigg\{ |v_{K}(\eta_H)|^2 \mathcal{D}_{\mathbf{K}}^{\text{vac}} (1;2) + \frac{1}{2} \mathcal{D}^{\text{ts-corr}}_{\mathbf{K}}(R_{1},t; R_{2},t) \bigg\} \,. \nonumber
\end{eqnarray}
together with
\begin{eqnarray}\label{4.15}
\mathcal{C}^{\text{vac}}_{\mathbf{K}}(R_{1},t; R_{2},t) &\equiv& \Im \Big[ \eta_{\mathbf{K}}(R_{1},t) \eta^{\ast}_{\mathbf{K}}(R_{2},t) \Big], \\
\mathcal{D}^{\text{vac}}_{\mathbf{K}}(R_{1},t; R_{2},t) &\equiv& \big\vert \eta_{\mathbf{K}}(R_{1},t) - \eta_{\mathbf{K}}(R_{2},t)\big\vert ^2, \nonumber \\
\mathcal{D}_{\mathbf{K}}^{\text{ts-corr}} (R_{1},t; R_{2},t) &\equiv& 2 \Re\Big[ u^{\ast}_K(\eta_H) v^{\ast}_{K}(\eta_H) \Big( \eta_{\mathbf{K}}(R_{1},t) - \eta_{\mathbf{K}}(R_{2},t) \Big) \Big( \eta_{-\mathbf{K}}(R_{1},t) - \eta_{-\mathbf{K}}(R_{2},t) \Big) \Big]. \nonumber
\end{eqnarray}
In these expressions, $u_K(\eta_H)$ and $v_K(\eta_H)$ are determined by Eq.~(\ref{2.3}), and other quantities such as $\kappa(K)$ and $\eta_{\mathbf{K}}(R,t)$ are defined in App.~\ref{app:B} by Eq.~(\ref{b.5}) and Eq.~(\ref{b.17}), respectively. Eq.~(\ref{4.13}) implies that PGWs contribute to the loss of spatial correlations. Although vacuum fluctuations induce both coherent and incoherent contributions, embarked in $\mathcal{C}^{\text{vac}}$ and $\mathcal{D}^{\text{vac}}$, the effect of squeezed gravitons appears only in the incoherence mechanism and is determined by $\mathcal{D}^{\text{ts}}$. Due to the negligible factor $\kappa^2(K)\propto (\hbar\omega_k/E_{\text{Pl}})^2$ appearing in all kernels, the magnitude of the incoherence effect is automatically vanishing. However, for highly squeezed PGWs possessing an enormously large number of gravitons (as implied by Eq.~(\ref{2.6})), there is a reasonable chance to overcome the tiny coupling strength $\kappa(K)$ and produce a sensible effect. In the following, we evaluate the kernel $\mathcal{D}^{\text{ts}}(R_{1},t;R_{2},t)$ and obtain the expression of $g^{(1)}_{ts}(R_{1},t;R_{2},t)$, from which the intensity $I(\mathbf{r}_C,t)$ can be found. Note that the contribution of vacuum is negligibly small and can be disregarded safely. With the help of definition Eq.~(\ref{b.17}) for $\eta_{\mathbf{K}}(R_{i},t)$ one finds
\begin{eqnarray}\label{4.16}
\mathcal{D}_{\mathbf{K}}^{\text{vac}}(R_{1},t;R_{2},t) &=& 4 \sin^2\Big( \frac{\Omega_K t}{2}\Big)\, \big\vert \mathcal{G}^{\lambda}_{\mathbf{K}}(R_{1}) - \mathcal{G}^{\lambda}_{\mathbf{K}}(R_{2}) \big\vert^2 \, ,\\
\mathcal{D}_{\mathbf{K}}^{\text{ts-corr}}(R_{1},t;R_{2},t) &=& \sinh 2r_{K} \, \Big( \cos(2\phi_K) -2\cos(\Omega_K t -2\phi_K) + \cos(2\Omega_K t - 2\phi_K) \Big) \nonumber\\
&\times& \, \big\vert \mathcal{G}^{\lambda}_{\mathbf{K}}(R_{1}) - \mathcal{G}^{\lambda}_{\mathbf{K}}(R_{2}) \big\vert^2 \, . \nonumber
\end{eqnarray}
For highly squeezed PGWs, we can proceed by $\sinh^2 r_K \simeq e^{2r_K}/4$ and $\sinh 2r_K \simeq e^{2r_K}/2$.
Combining Eq.~(\ref{4.15}) and Eq.~(\ref{4.17}) yields
\begin{eqnarray}\label{4.17}
\mathcal{D}^{\text{ts}}(R_{1},t;R_{2},t) &=& \frac{1}{(2\pi)^2} \Big( \frac{\hbar\omega_k}{E_{\text{Pl}}}\Big)^2 \int \frac{dK}{K} \Big(\frac{e^{2r_K}}{4} \Big)\,\Big( 8 \, \sin^2\big( \frac{\Omega_K t}{2} \big)\, \sin^2\big( \frac{2\phi_K -\Omega_K t}{2}\big) \Big) \nonumber\\
&\times& \sum_{\lambda=+,\times}\int_0^{2\pi} d\Phi_K \int_0^{\pi} d(\cos\Theta_K) \, \big\vert \mathcal{G}^{\lambda}_{\mathbf{K}}(R_{1}) - \mathcal{G}^{\lambda}_{\mathbf{K}}(R_{2}) \big\vert^2\, . 
\end{eqnarray}
To proceed, we first compute the spatial factor. This step is left to App.~\ref{app:C.2}, where it is shown that
\begin{eqnarray}\label{4.18}
\big\vert \mathcal{G}_{\mathbf{K}}^{\lambda}(R_{1}) - \mathcal{G}_{\mathbf{K}}^{\lambda}(R_{2})\big\vert^2 &=& \Big\vert \sum_{\ell=0}^{\infty} ( i^{\ell}) \, \frac{2\ell+1}{2\pi}\Big( \int d\Omega_{\hat{r}} \, e_{ij}^{\lambda}[\hat{\mathbf{K}}] \, \hat{r}_i \, \hat{r}_j \, P_{\ell}(\cos\gamma) \Big) \\
&\times& \Big( \frac{j_{\ell}(KR)}{(KR)} - \frac{\int_{0}^{KR} j_{\ell}(u) \, du}{(KR)^2} \Big) \Big\vert^2 \, \Big( K(R_{1} - R_{2}) \Big)^2\nonumber\,.
\end{eqnarray}
In the above equation, $P_{\ell}(\cos\gamma)$ stands for the Legendre polynomial, and $\gamma$ shows the angle between $\hat{\mathbf{K}}$ and $\hat{r}$. Note that $\mathbf{r}$ is an integration variable inside the interaction volume $V$ where the interaction between GWs and spherical EM waves is non-vanishing (see Fig.~\ref{fig2}). Moreover, $ j_{\ell}(u) $ shows a spherical Bessel function of the first kind. The distance $R_{1}-R_{2}$ can be calculated from Fig.~\ref{fig4}, according to
\begin{eqnarray}\label{4.19}
R_{1} &=& r_1 - \hat{\mathbf{s}}_1 \cdot \mathbf{r}'\, ,\\
R_{2} &=& r_2 - \hat{\mathbf{s}}_2 \cdot \mathbf{r}'\, , \nonumber 
\end{eqnarray}
where $r_i$ shows the distance of the $i$-th detector with respect to some origin, $\mathbf{r}'$ shows the position of the source element from the origin, and $\hat{\mathbf{s}}_i$ shows the unit vector from the source element to the $i$-th detector. Since the PGW background is homogeneous and isotropic, we may proceed by taking the center of the planar disc as the origin of the coordinate system. Hence,
\begin{eqnarray}\label{4.20}
R_{1} - R_{2} &=& r_1 - r_2 + (\hat{\mathbf{s}}_2 - \hat{\mathbf{s}}_1)\cdot \mathbf{r}'\, .
\end{eqnarray}
The path difference $r_1 - r_2$ produces a constant phase factor $e^{-ik(r_1-r_2)}$, which is irrelevant in subsequent calculations. One can assume $r_1 \simeq r_2 \equiv R$ for simplification. Now, one can define polar coordinates $\mathbf{r}'_{k} \equiv (\rho'\cos\phi', \rho\sin\phi')$ where $0\leq \rho' \leq a$ and $0 \leq \phi' \leq 2\pi$. Eq.~(\ref{4.20}) implies that only the projection of $\hat{\mathbf{s}}_2 - \hat{\mathbf{s}}_1$ onto the planar disc is encountered, and we can write $ (\hat{\mathbf{s}}_{2\perp} - \hat{\mathbf{s}}_{1\perp}) \equiv (w\cos\psi, w\sin\psi)$ where $w=|\hat{\mathbf{s}}_{2\perp} - \hat{\mathbf{s}}_{1\perp}|$. From Fig.~\ref{fig4} one can write the projected unit vectors as $\hat{\mathbf{s}}_{i\perp} = (\frac{x_i}{R},\frac{y_i}{R}, 0)$ with $i=1,2$ and $(x_i,y_i)$ representing the coordinates of detectors (projected to the source plane); hence $w = \frac{d}{R}$. Consequently, we can write
\begin{eqnarray}\label{4.21}
R_{1} - R_{2} &=& (\hat{\mathbf{s}}_{2\perp} - \hat{\mathbf{s}}_{1\perp}) \cdot (\rho'\cos\phi', \rho\sin\phi') = \rho'\, w \, \cos(\phi'-\psi)\, .
\end{eqnarray}
Combining Eqs.~(\ref{4.17}, \ref{4.18}, \ref{4.21}) yields 
\begin{eqnarray}\label{4.22}
\mathcal{D}^{\text{ts}}(R_{1},t;R_{2},t) &=& \frac{1}{(2\pi)^2} \Big( \frac{\hbar\omega_k}{E_{\text{Pl}}}\Big)^2 \int dK \Big(\frac{K \, e^{2r_K}}{4} \Big)\,\Big( 8 \, \sin^2\big( \frac{\Omega_K t}{2} \big)\, \sin^2\big( \frac{2\phi_K -\Omega_K t}{2}\big) \Big) \\
&\times& \int_0^{2\pi} d\Phi_K \int_0^{\pi} d(\cos\Theta_K) \, \sum_{\lambda=+,\times} \Big\vert \sum_{\ell=0}^{\infty} ( i^{\ell}) \, \frac{2\ell+1}{2\pi}\Big( \int d\Omega_{\hat{r}} \, e_{ij}^{\lambda}[\hat{\mathbf{K}}] \, \hat{r}_i \, \hat{r}_j \, P_{\ell}(\cos\gamma) \Big) \nonumber\\
&\times& \Big( \frac{j_{\ell}(KR)}{(KR)} - \frac{\int_{0}^{KR} j_{\ell}(u) \, du}{(KR)^2} \Big) \Big\vert^2\, \Big( \rho' \, w \, \cos(\phi'-\psi) \Big)^2 \nonumber\\
&\equiv& \xi_{\text{ts}}^{-2}(R,t) \, \Big( \rho'\, w\, \cos(\phi'-\psi) \Big)^2\, . \nonumber
\end{eqnarray}
In the last line, we have defined the incoherence length induced by two-mode squeezed PGWs according to
\begin{eqnarray}\label{4.23}
\xi_{\text{ts}}(R,t) &=& \bigg[ \frac{1}{(2\pi)^2} \Big( \frac{\hbar\omega_k}{E_{\text{Pl}}}\Big)^2 \int dK \Big(\frac{K \, e^{2r_K}}{4} \Big)\,\Big( 8 \, \sin^2\big( \frac{\Omega_K t}{2} \big)\, \sin^2\big( \frac{2\phi_K -\Omega_K t}{2}\big) \Big) \\
&\times& \int_0^{2\pi} d\Phi_K \int_0^{\pi} d(\cos\Theta_K) \, \sum_{\lambda=+,\times} \Big\vert \sum_{\ell=0}^{\infty} ( i^{\ell}) \, \frac{2\ell+1}{2\pi}\Big( \int d\Omega_{\hat{r}} \, e_{ij}^{\lambda}[\hat{\mathbf{K}}] \, \hat{r}_i \, \hat{r}_j \, P_{\ell}(\cos\gamma) \Big) \nonumber\\
&\times& \Big( \frac{j_{\ell}(KR)}{(KR)} - \frac{\int_{0}^{KR} j_{\ell}(u) \, du}{(KR)^2} \Big) \Big\vert^2 \bigg]^{-1/2} \, ,\nonumber
\end{eqnarray}
which is a function of the distance of the source to the Earth, $R$, and the interaction time $t$. Plugging Eq.~(\ref{4.22}) into Eq.~(\ref{4.13}) yields
\begin{eqnarray}\label{4.24}
g^{(1)}_{\text{ts}}(R_{1},t;R_{2},t) &=& e^{-\xi_{\text{ts}}^{-2} \, \big( \rho'\, w\, \cos(\phi'-\psi) \big)^2 }\,
\end{eqnarray}
and Eq.~(\ref{4.9}) for the intensity $I(\mathbf{r}_C,t)$ becomes
\begin{eqnarray}\label{4.25}
\hspace*{-0.7cm}I(\mathbf{r}_C,t) &=& \frac{2\, |\mathcal{E}|^2 \big\langle \hat{n}_k \big\rangle}{R^2} \bigg\{ 1 + \frac{1}{\pi a^2}\, \Re\,\Big[ \int_0^{a} d\rho' \, \rho' \, \int_0^{2\pi} d\phi' \, e^{-\xi_{\text{ts}}^{-2} \, \big( \rho'\, w\, \cos(\phi'-\psi) \big)^2 } \, e^{-i\,k\,\rho'\, w\, \cos(\phi'-\psi) } \Big] \bigg\}\,. \qquad\quad
\end{eqnarray}

According to the definition, the \textit{visibility}, which is a measure of the sharpness of the interference fringes, is defined by \cite{scully1997quantum}
\begin{eqnarray}\label{4.26}
\mathcal{V}(\mathbf{r}_C,t) \equiv \frac{I_{\text{max}}(\mathbf{r}_C,t) - I_{\text{min}}(\mathbf{r}_C,t)}{I_{\text{max}}(\mathbf{r}_C,t) + I_{\text{min}}(\mathbf{r}_C,t)}.
\end{eqnarray}
Here, $I_{\text{max}}(\mathbf{r}_C,t)$ and $I_{\text{min}}(\mathbf{r}_C,t)$ represent the maximum and minimum values that the intensity assumes at $\mathbf{r}_C$. The first term in Eq.~(\ref{4.25}) shows the constant intensity at the detectors, and the second term is conveniently written as the real part of a complex degree of coherence, namely
\begin{eqnarray}
\delta &\equiv& \frac{1}{\pi a^2}\, \Big[ \int_0^{a} d\rho' \, \rho' \, \int_0^{2\pi} d\phi' \, e^{-\xi_{\text{ts}}^{-2} \, \big( \rho'\, w\, \cos(\phi'-\psi) \big)^2 } \, e^{-i\,k\,\rho'\, w\, \cos(\phi'-\psi) } \Big] \,. \nonumber
\end{eqnarray}
Considering the complex number $\delta \equiv |\delta| \, e^{i\alpha_{\delta}}$ with $\alpha_{\delta}\equiv \text{arg}\,\delta$, one has $\Re[\delta] = |\delta|\, \cos\alpha_{\delta}$, and the maximum and minimum intensities are followed accordingly. Hence, with the help of Eq.~(\ref{4.25}), the visibility is determined by
\begin{eqnarray}\label{4.27}
\mathcal{V}_{\text{ts}}(\mathbf{r}_C,t) = \frac{1}{\pi a^2}\, \Big\vert \int_0^{a} d\rho' \, \rho' \, \int_0^{2\pi} d\phi' \, e^{-\xi_{\text{ts}}^{-2} \, \big( \rho'\, w\, \cos(\phi'-\psi) \big)^2 } \, e^{-i\,k\,\rho'\, w\, \cos(\phi'-\psi) } \Big\vert \,.
\end{eqnarray}
In App.~\ref{app:D} we compute the integration over $\phi'$ in Eq.~(\ref{4.27}). Eventually, the expression of the visibility in Eq.~(\ref{4.27}) is obtained as follows:
\begin{eqnarray}\label{4.28}
\hspace*{-1.2cm}\mathcal{V}_{\text{ts}}(\mathbf{r}_C,t) &=& 2\, \Big\vert \int_0^1 \, du \, u \, e^{-\Big( \frac{\nu^2}{2 ( k \xi)^2} \, u^2 \Big)} \Big( J_{0}\big(\nu u \big) \, J_0 \big(i\,\frac{\nu^2}{2 ( k \xi)^2} \, u^2 \big) \\
&+& 2 \, \sum_{n=1}^{\infty} (-i)^n \, J_{2n}\big( \nu u \big) \, J_{n} \big( i\,\frac{\nu^2}{2 ( k \xi)^2} \, u^2  \big) \Big) \Big\vert \,. \nonumber
\end{eqnarray}
In the above equation, we have defined $\nu = k\, d\, \theta/2$, where $d$ shows the distance between two detectors and $\theta$ stands for the angular diameter of the source. Moreover, $J_{n}(u)$ stands for the Bessel function of the first kind. Eq.~(\ref{4.28}) explicitly shows the \textit{blurring} of the visibility induced by PGWs that could completely suppress the interference pattern. The incoherence effect depends on the interaction time $t$, as well as the distance of the source to the Earth $R$. On top of that, the length scale $\xi_{\text{ts}}(R,t)$ depends on the squeezing parameters $(r_K,\phi_K)$, which on their own depend on the inflationary parameters such as the tensor-to-scalar ratio. Thus, the order of magnitude of the incoherence crucially depends on the level of generated PGWs during inflation. On the other hand, successful measurement of spatial correlations of most distant sources is convincing evidence that spatial correlations have survived during their journey to the Earth. In the following, we describe the very long baseline interferometry measurements of the angular diameter $\theta$ and its implications on the level of incoherence induced by the underlying PGWs background. 


\section{Implications of VLBI measurements on the level of PGWs background}\label{sec:5}

In this section, we discuss how the van Citter-Zernike correlations are affected by the presence of quantum GWs. We evaluate the incoherence length induced by PGWs, $\xi_{\text{ts}}$, and try to explore it with the help of the angular size-redshift $\theta-z$ measurements made by VLBI means. As we shall see, inflation-generated PGWs are feeble to ruin the van Citter-Zernike correlations, provided that the mean number of gravitons in TS state is determined by Eq.~(\ref{2.6}).


\subsection{van Citter-Zernike theorem and Very Long Baseline Interferometry (VLBI)}\label{subsec:5.1}

In this section, we first ignore the presence of GW background and describe the way the VLBI method measures the angular size of an object by exploiting the spatial coherence of a distant source. In this case, the van Citter-Zernike theorem expresses the field correlations at two points in space, generated by a spatially incoherent, quasi-monochromatic, planar source, according to \cite{mandel1995optical}
\begin{eqnarray}\label{5.1}
j(\mathbf{r}_1,\mathbf{r}_2) = \frac{2J_{1}(\nu)}{\nu}.
\end{eqnarray}
\begin{figure*}[htb]
\centering
\includegraphics[
width=0.6\columnwidth]{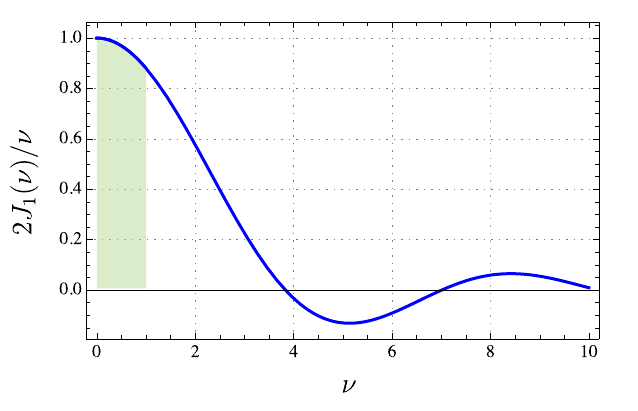}
\caption{The mutual intensity $j(\mathbf{r}_1,\mathbf{r}_2) = 2J_{1}(\nu)/\nu$ versus the dimensionless parameter $\nu = k \left( \frac{a}{R} \right) d$. The colored region shows $x\leq 1$ corresponding to the coherence length $0 \leq d \leq \frac{2}{k\theta}$, where correlations decrease from unity to a value of $0.88$. The mutual intensity vanishes for the first time for $\nu\simeq3.83$.}
\label{fig5}
\end{figure*}
The function $j(\mathbf{r}_1,\mathbf{r}_2)$ is previously defined in Eq.~(\ref{4.11}) and expresses the equal-time spatial correlations of the EM field at two points $\mathbf{r}_1$ and $\mathbf{r}_2$. In Eq.~(\ref{5.1}), the dimensionless parameter $\nu$ is defined as $\nu = k \left( \frac{a}{R} \right) d = k\, d\, \theta/2$ with $k$ being the central wave number of the quasi-monochromatic radiation, $d$ standing for the separation of two telescopes (as shown in Fig.~\ref{fig3}), $a$ the radius of the planar source $\sigma$, and $R$ its distance to the Earth. The behavior of the function $j(\mathbf{r}_1,\mathbf{r}_2)$ versus the dimensionless parameter $x$ is depicted in Fig.~\ref{fig5}. Starting from complete coherence $j(\mathbf{r}_1,\mathbf{r}_2)=1$ at zero separation distance, correlations decrease steadily by increasing the distance between the detectors from $d = 0$ to $d = (3.83) k^{-1} (R/a)$ where complete incoherence $j(\mathbf{r}_1,\mathbf{r}_2)=0$ happens for the first time. Then, correlations slightly increases again before vanishing for the second time. Practically, a drop from unity, which does not exceed about $12\%$, is not considered very significant \cite{mandel1995optical}. This happens when $\nu < 1$, i.e.,
\begin{eqnarray}\label{5.2}
 k\, d\,\theta/2 \leq 1\, ,
\end{eqnarray}
when correlations get a value of $0.88$. Here, $\theta = 2a/R$ represents the angular diameter subtended by the source when viewed from the Earth. In practice, by varying the separation between two telescopes, one monitors the variation of the visibility, which gives rise to the measurement of the angular size of the source. Successful measurement of the angular size of distant objects based on visibility implies that the correlation length of the source is at least equal to the projected baseline of the VLBI instrument. 


\subsection{VLBI constraint on PGWs}\label{subsec:5.2}

\begin{figure*}[htb]
\centering
\includegraphics[
width=0.6\columnwidth]{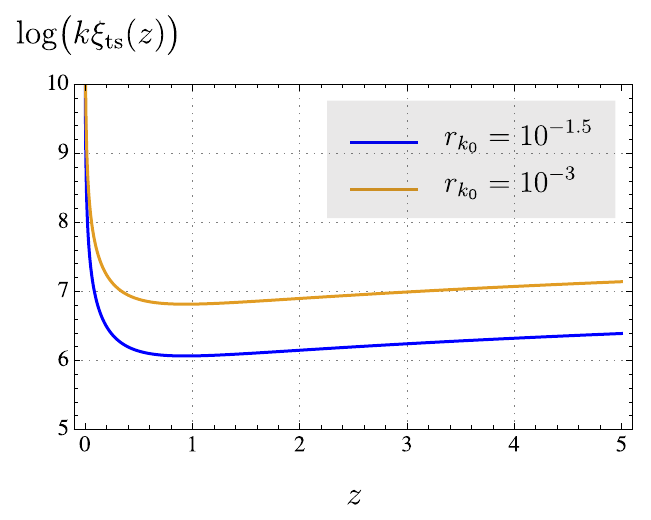}
\caption{The dimensionless incoherence length $k\,\xi_{\text{ts}}(z)$ in logarithmic scale versus the redshift $z$, based on Eq.~(\ref{4.23}). The blue and orange curves correspond to two values, $r_{k_0}=0.032\simeq 10^{-1.5}$ and $r_{k_0} = 10^{-3}$, for the tensor-to-scalar ratio, respectively, corresponding to existing upper limits on this parameter made by \textit{Planck PR4} \cite{tristram2022improved}. Other parameters are chosen as $\beta=-2$, $\beta_s=1$, and $T_{\text{reh}}=10^{8}\,$GeV. For the whole range of the redshift, $k\,\xi_{\text{ts}}$ remains larger than $10^{6}$.}
\label{fig6}
\end{figure*}

The incoherence induced by two-mode squeezed PGWs is encapsulated in the visibility $\mathcal{V}_{ts}(\mathbf{r}_C,t)$ given by Eq.~(\ref{4.28}). To visualize the behavior of the visibility, we first evaluate the dimensionless parameter $k\,\xi_{\text{ts}}$, as implied by Eq.~(\ref{4.23}). Note that the distance of the source and the interaction time can be expressed in terms of the redshift of the source, $z$ (see App.~\ref{app:C.3}). Thus, the plot of $k\, \xi_{\text{ts}}(z)$ in logarithmic scale versus the redshift $z$ is depicted in Fig.~\ref{fig6}. In this plot, the blue and orange curves correspond to two values, $r_{k_0}=0.032\simeq 10^{-1.5}$ and $r_{k_0}\simeq 10^{-3}$ for the tensor-to-scalar ratio, corresponding to existing upper limits on this parameter made by \textit{Planck PR4} \cite{tristram2022improved}. At very small distances and interaction times $z\rightarrow 0$, the incoherence length takes a large value so that $g^{(1)}_{k}(R_{1},t;R_{2},t) \rightarrow e^{-ik(R_1-R_2)}$ as it should. By increasing the redshift, the incoherence length takes smaller values, down to $k \, \xi_{\text{ts}} \simeq 10^{6}$ for $r_{k_0}=10^{-1.5}$. For the whole redshift range $0\leq z \leq 5$, we have $k\, \xi_{\text{ts}} \gtrsim 10^{6}$.

\begin{figure*}[htb]
\centering
\includegraphics[
width=0.6\columnwidth]{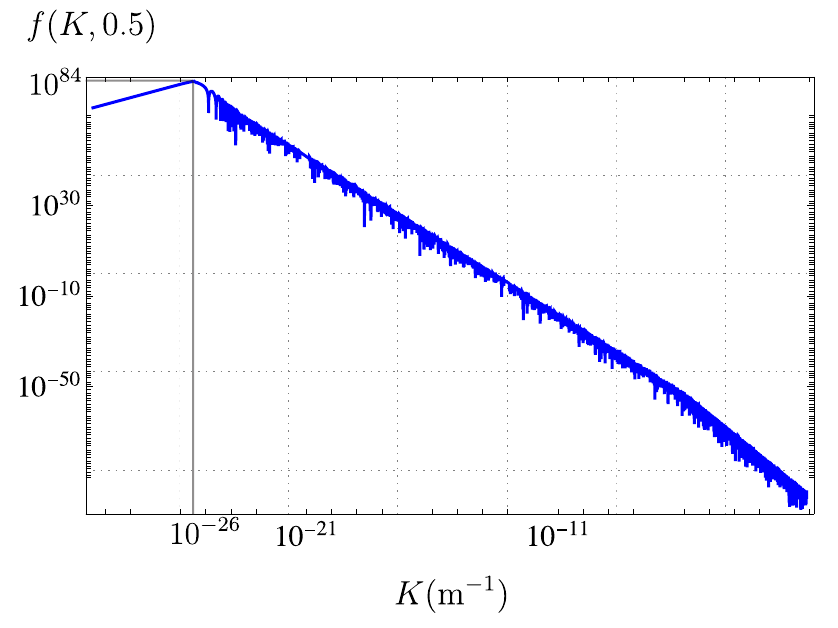}
\caption{Behavior of the integrand $f(K,z)$ for $z=0.5$ versus the wave number $K$ based on Eq.~(\ref{5.3}). Scales with $K\simeq 4\times 10^{-26}\, \text{m}^{-1} \sim K_{H}$ have the highest contribution in the incoherence length $\xi_{\text{ts}}(z)$, and the effect of longer and shorter waves automatically cancels out.}
\label{fig7}
\end{figure*}

These plots are numerically calculated from Eq.~(\ref{4.23}), where only the first three terms corresponding to $\ell=0, 1, 2$ in the summation $\sum_{\ell}$ are encountered, since the effect of higher-order terms is negligible and can be disregarded safely. Moreover, note that the integration over the GWs wave number $K$ should be performed between lower and upper limits, namely $K_{\text{low}}$ and $K_{\text{up}}$. In order to find the effective range of $K$, which has the prominent contribution in integration, it is instructive to plot the integrand in Eq.~(\ref{4.23}), namely
\begin{eqnarray}\label{5.3}
f(K,z) &\equiv& \frac{1}{(2\pi)^2} \Big( \frac{\hbar\omega_k}{E_{\text{Pl}}}\Big)^2 \Big(\frac{K \, e^{2r_K}}{4} \Big)\,\Big( 8 \, \sin^2\big( \frac{\Omega_K t(z)}{2} \big)\, \sin^2\big( \frac{2\phi_K -\Omega_K t(z)}{2}\big) \Big) \\
&\times& \int_0^{2\pi} d\Phi_K \int_0^{\pi} d(\cos\Theta_K) \, \sum_{\lambda=+,\times} \Big\vert \sum_{\ell=0}^{\infty} ( i^{\ell}) \, \frac{2\ell+1}{2\pi}\Big( \int d\Omega_{\hat{r}} \, e_{ij}^{\lambda}[\hat{\mathbf{K}}] \, \hat{r}_i \, \hat{r}_j \, P_{\ell}(\cos\gamma) \Big) \nonumber\\
&\times& \Big( \frac{j_{\ell}\big(KR(z)\big)}{\big( KR(z) \big)} - \frac{\int_{0}^{KR(z)} j_{\ell}(u) \, du}{\big(KR(z)\big)^2} \Big) \Big\vert^2 \, ,\nonumber
\end{eqnarray} 
where $t(z)$ and $R(z)$ are given by Eqs.~(\ref{c.30}, \ref{c.31}). The plot of $f(K,z)$ is sketched for $z=0.5$ in Fig.~\ref{fig7}, for $r_{k_0}=0.032$. It can be seen that the major contribution of the integrand $f(K,0.5)$ comes from those waves with $K \simeq 4\times 10^{-26}\,\text{m}^{-1} \sim K_{H}$, and the contribution of very low-frequency or very high-frequency GWs becomes negligible. In this way, depending on the length of the probe, the lower and upper wave numbers $K_{\text{low}}$ and $K_{\text{up}}$, which have the major contribution, can be found.
According to Eq.~(\ref{4.28}) for the visibility, one can check that for very large values of $k\, \xi_{\text{ts}} \gg \nu$, the decaying factor goes to unity and the visibility reduces to
\begin{eqnarray}\label{5.4}
\hspace*{-1.2cm}\mathcal{V}_{\text{ts}}(\mathbf{r}_C,t) &\simeq& 2\, \Big\vert \int_0^1 \, du \, u \, J_{0}\big(\nu u \big) \Big\vert = \Big\vert \frac{2 J_1(\nu)}{\nu} \Big\vert \quad , \quad k\,\xi_{\text{ts}} \gg k\,d\,\theta/2 \, 
\end{eqnarray} 
which is nothing but the van Citter-Zernike correlation Eq.~(\ref{5.1}). Here, we have used the identities $J_0(0) =1$, $J_n(0) = 0$ for $n\geq 1$, and $ \int_0^{a} J_0(u)\, u\, du = a\, J_1(a)$. In a VLBI measurement, the visibility is non-vanishing provided that $\nu \simeq 1$ or, equivalently, $k\, d\, \theta/2 \simeq 1$ (see Fig.~\ref{fig5}). For typical values of the tensor-to-scalar ratio, namely $r_{k_0} \leq 0.032\simeq 10^{-1.5}$ as constrained by \textit{Planck PR4} \cite{tristram2022improved}, the incoherence induced by PGWs takes large values $k \xi_{\text{ts}} \geq 10^{6}$, which is much higher than $\nu \simeq 1$. As a result, PGWs are too weak to be detected by VLBI means. This is mainly due to the very small coupling strength $\propto \big(\frac{\hbar\omega_k}{E_{\text{Pl}}} \big)^2$ between the EM field and GWs, despite the large graviton content of the two-mode squeezed PGWs. 

The situation can change whenever $k\, \xi_{\text{ts}} \sim \nu = k\, d\, \theta/2$. This is possible whenever the coupling between the EM field and the underlying background is stronger, for instance, when the EM field suffers from material fields such as primordial density perturbations. In this case, it is natural to think of the gravitational lensing created by matter fluctuations as a source of incoherence of the EM phase. In this regard, the present formalism can be straightforwardly generalized to encounter the incoherence induced by the primordial density perturbations.

The other issue concerns the quantum-to-classical transition \cite{burgess2023minimal,martin2022discord}. The possible decoherence of PGWs may alter the incoherence of the EM field. Basically, in a given decoherence schema for PGWs, it is crucial to follow the dynamics of the \textit{mean number of gravitons} and \textit{quantum correlations between gravitons} in the course of Universe expansion. This information is encoded in the diagonal and off-diagonal elements (in the Fock space representation) of the density matrix describing the evolution of PGWs, namely $\rho_{\text{pgws}}^{mn}(\eta)$. On the other hand, the interaction of PGWs with matter fields in subsequent epochs after they have been generated can alter the quantum state of PGWs and turn them into coherent states \cite{kanno2019nonclassical}. Hence, the effect of PGWs subjected to decoherence or subsequent interactions deserves further assay that can be addressed within the present formalism. 

The main goal of the present study is to introduce and evaluate a new way to look forward to quantum gravitational effects using the well-established interferometric techniques and to infer the corresponding model parameters based on that.


\section{Summary and conclusion} \label{sec:6}

In a nutshell, this work promotes the idea of using gravitational-induced spatial incoherence of the EM field radiated from distant objects as a new approach to probe the quantum gravitational universe. We present a detailed formulation of the field theory of interaction between quantum GWs and an EM probe. As an application, we investigate the spatial coherence of the electromagnetic field emitted from distant objects interacting with the background of primordial gravitational waves, which encode historical features of the expanding universe in itself. It turns out that spatial correlation of the EM field, which in the absence of PGWs is simply governed by the van Citter-Zernike theorem, is modified by the presence of PGWs such that it blurs the visibility in an interferometric setup. However, successful observation of the visibility pattern in VLBI experiments implies that the incoherence induced by two-mode squeezed PGWs is so weak that it cannot ruin the van Citter-Zernike correlations. This is a direct consequence of the very weak interaction between GWs and an EM probe, despite the existence of a high graviton content in the two-mode squeezed PGWs. Nevertheless, the present method can be generalized to account for the effect established by primordial density fluctuations, which couple to the EM field more vigorously, on the phase correlations of the EM field.

The aim of the present study is to highlight the capability of interferometric methods, in particular the very long baseline interferometry technique, to inquire about the primordial universe through the gravitationally induced phase change of the electromagnetic field. Altogether, the VLBI method seems promising for use as a new observatory to search for quantum features of spacetime.


\section{Supplementary materials} \label{sec:7}

\appendix

\section{PGWs spectrum and related parameters} \label{app:A}

\subsection{Increase parameters $\zeta_E, \zeta_2, \zeta_s, \zeta_1$} \label{app:A.1}

In the $\Lambda CDM$ framework, one can show that the increase parameter during the $\Lambda$-dominated stage is given by 
\begin{eqnarray}\label{a.1}
\zeta_E=1+z_E=(\Omega_{\Lambda}/\Omega_m) ^{1/3} \simeq 1.33,
\end{eqnarray}
\noindent given the current values $\Omega_{\Lambda} \sim 0.7$ and $\Omega_m \sim 0.3$. Here, $z_E$ is the redshift at the matter-dark energy equality. The increase factor at matter-radiation equality is
\begin{eqnarray}\label{a.2}
\zeta_2= \left( 1+z_{eq} \right) \zeta_E^{-1} \simeq 2547,
\end{eqnarray}
\noindent where $z_{\text{eq}}=3387$ \cite{collaboration2020planck}. The increase of scale factor during the reheating, namely $\zeta_s$, depends on the reheating temperature $T_{\text{reh}}$ through \cite{tong2013relic}
\begin{eqnarray}\label{a.3}
\zeta_s = \frac{T_{\text{reh}}}{T_{\text{CMB}}(1+z_{eq})}\left( \frac{g_{\ast s}}{g_{\star s}} \right)^{1/3},
\end{eqnarray}
\noindent where $T_{\text{CMB}}=2.348\cdot 10^{-13}$ GeV \cite{fixsen2009temperature}. Here, $g_{\ast s}$ and $g_{\star s}$ count the effective number of relativistic species contributing to the entropy during the reheating and recombination, respectively, and will be taken as $g_{\ast s}\simeq 200$ and $g_{\star s}=3.91$ \cite{watanabe2006improved}, as was also employed in \cite{tong2013relic}. Moreover, under the quantum normalization condition, the increase parameter $\zeta_1$ is expressed in terms of $\zeta_s, \zeta_2, \zeta_E$ according to Eq.~(\ref{2.8}).


\subsection{Spectral amplitude of PGWs}\label{app:A.2}

The dimensionless \textit{spectral amplitude} of PGWs, $h(k,\eta)$, is usually defined by \cite{tong2013using}
\begin{eqnarray}\label{a.4}
\int_{K_E}^{K_1} h^2(K,\eta) \frac{dK}{K} \equiv \langle \text{TS}(\eta)| \hat{h}_{ij}(\mathbf{r},\eta_{\text{ini}}) \, \hat{h}^{ij}(\mathbf{r},\eta_{\text{ini}}) | \text{TS}(\eta) \rangle,
\end{eqnarray}

\noindent where the r.h.s. is the variance of the field in the two-mode squeezed state $\big\vert \text{TS} \big\rangle$, given by Eq.~(\ref{c.1}). The tensor field $\hat{h}_{ij}(\mathbf{r},\eta_{\text{ini}})$ is given by Eq.~(\ref{3.3}). Note that, as we discussed in Sec.~(\ref{subsec:2.2}), in the Schr\"{o}dinger picture the evolution of perturbations is included in the state of the system, while the field operator is given by its initial expression. Moreover, the lower and upper limits of the integral are considered as $K_{\text{low}} = K_E$ and $K_{\text{up}} = K_1$. Inserting Eq.~(\ref{3.3}) into Eq.~(\ref{a.4}) one may show that
\begin{eqnarray}\label{a.5}
h(K,\eta) = \frac{4 \ell_{\text{Pl}}}{\sqrt{\pi}a(\eta)} K \Big( |u_K(\eta)|^2 + |v_K(\eta)|^2 + u_K(\eta) v_K(\eta) + u^{\ast}_K(\eta) v^{\ast}_K(\eta) \Big)^{1/2}.
\end{eqnarray}
where the functions $u_{K}(\eta)$ and $v_{K}(\eta)$ are given by Eq.~(\ref{2.3}) with the initial conditions $u_K(\eta_{\text{ini}}) = 1$ and $v_K(\eta_{\text{ini}})=0$ \cite{martin2016quantum}. Thus, using the expression Eq.~(\ref{a.5}), one may find the evolution of the spectral amplitude $h(K,\eta)$ easily, as soon as the evolution of the functions $u_{K}(\eta)$ and $v_{K}(\eta)$ is determined. Another important application of Eq.~(\ref{a.5}) is to build the relationship between the squeezing parameters $(r_K,\phi_K)$ and the spectral amplitude $h(K,\eta)$. Especially, by substitution of Eq.~(\ref{2.3}) for the functions $u_K$ and $v_K$ into Eq.~(\ref{a.5}) and assuming $r_K\gg1$ for super-Hubble modes ($K \leq 2\pi\mathcal{H}$), one obtains the following relationship
\begin{eqnarray}\label{a.6}
h(K,\eta) = 8\sqrt{\pi} \left( \frac{\ell_{\text{Pl}}}{\ell_H} \right) \left( \frac{K}{K_H} \right) e^{r_K(\eta)} \cos\phi_K(\eta). 
\end{eqnarray}
Eq.~(\ref{a.6}) is in accordance with Eq.~(31) of \cite{grishchuk2001relic}. Note that, according to Eq.~(\ref{2.3}), the dynamics of variable $\phi_k$ are determined once the dynamics of $r_K$ are specified. Hence, the evolution of $h(K,\eta)$ can be solely given by the evolution of the squeezing amplitude $r_K$. Inversely, one may obtain the approximate expression of the squeezing factor $e^{r_K}$ from $h(K,\eta)$ according to
\begin{eqnarray}\label{a.7}
e^{r_K(\eta)}= \frac{1}{8\sqrt{\pi}} \left( \frac{\ell_H}{\ell_{\text{Pl}}} \right) \left( \frac{K_H}{K} \right) h(K,\eta). 
\end{eqnarray}
\noindent in the super-Hubble regime, where the oscillatory factor $\cos\phi_k$ tends to 1 (see Eq.~(\ref{2.4})). Approximate solutions for $h(K,\eta)$ can be found in the super-Hubble regime. Hence, it turns out that the spectral amplitude is determined by \cite{tong2012revisit}
\begin{eqnarray}\label{a.8}
h(K,\eta) =
\begin{cases}
    A \left( \frac{K}{K_H} \right)^{2+\beta} \quad,\quad K\leq K_E\\
    A \left( \frac{K}{K_H} \right)^{\beta-\gamma} \zeta_E^{-\frac{2+\gamma}{\gamma}}\quad,\quad K_E \leq K \leq K_H\\
    A \left( \frac{K}{K_H} \right)^{\beta} \zeta_E^{-\frac{2+\gamma}{\gamma}}\quad,\quad K_H \leq K \leq K_2\\
    A \left( \frac{K}{K_H} \right)^{1+\beta} \left( \frac{K_H}{K_2} \right) \zeta_E^{-\frac{2+\gamma}{\gamma}}\quad,\quad K_2 \leq K \leq K_s\\
    A \left( \frac{K}{K_H} \right)^{1+\beta-\beta_s} \left( \frac{K_s}{K_H} \right)^{\beta_s} \left( \frac{K_H}{K_2} \right) \zeta_E^{-\frac{2+\gamma}{\gamma}}\quad,\quad K_s \leq K \leq K_.
  \end{cases}
\end{eqnarray}
where the coefficient $A$ is discussed in Sec.~\ref{subsec:2.3}.


\subsection{Characteristic wave numbers $K_E, K_H, K_2, K_s$ and $K_1$} \label{app:A.3}

The conformal wave number at a given jointing time $\eta_{x}$ is defined as $K_{x} \equiv K(\eta_{x}) = 2\pi \mathcal{H}(\eta)$, assuming the wave mode crosses the horizon when $\gamma = 1/\mathcal{H}$ with $\mathcal{H}$ being the Hubble radius (this definition for horizon-crossing is also used in \cite{grishchuk2001relic,miao2007analytic}). Thus $K_{H} = 2\pi \mathcal{H}(\eta_H) = 2\pi \gamma$ is the conformal Hubble wave number. The physical wave number at present is related to the conformal wave number $K$ according to $K^{\text{ph}} = K/a(\eta_H) = K/ \ell_{H}$. Hence, $K_H^{\text{ph}} = 2\pi \gamma/ \ell_H = 2\pi H_0/c$ with $H_0$ being today's Hubble frequency. One can show that characteristic wave numbers at different jointing points $\eta_1$, $\eta_{s}$, $\eta_{2}$, and $\eta_{E}$ are related to the increase parameters according to \cite{tong2013using}
\begin{eqnarray}\label{a.10}
\frac{K_E}{K_H} = \zeta_E^{-\frac{1}{\gamma}}\quad,\quad \frac{K_2}{K_E} = \zeta_2^{\frac{1}{2}}\quad , \quad\frac{K_s}{K_2} = \zeta_s \quad,\quad \frac{K_1}{K_s} = \zeta_1^{\frac{1}{1+\beta_s}}.  
\end{eqnarray}
Similar expressions hold for the PGWs frequencies, $\Omega_K = cK$. Given $H_0=67.4\,$km s$^{-1}$ Mpc$^{-1}$, one has $K_{H}^{ph}\sim 4.52 \times 10^{-26}\,$, $K_E^{ph} = 3.4 \times 10^{-26}\,$ and $K_2 = 1.71 \times 10^{-24}\,$ in $m^{-1}$ units. The values of $K_s$ and $K_1$, corresponding to waves that crossed the horizon at the end of the reheating and inflationary stages, respectively, are determined by $(\zeta_s, \zeta_1, \beta_s)$. For $T_{\text{reh}}=10^{8}\,$GeV, $K_s = 7.9$ m$^{-1}$ and $K_1$ also depends on the values of $(\beta, \text{r}_{k_0})$ through Eq.~(\ref{2.8}). For $\beta=-2$ and $\text{r}_{k_0} = 10^{-1.5}$ adopted in this study, one has $K_1 = 0.08$ m$^{-1}$.

Basically, $\beta_s$ and $T_{\text{reh}}$ determine characteristic features of the expanding Universe during the reheating stage, so they only affect the frequencies $\Omega_s$ and $\Omega_1$, e.g., waves re-entering the horizon during the reheating stage (see panel $(a,b)$ of Fig~\ref{fig1}). However, the main contribution to the incoherence mechanism comes from ultra-low frequency PGWs, say $\Omega_K \leq \Omega_s$, which possess higher squeezing amplitudes. Throughout the paper, unless it is stated, we mostly take $\beta_s=1$ and $T_{\text{reh}}=10^{8}\,$GeV in our calculations.

The value of the upper frequency of PGWs, $\Omega_{K_1} =c K_1$, depends on the choice of the increase parameter $\zeta_1$. Modes with frequency higher than $\Omega_{K_1}$ have been decayed by the expansion of the universe and have not been squeezed at all. In any case, the value of $\Omega_{K_1}$ should be below the constraint from the rate of the primordial nucleosynthesis, i.e., $\Omega_{K_1} \lesssim 10^{10}\,$Hz \cite{zhang2005relic,tong2013relic}, which is the case in our study.


\section{Heisenberg equation of the EM field in the presence of GWs }\label{app:B}

In order to solve the Heisenberg equation, we first initialize the \textit{time evolution operator} generated by the Hamiltonian Eq.~(\ref{3.51}). When written in the Schr\"{o}dinger picture, the total Hamiltonian becomes time-independent, which lets us write the time evolution operator of the total system as
\begin{eqnarray}\label{b.1}
\hat{U}(R,t) &=& e^{-\frac{i}{\hbar} \int_{0}^{t} dt' \hat{H}^{(S)}_{\text{tot}}(R)} = e^{-\frac{i}{\hbar} t \hat{H}^{(S)}_{\text{tot}}(R)} \, ,
\end{eqnarray}
where the total Hamiltonian in the Schr\"{o}dinger picture is considered as
\begin{eqnarray}\label{b.2}
\hat{H}^{(S)}_{\text{tot}} = \hat{H}^{(0)}_{\text{gw}} + \hat{H}^{(0)}_{\text{em}} + \hat{H}^{(S)}_{\text{int}} \,, 
\end{eqnarray}
and $\hat{H}^{(0)}_{\text{gw}}$ and $ \hat{H}^{(0)}_{\text{em}}$ are free Hamiltonians describing free GWs and EM fields given by Eqs.~(\ref{3.7}, \ref{3.48}). Thus $\hat{U}(R,t)$ is written as
\begin{eqnarray}\label{b.3}
\hat{U}(R,t) &=& \exp\bigg[ -i(\omega_k t)\hat{a}^{\dagger}_k \hat{a}_k - i \sum_{\lambda=+,\times} \int d^3\mathbf{K} \, (\Omega_K t) \, \hat{b}^{\dagger}_K \hat{b}_K \\
&+& \frac{i}{2\pi} \Big(\frac{\hbar\omega_k}{E_{\text{Pl}}} \Big) \sum_{\lambda=+,\times} \int \frac{d^3\mathbf{K}}{K^{3/2}} \, \Big( \hat{b}_K \, \mathcal{G}_{\mathbf{K}}^{\lambda} + \hat{b}^{\dagger}_K \, \mathcal{G}_{\mathbf{K}}^{\lambda\ast} \Big) \big( \Omega_K t\big) \bigg]\,. \nonumber
\end{eqnarray}
Here, for the sake of abbreviation, we use the notation $K\equiv(\mathbf{K},\lambda)$. The spatial function $\mathcal{G}_{\mathbf{K}}^{\lambda}$ is defined in Eq.~(\ref{3.52}). To break up $\hat{U}$ into separate exponential factors, we follow the trick introduced in \cite{bose1997preparation} for a single-mode optomechanical interaction and generalize it to the case of a continuum of GW modes with a space-dependent Hamiltonian. We define the unitary operator $\hat{T}$ as 
\begin{eqnarray}\label{b.4}
\hat{T} \equiv e^{-\hat{a}_{k}^{\dagger}\hat{a}_{k} \sum_{\lambda}\int d^3\mathbf{K} \kappa(K) \Big( \hat{b}_K^{\dagger} \mathcal{G}_{\mathbf{K}}^{\lambda\ast}(R) - \hat{b}_{K} \mathcal{G}_{\mathbf{K}}^{\lambda}(R) \Big)},
\end{eqnarray}
Here, we have defined
\begin{eqnarray}\label{b.5}
\kappa(K) \equiv \frac{1}{2\pi} \left( \frac{\hbar\omega_{k}}{E_{\text{Pl}}} \right) K^{-3/2}\, ,
\end{eqnarray}
which can be interpreted as a measure of the EM-GWs coupling strength. By performing straightforward algebra, one may show that 
\begin{eqnarray}\label{b.6}
&&\hat{T} \hat{b}_K \hat{T}^{\dagger} = \hat{b}_K + \hat{a}^{\dagger}_{k} \hat{a}_{k} \, \kappa(K) \, \mathcal{G}^{\lambda \ast}(R), \nonumber\\
&& \hat{T}\hat{b}_K^{\dagger} \hat{T}^{\dagger} = \hat{b}_K^{\dagger} + \hat{a}^{\dagger}_{k} \hat{a}_{k} \, \kappa(K) \, \mathcal{G}_{\mathbf{K}}^{\lambda}(R), \\
&&\hat{T}\hat{a}_{k}^{\dagger} \hat{a}_{k} \hat{T}^{\dagger} = \hat{a}_{k}^{\dagger} \hat{a}_{k}, \nonumber
\end{eqnarray}
which is a generalization of the result of \cite{guerreiro2020quantum} to the case of a detector with arbitrary size $R$ for which $KR \sim 1 $, and for spherical EM waves. For a unitary operator $\hat{T}$ and a given function $\hat{f}(\{\hat{X}_i\})$ of some operators $\hat{X}_i$, one has the property $\hat{T} \hat{f}(\{\hat{X}_i\}) \hat{T}^{\dagger} = \hat{f}(\{\hat{T}\hat{X}_i \hat{T}^{\dagger} \})$. With the help of this property, the time evolution operator may be written as
\begin{eqnarray}\label{b.7}
\hat{U} = \hat{T}^{\dagger} \Big( \hat{T} \hat{U} \hat{T}^{\dagger} \Big) \hat{T} &=& e^{-i\omega_{k} t \, \hat{a}^{\dagger}_{k} \hat{a}_{k}}  \, e^{i(\hat{a}^{\dagger}_{k} \hat{a}_{k} )^2 \sum_{\lambda=+,\times}\int d^3\mathbf{K} \, \kappa^2(K) \big\vert \mathcal{G}_{\mathbf{K}}^{\lambda}(R)\big\vert^2 \big( \Omega_K t \big)} \\
&\times& e^{ \hat{a}_{k}^{\dagger} \hat{a}_{k} \sum_{\lambda=+,\times} \int d^3\mathbf{K} \, \kappa(K) \Big( \hat{b}_K^{\dagger} \,\mathcal{G}_{\mathbf{K}}^{\lambda\ast}(R) - \hat{b}_{K} \, \mathcal{G}_{\mathbf{K}}^{\lambda}(R) \Big)} \nonumber\\
&\times& e^{-i\sum_{\lambda=+,\times} \int d^3\mathbf{K} \hat{b}_{K}^{\dagger} \hat{b}_K (\Omega_K t)} \, e^{-\hat{a}_{k}^{\dagger}\hat{a}_{k} \sum_{\lambda=+,\times} \int d^3\mathbf{K} \, \kappa(K) \Big( \hat{b}_K^{\dagger} \mathcal{G}_{\mathbf{K}}^{\lambda\ast}(R) - \hat{b}_{K} \mathcal{G}_{\mathbf{K}}^{\lambda}(R) \Big)} \nonumber\\
&\times& e^{i \int d^3\mathbf{K} \, \hat{b}_K^{\dagger}\hat{b}_K (\Omega_K t)}\,  e^{-i \int d^3\mathbf{K} \, \hat{b}_K^{\dagger}\hat{b}_K (\Omega_K t)}. \nonumber
\end{eqnarray}
Note that terms containing $\hat{a}^{\dagger}_{k} \hat{a}_{k}$ commute with each other, and in the last line we inserted the unit operator $\mathcal{I} = e^{i \int d^3\mathbf{K} \, \hat{b}_K^{\dagger}\hat{b}_K (\Omega_K t)} e^{-i \int d^3\mathbf{K} \, \hat{b}_K^{\dagger}\hat{b}_K (\Omega_K t)}$. By rearranging the terms and using the Baker–Campbell–Hausdorff formula, one may find the time evolution operator as follows:
\begin{eqnarray}\label{b.8}
\hspace*{-1.2cm}\hat{U}(R,t) &=& e^{i E(R,t)\, (\hat{a}_{k}^{\dagger} \hat{a}_{k})^2} \, e^{\hat{a}^{\dagger}_{k} \hat{a}_{k} \sum_{\lambda=+,\times} \int d^3 \mathbf{K} \, \kappa(K) \Big(\hat{b}^{\dagger}_{K} \nu_{\mathbf{K}}(R,t) - \hat{b}_{K}\nu^{\ast}_{\mathbf{K}}(R,t)\Big)} \\
&\times& e^{-i\omega_{k} t \, \hat{a}^{\dagger}_{k} \hat{a}_{k}} \, e^{-i \sum_{\lambda}\int d^3 \mathbf{K} \, \hat{b}_{K}^{\dagger} \hat{b}_{K} (\Omega_{K} t)}, \nonumber
\end{eqnarray}
In the above expression, we have defined
\begin{eqnarray} \label{b.9}
\nu_{\mathbf{K}}(R,t) &\equiv& \Big( 1-e^{-i\Omega_K t} \Big) \mathcal{G}_{\mathbf{K}}^{\lambda\ast}(R)\, ,\\
E(R,t) &\equiv& \sum_{\lambda=+,\times} \int d^3\mathbf{K} \, \kappa^2(K) \, \big\vert \mathcal{G}_{\mathbf{K}}^{\lambda}(R) \big\vert^2 \Big( \Omega_K t - \sin(\Omega_K t) \Big).\nonumber
\end{eqnarray}
Note that the spatial dependence of the time evolution operator is inherited from the spatial dependence of the underlying GWs background through the spatial factors $\mathcal{G}_{\mathbf{K}}^{\lambda}(R)$. In App.~\ref{app:C.2} we find the explicit expression of $\mathcal{G}_{\mathbf{K}}^{\lambda}(R)$ (see Eq.~(\ref{c.13})). 

With the help of $\hat{U}(R,t)$, both Schr\"{o}dinger and Heisenberg dynamics can be perused. We proceed by choosing the Heisenberg picture, where the equation of motion for the EM field operators can be solved without effort, thanks to the well-established cavity opto-mechanical investigations. The time evolution of the EM field operator $\hat{a}_{k}(R,t)$ can be obtained from $\hat{U}(R,t)$ according to
\begin{eqnarray}\label{b.10}
\hat{a}_{k} (R,t) &=& \hat{U}^{\dagger}(R,t) \, \hat{a}_{k} \hat{U}(R,t) \\
&=& \bigg( e^{i \int d^3\mathbf{K} \, \hat{b}_{K}^{\dagger} \hat{b}_{K} (\Omega_{K} t)} \, e^{i\omega_{k} t \, \hat{a}_{k}^{\dagger} \hat{a}_{k} } \, e^{-\hat{a}_{k}^{\dagger} \hat{a}_{k} \sum_{\lambda=+,\times} \int d^3 \mathbf{K} \, \kappa(K) \Big(\hat{b}^{\dagger}_{K} \nu_{\mathbf{K}}(R,t) - \hat{b}_{K} \nu^{\ast}_{\mathbf{K}}(R,t) \Big)} \nonumber\\
&\times& e^{-i E(R,t) (\hat{a}_{k} ^{\dagger} \hat{a}_{k})^2} \bigg) \, \hat{a}_{k} \, \bigg( e^{i E(R,t) (\hat{a}_{k}^{\dagger}\hat{a}_{k})^2} \nonumber\\
&\times& e^{\hat{a}_{k}^{\dagger} \hat{a}_{k} \sum_{\lambda=+,\times} \int d^3 \mathbf{K} \, \kappa(K) \Big(\hat{b}^{\dagger}_{K} \nu_{\mathbf{K}}(R,t) - \hat{b}_{K}\nu^{\ast}_{\mathbf{K}}(R,t) \Big)} \, e^{-i\omega_{k} t \, \hat{a}_{k}^{\dagger} \hat{a}_{k} } \,  e^{-i \sum_{\lambda=+,\times} \int d^3\mathbf{K} \, \hat{b}_{K}^{\dagger} \hat{b}_{K}\Omega_{K} t} \bigg)\nonumber.
\end{eqnarray}
In the above equation, $\hat{a}_{k}$ stands for the annihilation operator of photons at the initial time $t=0$ when EM-GWs interaction initiates. Using the formula $e^{\hat{A}} \hat{B} e^{-\hat{A}} = \hat{B} + [\hat{A},\hat{B}] + \frac{1}{2!}[\hat{A},[\hat{A},\hat{B}]] + \cdot\cdot\cdot$, one may show that
\begin{eqnarray}\label{b.11}
&(1):& e^{-i E(R,t)(\hat{a}_{k}^{\dagger} \hat{a}_{k})^2} \, \hat{a}_{k} \, e^{i E(R,t)(\hat{a}_{k}^{\dagger} \hat{a}_{k})^2} = e^{2iE(R,t) \Big( \hat{a}^{\dagger}_{k} \hat{a}_{k} + \frac{1}{2}\Big)} \hat{a}_{k} \;,\\
&(2):& e^{-\hat{a}_{k}^{\dagger} \hat{a}_{k} \sum_{\lambda=+,\times} \int d^3 \mathbf{K} \, \kappa(K) \Big( \hat{b}^{\dagger}_{K} \nu_{\mathbf{K}}(R,t) - \hat{b}_{K}\nu^{\ast}_{\mathbf{K}}(R,t) \Big) }  e^{2iE(R,t) \Big(\hat{a}^{\dagger}_{k} \hat{a}_{k} + \frac{1}{2} \Big)} \, \hat{a}_{k} \nonumber\\
&&\times \, e^{\hat{a}_{k}^{\dagger} \hat{a}_{k} \sum_{\lambda=+,\times} \int d^3 \mathbf{K} \, \kappa(K) \Big( \hat{b}^{\dagger}_{K} \nu_{\mathbf{K}}(R,t)-\hat{b}_{K}\nu^{\ast}_{\mathbf{K}}(R,t)\Big)} \nonumber\\
&&= e^{2iE(R,t) \Big( \hat{a}^{\dagger}_{k} \hat{a}_{k} + \frac{1}{2}\Big)} \, \hat{a}_{k} \, e^{\sum_{\lambda=+,\times} \int d^3\mathbf{K} \, \kappa(K) \Big(\hat{b}^{\dagger}_{K} \nu_{\mathbf{K}}(R,t) - \hat{b}_{K}\nu^{\ast}_{\mathbf{K}}(R,t) \Big)}\;, \nonumber\\
&(3):& e^{i\omega_{k} t\, \hat{a}_{k}^{\dagger} \hat{a}_{k}} \, e^{2i E(R,t) \Big( \hat{a}_{k}^{\dagger} \hat{a}_{k} + \frac{1}{2}\Big)} \, \hat{a}_{k} \, e^{\sum_{\lambda=+,\times} \int d^3 \mathbf{K} \, \kappa(K)\Big(\hat{b}^{\dagger}_{K} \nu_{\mathbf{K}}(R,t) - \hat{b}_{K}\nu^{\ast}_{\mathbf{K}}(R,t)\Big)} \, e^{-i \omega_k t \, \hat{a}^{\dagger}_{k} \hat{a}_{k}} \nonumber \\
&& = e^{2i E(R,t) \Big( \hat{a}_{k}^{\dagger} \hat{a}_{k} + \frac{1}{2}\Big)} \, \hat{a}_{k} e^{-i\omega_{k} t} e^{\sum_{\lambda=+,\times} \int d^3 \mathbf{K} \, \kappa(K) \Big(\hat{b}^{\dagger}_{K} \nu_{\mathbf{K}}(R,t) - \hat{b}_{K} \nu^{\ast}_{\mathbf{K}}(R,t)\Big)}\;. \nonumber
\end{eqnarray}
\noindent Thus, Eq.~(\ref{b.10}) can be written as follows:
\begin{eqnarray}\label{b.12}
\hat{a}_{k}(R,t)=e^{-\sum_{\lambda=+,\times} \int d^3K \, \kappa(K) \Big( \hat{b}_{K}^{\dagger} \eta_{\mathbf{K}}(R,t) - \hat{b}_{K} \eta_{\mathbf{K}}^{\ast} (R,t) \Big)} \, e^{2i E(R,t) \Big( \hat{a}_{k}^{\dagger} \hat{a}_{k} + \frac{1}{2} \Big)} \, e^{-i\omega_k t} \hat{a}_{k}\, , \quad
\end{eqnarray}
where we have defined
\begin{eqnarray}\label{b.13}
\eta_{\mathbf{K}}(R,t) \equiv \Big( 1-e^{i\Omega_K t}\Big) \mathcal{G}_{\mathbf{K}}^{\lambda\ast}(R) \, .
\end{eqnarray}
Note that in the above equations $(\hat{a}_{k}, \hat{a}^{\dagger}_{k})$ stand for the EM ladder operators at the initial moment (before the EM-GWs interaction takes place). Expression (\ref{b.12}) is a generalization of the result of \cite{guerreiro2020quantum, bose1997preparation} to the case of a continuum of GWs modes interacting with spherical EM waves, where the spatial dependence of the field operators is included, and with a general geometric configuration encoded in the spatial function $\mathcal{G}_{\mathbf{K}}^{\lambda}(R)$.

Now, from the general definition of the displacement operator corresponding to a bosonic oscillator, $\hat{D}(\alpha)\equiv e^{\alpha \hat{b}^{\dagger} - \alpha^{\ast} \hat{b}}$, one may identify the \textit{displacement operator} associated to GW of mode $K$ in the time evolution of $\hat{a}_{k}(R,t)$ as follows:
\begin{eqnarray}\label{b.14}
\hat{D}_{K}\big(-\kappa(K) \, \eta_{\mathbf{K}} (R,t) \big) \equiv \exp\Big(-\kappa(K) \, \eta_{\mathbf{K}}(R,t) \, \hat{b}^{\dagger}_{K} + \kappa(K) \, \eta^{\ast}_{\mathbf{K}}(R,t) \, \hat{b}_{K}\Big). \qquad
\end{eqnarray}
Now, if one formally discretizes the integration over GWs modes in Eq.~(\ref{b.12}) according to $\int d^3 \mathbf{K} \rightarrow \frac{(2\pi)^3}{V} \sum_{\mathbf{K}}$, then $\hat{a}_{k}(R,t)$ is identified as follows
\begin{eqnarray}\label{b.15}
\hat{a}_{k} (R,t)= \Big[ \prod_{\lambda,\mathbf{K}} \hat{D}_{K}\big(-\kappa(K) \, \eta_{\mathbf{K}} (R,t) \big) \Big] \, e^{2i E(R,t) \Big( \hat{a}_{k}^{\dagger} \hat{a}_{k} + \frac{1}{2} \big)} \, e^{-i\omega_{k} t} \hat{a}_{k}\, , \quad
\end{eqnarray}
The quantity inside the brackets is nothing but the tensor product of the displacement operator of each GW mode. Hence, each mode interacts with the EM field independent of the other modes. The appearance of the displacement operator of the tensor field in the EM field dynamics can be justified by our intuition about the opto-mechanical system. In a cavity opto-mechanical system of length $L$, a single-mode EM field couples with the vibrational mode of the end mirror of the cavity, and the radiation pressure leads the mirror to vibrate. Vibration of the end mirror causes a change in the cavity length, $\Delta L$, which induces the resonant frequency of the cavity to vary as
\begin{eqnarray}\label{b.16}
\omega_{k} = \frac{n \pi}{L} \quad \rightarrow \quad \frac{n \pi}{L+\Delta L} = \frac{n \pi}{L \big( 1+\frac{\Delta L}{L}\big)} = \omega_k \big( 1 - \frac{\Delta L}{L} \big)\, ,
\end{eqnarray}
In the last equality, one may identify the strain field, $h \equiv \frac{\Delta L}{L}$, and the analogy becomes evident. In the EM-GWs analogous case, the strain field of GWs leads the EM frequency to change, and the appearance of the displacement operator of GWs in the evolution of the EM field can be understood from the analogy with the vibrational mode in the cavity opto-mechanical system.

It is beneficial to separate the contribution of the opposite momenta $\mathbf{K}$ and $-\mathbf{K}$ in Eq.~(\ref{b.15}). This step makes the investigation of the two-mode squeezed PGWs easier. To separate the contribution of the opposite momenta, one may start from the interaction Hamiltonian and rewrite it as a sum of two commuting Hamiltonians corresponding to $\mathbf{K} \in \Re^{3\pm}$. 
One ends up with the following expression for $\hat{a}_{k}(R,t)$:
\begin{eqnarray}\label{b.17}
\hat{a}_{k}(R,t) &=& \Big[ \prod_{\lambda,\mathbf{K}\in \Re^{3+}} \hat{D}_{\mathbf{K},\lambda}\big(-\kappa(K) \, \eta_{\mathbf{K}} (R,t) \big) \otimes \hat{D}_{-\mathbf{K},\lambda} \big(-\kappa(K) \, \eta_{-\mathbf{K}} (R,t) \big) \Big] \nonumber\\
&\times& e^{2i E(R,t) \Big( \hat{a}_{k}^{\dagger} \hat{a}_{k} + \frac{1}{2} \Big)} \, e^{-i\omega_{k} t} \hat{a}_{k} \, ,
\end{eqnarray}
which is a direct product of $\mathbf{K}$ and $-\mathbf{K}$ modes, as one could expect from Eq.~(\ref{b.15}).

With the help of Eq.~(\ref{b.17}), we are equipped to investigate spatial correlations of the EM field emitted from distant objects in the presence of the two-mode squeezed PGWs background.


\section{Preliminaries for evaluation of $g^{(1)}(R_{1},t;R_{2},t)$} \label{app:C}

\subsection{Mutual intensity of a planar source in the presence of PGWs} \label{app:C.1}

In this section, we derive Eq.~(\ref{4.13}). Firstly, the two-mode squeezed state of PGWs $\big\vert \text{TS} \big\rangle$ can be written as \cite{martin2016quantum}
\begin{eqnarray}\label{c.1}
|{\text{TS}} (\eta)\rangle \equiv \prod_{\mathbf{K}\in \Re^{3+}}\hat{S}_{\mathbf{K}} (\zeta_K(\eta) ) \hat{R}_{\mathbf{K}}(\theta_K(\eta))| 0_{\mathbf{K}} , 0_{-\mathbf{K}} \rangle\, ,
\end{eqnarray}
where $| 0_{\mathbf{K}} , 0_{-\mathbf{K}} \rangle$ denotes the vacuum state of gravitons of opposite momenta and $\eta$ stands for the conformal time. Here, because the opposite momenta are not independent degrees of freedom, the multiplication is taken over half Fourier space $\mathbf{K} \in \Re^{3+}$. The two-mode squeezing operator $\hat{S}_{\mathbf{K}}\big(\zeta_K(\eta)\big)$ and the rotation operator $\hat{R}_{\mathbf{K}}\big(\theta_K(\eta)\big)$ are defined by (see \cite{martin2016quantum} for details)
\begin{eqnarray}\label{c.2}
\hat{S}_{\mathbf{K}}(\zeta_K(\eta)) &\equiv& \exp \left[\zeta_K^{\ast}(\eta) \hat{b}_{-\mathbf{K}} \hat{b}_{\mathbf{K}} - \zeta_K(\eta) \hat{b}^{\dagger}_{-\mathbf{K}} \hat{b}^{\dagger}_{\mathbf{K}} \right],\nonumber\\
\hat{R}_{\mathbf{K}}(\theta_{K}(\eta)) &\equiv & \exp \left[ -i \theta_{K}(\eta) \Big( \hat{b}^{\dagger}_{\mathbf{K}} \hat{b}_{\mathbf{K}} + \hat{b}^{\dagger}_{-\mathbf{K}} \hat{b}_{-\mathbf{K}} \Big) \right]\, ,
\end{eqnarray}
and $\zeta_K(\eta)$ and $\theta_K(\eta)$ are the squeezing and rotation parameters, respectively. Starting from Eq.~(\ref{4.12}), it can be written as follows
\begin{eqnarray}\label{c.3}
\hspace*{-1.5cm} g_{\text{ts}}^{(1)}(1; 2) &=& \prod_{\lambda,\mathbf{K}\in \Re^{3+}} \langle 0_{\mathbf{K}},0_{-\mathbf{K}} | \hat{R}^{\dagger}_{\mathbf{K}} \hat{S}^{\dagger}_{\mathbf{K}} \hat{D}^{\dagger}_{\mathbf{K},\lambda} \Big(-\kappa(K) \, \eta_{\mathbf{K}}(1)\Big) \hat{D}_{\mathbf{K},\lambda} \Big(-\kappa \eta_{-\mathbf{K}}(2) \Big) \nonumber\\
&\times& \hat{D}^{\dagger}_{-\mathbf{K},\lambda} \Big(-\kappa(K) \, \eta_{-\mathbf{K}}(1)\Big) \hat{D}_{-\mathbf{K},\lambda} \Big(-\kappa(K) \, \eta_{-\mathbf{K}}(2) \Big) \hat{S}_{\mathbf{K}} \hat{R}_{\mathbf{K}} | 0_{\mathbf{K}},0_{-\mathbf{K}}\rangle\nonumber\\
&=& e^{-i \sum_{\lambda} \int_{ \mathbf{K} \in \Re^{3} } \kappa^2(K) \, \Im \Big( \eta_{\mathbf{K}}(1) \eta^{\ast}_{\mathbf{K}} (2) \Big) } e^{-i \omega_k (t_1 - t_2)} \\
&\times& \prod_{\lambda, \mathbf{K} \in \Re^{3+}} \Big\langle 0_{\mathbf{K}} , 0_{-\mathbf{K}} \Big\vert \hat{R}^{\dagger} \hat{S}^{\dagger} \hat{D}_{\mathbf{K},\lambda} \Big( \kappa(K) \, [\eta_{\mathbf{K}}(1) - \eta_{\mathbf{K}}(2) ] \Big) \hat{S} \hat{R} \nonumber\\
&\times& \hat{R}^{\dagger} \hat{S}^{\dagger} \hat{D}_{-\mathbf{K},\lambda} \Big( \kappa(K) \, [\eta_{-\mathbf{K}}(1) - \eta_{-\mathbf{K}} (2) ] \Big) \hat{S} \hat{R} \Big\vert 0_{\mathbf{K}}, 0_{-\mathbf{K}} \Big\rangle. \nonumber
\end{eqnarray}
Here, for the sake of abbreviation, we used $(1,2)$ in place of $(R_{1},t_1)$ and $(R_{2},t_2)$, respectively. The opposite momenta $\mathbf{K}$ and $-\mathbf{K}$ are independent in the half Fourier space $\Re^{3+}$; consequently, their associated displacement operators commute. Using the multiplication property of the displacement operator $\hat{D}$, namely $\hat{D}(\alpha) \hat{D}(\beta) = e^{i\Im(\alpha\beta^{\ast})} \hat{D}(\alpha+\beta)$, one has
\begin{eqnarray}\label{c.4}
\hat{D}_{\mathbf{K}}^{\dagger}\big( -\kappa(K) \, \eta_{\mathbf{K}}(1) \big) \hat{D}_{\mathbf{K}}\big( -\kappa(K) \, \eta_{\mathbf{K}}(2) \big) = e^{-i\kappa^2(K) \, \Im \left( \eta_{\mathbf{K}}(1) \eta^{\ast}_{\mathbf{K}}(2) \right)} \hat{D}_{\mathbf{K}}\big(\kappa(K) \, [\eta_{\mathbf{K}}(1) - \eta_{\mathbf{K}}(2)] \big)\,. \nonumber\\
\end{eqnarray}
together with a similar expression for $-\mathbf{K}$. Note that $\Im$ stands for the imaginary part. Moreover, the two-mode squeezing operator $\hat{S}_{\mathbf{K}}$ and the rotation operator $\hat{R}_{\mathbf{K}}$ involve opposite modes $\mathbf{K}$ and $-\mathbf{K}$ (see the definitions in Eq.~(\ref{c.2})), so in order to write Eq.~(\ref{c.3}) in a separated form, we may use the following identities:
\begin{eqnarray}\label{c.5}
\hat{R}^{\dagger}_{\mathbf{K}} \hat{S}^{\dagger}_{\mathbf{K}} \hat{D}_{\mathbf{K}} \big(-\kappa(K)\,  \eta_{\mathbf{K}} \big) \hat{S}_{\mathbf{K}} \hat{R}_{\mathbf{K}} 
&=& \hat{D}_{\mathbf{K}} \big( -\kappa(K) \, \eta_{\mathbf{K}} u_{K}^{\ast}(\eta_H) \big) \otimes \hat{D}_{-\mathbf{K}} \big( \kappa(K)\, \eta^{\ast}_{\mathbf{K}} v_{K}(\eta_H) \big), \nonumber\\
\hat{R}^{\dagger}_{\mathbf{K}} \hat{S}^{\dagger}_{\mathbf{K}} \hat{D}_{-\mathbf{K}} \big( -\kappa(K) \, \eta_{-\mathbf{K}} \big) \hat{S}_{\mathbf{K}} \hat{R}_{\mathbf{K}} 
&=& \hat{D}_{-\mathbf{K}} \big( -\kappa(K) \, \eta_{-\mathbf{K}} u_{K}^{\ast}(\eta_H) \big) \otimes \hat{D}_{\mathbf{K}} \big( \kappa(K) \, \eta^{\ast}_{-\mathbf{K}} v_{K}(\eta_H) \big). \nonumber\\
\end{eqnarray}
\noindent where $u_K(\eta_H)$ and $v_{K}(\eta_H)$ are defined by Eq.~(\ref{2.3}). Note that the quantities related to the PGWs are evaluated at the present time $\eta_H$. With the help of Eq.~(\ref{c.5}), Eq.~(\ref{c.3}) can be written as follows:
\begin{eqnarray}\label{c.6}
g_{\text{ts}}^{(1)}(1;2) &=& e^{-i \sum_{\lambda} \int_{ \mathbf{K} \in \Re^{3} } \kappa^2(K) \, \Im \Big( \eta_{\mathbf{K}}(1) \eta^{\ast}_{\mathbf{K}} (2) \Big) } e^{-i \omega_k (t_1 - t_2)} \\
&\times& \prod_{\lambda, \mathbf{K} \in \Re^{3+}} \Big\langle 0_{\mathbf{K}} , 0_{-\mathbf{K}} \Big\vert \hat{D}_{\mathbf{K}} \Big( \kappa(K) \, u_K^{\ast}(\eta_H) [ \eta_{\mathbf{K}}(1) - \eta_{\mathbf{K}}(2) ] \Big) \nonumber\\
&\otimes &\hat{D}_{-\mathbf{K}} \Big( -\kappa(K) \, v_K(\eta_H) [ \eta^{\ast}_{\mathbf{K}}(1) - \eta^{\ast}_{\mathbf{K}}(2) ] \Big) \nonumber\\
&\times& \hat{D}_{-\mathbf{K}} \Big( \kappa(K)\, u_K^{\ast}(\eta_H) [ \eta_{-\mathbf{K}}(1) - \eta_{-\mathbf{K}}(2) ] \Big) \nonumber\\
&\otimes& \hat{D}_{\mathbf{K}} \Big( -\kappa(K)\, v_K(\eta_H) [ \eta^{\ast}_{-\mathbf{K}}(1) - \eta^{\ast}_{-\mathbf{K}}(2) ] \Big) \Big\vert 0_{\mathbf{K}}, 0_{-\mathbf{K}} \Big\rangle \nonumber\\
&=& e^{-i \sum_{\lambda} \int_{ \mathbf{K} \in \Re^{3} } \kappa^2(K)\, \Im \Big( \eta_{\mathbf{K}}(1) \eta^{\ast}_{\mathbf{K}} (2) \Big) } e^{-i \omega_k (t_1 - t_2)} \nonumber \\
&\times& \prod_{\lambda, \mathbf{K} \in \Re^{3+}} \Big\langle 0_{\mathbf{K}} \Big\vert \hat{D}^{\dagger}_{\mathbf{K}} \Big( -\kappa(K)\, u_K^{\ast}(\eta_H) [ \eta_{\mathbf{K}}(1) - \eta_{\mathbf{K}}(2) ] \Big) \nonumber\\
&\times& \hat{D}_{\mathbf{K}} \Big( -\kappa(K)\, v_K(\eta_H) [ \eta^{\ast}_{-\mathbf{K}}(1) - \eta^{\ast}_{-\mathbf{K}}(2) ] \Big) \Big\vert 0_{\mathbf{K}} \Big\rangle \nonumber\\
&\times& \Big\langle 0_{-\mathbf{K}} \Big\vert \hat{D}^{\dagger}_{-\mathbf{K}} \Big( \kappa(K)\, v_K(\eta_H) [ \eta^{\ast}_{\mathbf{K}}(1) - \eta^{\ast}_{\mathbf{K}}(2) ] \Big) \nonumber\\
&\times& \hat{D}_{-\mathbf{K}} \Big( \kappa(K)\, u_K^{\ast}(\eta_H) [ \eta_{-\mathbf{K}}(1) - \eta_{-\mathbf{K}}(2) ] \Big)\Big\vert 0_{-\mathbf{K}} \Big\rangle \nonumber \\
&=& e^{-i \sum_{\lambda} \int_{ \mathbf{K} \in \Re^{3} } \kappa^2(K)\, \Im \Big( \eta_{\mathbf{K}}(1) \eta^{\ast}_{\mathbf{K}} (2) \Big) } e^{-i \omega_k (t_1 - t_2)} \nonumber \\
&\times& \prod_{\lambda, \mathbf{K} \in \Re^{3+}} \Big\langle -\kappa(K)\, u_K^{\ast}(\eta_H) [ \eta_{\mathbf{K}}(1) - \eta_{\mathbf{K}}(2) ] \Big\vert -\kappa(K)\, v_K(\eta_H) [ \eta^{\ast}_{-\mathbf{K}}(1) - \eta^{\ast}_{-\mathbf{K}}(2) ] \Big\rangle \nonumber\\
&\times& \Big\langle \kappa(K)\, v_K(\eta_H) [ \eta^{\ast}_{\mathbf{K}}(1) - \eta^{\ast}_{\mathbf{K}}(2) ] \Big\vert \kappa(K)\, u_K^{\ast}(\eta_H) [ \eta_{-\mathbf{K}}(1) - \eta_{-\mathbf{K}}(2) ] \Big\rangle. \nonumber
\end{eqnarray}
Using the projection property of coherent states, $\langle \alpha|\alpha' \rangle = e^{-\frac{1}{2}|\alpha|^2 - \frac{1}{2}|\alpha'|^2 + \alpha^{\ast}\alpha'}$, Eq.~(\ref{c.6}) reads
\begin{eqnarray}\label{c.7}
g^{(1)}_{\text{ts}}(1;2) &=&  e^{-i \sum_{\lambda} \int_{ \mathbf{K} \in \Re^{3} } \kappa^2(K)\, \Im \Big( \eta_{\mathbf{K}}(1) \eta^{\ast}_{\mathbf{K}} (2) \Big) } e^{-i \omega_k (t_1 - t_2)} \\
&\times& \prod_{\lambda, \mathbf{K} \in \Re^{3+}} \exp \Big[ -\frac{1}{2} \kappa^2(K)\, |u_K(\eta_H)|^2 \Big\vert \eta_{\mathbf{K}}(1) - \eta_{\mathbf{K}}(2) \Big\vert^2 \Big] \nonumber\\
&\times& \exp \Big[ -\frac{1}{2} \kappa^2(K)\, |v_K(\eta_H)|^2 \Big\vert \eta^{\ast}_{-\mathbf{K}}(1) - \eta^{\ast}_{-\mathbf{K}}(2) \Big\vert^2 \Big] \nonumber\\
&\times& \exp \Big[ \kappa^2(K)\, u_K(\eta_H) v_K(\eta_H) \Big( \eta^{\ast}_{\mathbf{K}}(1) - \eta^{\ast}_{\mathbf{K}}(2) \Big) \Big( \eta^{\ast}_{-\mathbf{K}}(1) - \eta^{\ast}_{-\mathbf{K}}(2) \Big) \Big] \nonumber\\
&\times& \exp\Big[ -\frac{1}{2} \kappa^2(K)\, |v_K(\eta_H)|^2 \Big\vert \eta^{\ast}_{\mathbf{K}}(1) - \eta^{\ast}_{\mathbf{K}}(2) \Big\vert^2 \Big] \exp\Big[ -\frac{1}{2} \kappa^2(K)\, |u_K(\eta_H)|^2 \Big\vert \eta_{-\mathbf{K}}(1) - \eta_{-\mathbf{K}}(2) \Big\vert^2 \Big] \nonumber\\
&\times& \exp \Big[ \kappa^2(K)\, u_K^{\ast}(\eta_H) v_K^{\ast}(\eta_H) \Big( \eta_{\mathbf{K}}(1) - \eta_{\mathbf{K}}(2) \Big) \Big( \eta_{-\mathbf{K}}(1) - \eta_{-\mathbf{K}}(2) \Big) \Big] \nonumber \\
&=& e^{-i \sum_{\lambda} \int_{ \mathbf{K} \in \Re^{3} } \kappa^2(K)\, \Im \Big( \eta_{\mathbf{K}}(1) \eta^{\ast}_{\mathbf{K}} (2) \Big) } e^{-i \omega_k (t_1 - t_2)} \nonumber\\
&\times& \exp \Big[ -\frac{1}{2} \sum_{\lambda} \int_{\mathbf{K} \in \Re^{3}} d^3\mathbf{K} \kappa^2(K)\, \Big( |u_K(\eta_H)|^2 + |v_K(\eta_H)|^2 \Big) \Big\vert \eta_{\mathbf{K}}(1) - \eta_{\mathbf{K}}(2) \Big\vert^2 \Big] \nonumber\\
&\times& \exp \Big[ \sum_{\lambda} \int_{\mathbf{K} \in \Re^{3}} d^3\mathbf{K} \kappa^2(K)\, \Re\Big\{ u_K(\eta_H) v_K(\eta_H) \Big( \eta^{\ast}_{\mathbf{K}}(1) \eta^{\ast}_{-\mathbf{K}}(1) - \eta^{\ast}_{\mathbf{K}}(2) \eta^{\ast}_{-\mathbf{K}}(2) \Big) \Big\} \Big]\quad . \nonumber
\end{eqnarray}
In the last equality we transformed to the continuum notation and used the normalization property $|u_K(\eta_H)|^2-|v_K(\eta_H)|^2=1$ for two-mode squeezed states (see Eq.~(\ref{2.3})).

If one resumes the calculations for GWs in a vacuum state, namely $\hat{\rho}_{\text{vac}} = \prod_{K} |0_{K}\rangle \langle 0_K |$, it can be verified that the contribution of vacuum fluctuations is already involved in the contribution of the two-mode squeezed state. In this case, one has
\begin{eqnarray}\label{c.8}
\hspace*{-0.5cm}g_{\text{vac}}^{(1)}(R_{1},t;R_{2},t) &=& \prod_{\lambda,\mathbf{K}\in \Re^{3+}} \langle 0_{\mathbf{K}},0_{-\mathbf{K}} | \hat{D}_{\mathbf{K}}^{\dagger}\big( -\kappa(K)\, \eta_{\mathbf{K}}(R_{1},t) \big) \hat{D}_{\mathbf{K}} \big(-\kappa(K)\, \eta_{\mathbf{K}}(R_{2},t) \big) \nonumber\\
&\times& \hat{D}_{-\mathbf{K}}^{\dagger} \big( -\kappa(K)\, \eta_{-\mathbf{K}}(R_{1},t) \big) \hat{D}_{-\mathbf{K}} \big( -\kappa(K)\, \eta_{-\mathbf{K}}(R_{2},t) \big) | 0_{\mathbf{K}},0_{-\mathbf{K}} \rangle \nonumber \\
&=& \prod_{\lambda,\mathbf{K}\in \Re^{3+}} \langle 0_{\mathbf{K}}| \hat{D}_{\mathbf{K}}^{\dagger}\big( -\kappa(K)\, \eta_{\mathbf{K}}(R_{1},t) \big) \hat{D}_{\mathbf{K}} \big(-\kappa(K)\, \eta_{\mathbf{K}}(R_{2},t) \big) | 0_{\mathbf{K}} \rangle \nonumber\\
&\times& \langle 0_{-\mathbf{K}}| \hat{D}_{-\mathbf{K}}^{\dagger} \big( -\kappa(K)\, \eta_{-\mathbf{K}}(R_{1},t) \big) \hat{D}_{-\mathbf{K}} \big( -\kappa(K)\, \eta_{-\mathbf{K}}(R_{2},t) \big) |0_{-\mathbf{K}} \rangle \,. \nonumber\\
\end{eqnarray}
\noindent With the help of the property $\hat{D}^{\dagger}(\alpha) \hat{D}(\beta) = e^{-i\Im \alpha\beta^{\ast}} \hat{D}(\beta-\alpha)$ for displacement operator, Eq.~(\ref{c.8}) is rewritten as
\begin{eqnarray}\label{c.9}
\hspace*{-1.2cm}g^{(1)}_{\text{vac}}(R_{1},t; R_{2},t) &=& \prod_{\lambda,\mathbf{K}\in \Re^{3+}} \big\langle 0_{\mathbf{K}}\big\vert e^{-i\kappa^2\Im\big( \eta_{\mathbf{K}}(R_{1},t) \eta_{\mathbf{K}}^{\ast}(R_{2},t) \big)} \hat{D}_{\mathbf{K}} \big( \kappa(K)\, [ \eta_{\mathbf{K}}(R_{1},t) - \eta_{\mathbf{K}}(R_{2},t) ] \big) \big\vert 0_{\mathbf{K}} \big\rangle  \\
&\times& \big\langle 0_{-\mathbf{K}} \big\vert e^{-i\kappa^2 \Im\big( \eta_{-\mathbf{K}}(R_{1},t) \eta^{\ast}_{-\mathbf{K}}(R_{2},t) \big)} \hat{D}_{-\mathbf{K}} \big( \kappa(K)\, [ \eta_{-\mathbf{K}}(R_{1},t) - \eta_{-\mathbf{K}}(R_{2},t) ] \big) \big\vert 0_{-\mathbf{K}} \big\rangle \nonumber\\
&=& \prod_{\lambda,\mathbf{K}\in \Re^{3+}} 
e^{-i\kappa^2\Im\big( \eta_{\mathbf{K}}(R_{1},t) \eta_{\mathbf{K}}^{\ast}(R_{2},t) + \eta_{-\mathbf{K}}(R_{1},t) \eta^{\ast}_{-\mathbf{K}}(R_{2},t) \big)} \nonumber\\
&\times& \exp\big[ -\frac{1}{2} \kappa^2(K)\, |\eta_{\mathbf{K}}(R_{1},t) - \eta_{\mathbf{K}}(R_{2},t)|^2 \big] \, \exp\big[ -\frac{1}{2} \kappa^2(K)\, |\eta_{-\mathbf{K}}(R_{1},t) - \eta_{-\mathbf{K}}(R_{2},t)|^2 \big]\, . \nonumber
\end{eqnarray}
Using the continuous notation, Eq.~(\ref{c.9}) reads
\begin{eqnarray}\label{c.10}
g^{(1)}_{\text{vac}}(R_{1},t; R_{1},t) &=& e^{-i \sum_{\lambda} \int_{\mathbf{K}\in \Re^{3}} d^3\mathbf{K} \, \kappa(K)\, \Im \big( \eta_{\mathbf{K}}(R_{1},t) \, \eta_{\mathbf{K}}^{\ast}(R_{2},t) \big)} \\
&\times& \exp\big[ -\frac{1}{2} \sum_{\lambda}\int_{\mathbf{K}\in \Re^3}  d^3\mathbf{K} \, \kappa^2(K)\, |\eta_{\mathbf{K}}(R_{1},t) - \eta_{\mathbf{K}}(R_{2},t)|^2 \big]. \nonumber
\end{eqnarray}
With the help of definitions Eq.~(\ref{4.14}, \ref{4.15}), we may now rewrite Eq.~(\ref{c.10}) as follows:
\begin{eqnarray}\label{c.11}
g^{(1)}_{\text{vac}}(R_{1},t; R_{2},t) &=& e^{-i \, \mathcal{C}^{\text{vac}}(R_{1},t; R_{2},t)} \, e^{-\mathcal{D}^{\text{vac}}(R_{1},t; R_{2},t)} \,.
\end{eqnarray}
Combining Eqs.(\ref{c.7}, \ref{c.11}) and with the help of Eqs.~(\ref{4.14}, \ref{4.15}), Eq.~(\ref{4.13}) outcomes.


\subsection{Derivation of the spatial factor $\big\vert \mathcal{G}_{\mathbf{K}}^{\lambda}(R_{1}) - \mathcal{G}_{\mathbf{K}}^{\lambda}(R_{2}) \big\vert^2$} \label{app:C.2}

In order to calculate the expression $\big\vert \mathcal{G}_{\mathbf{K}}^{\lambda}(R_{1}) - \mathcal{G}_{\mathbf{K}}^{\lambda}(R_{2}) \big\vert^2$, first note that the phase factor $e^{i\mathbf{K}\cdot \mathbf{r}_s}$ in Eq.~(\ref{3.52}) vanishes and does not play a role as a result of the homogeneity of the PGWs background. Using the plane wave expansion in terms of the spherical Bessel functions, 
\begin{eqnarray}\label{c.12}
e^{i\mathbf{K}\cdot\mathbf{r}} &=& \sum_{\ell=0}^{\infty} (i^{\ell})\, (2\ell+1) \, j_{\ell}(Kr) \, P_{\ell}(\cos\gamma)\,,
\end{eqnarray}
The definition Eq.~(\ref{3.52}) for $\mathcal{G}_{\mathbf{K}}^{\lambda}(R)$ can be expanded as follows:
\begin{eqnarray}\label{c.13}
\mathcal{G}_{\mathbf{K}}^{\lambda}(R) &\equiv& \frac{e^{i \mathbf{K} \cdot \mathbf{r}_s}}{2\pi R} \, \int d\Omega_{\hat{r}} \, e_{ij}^{\lambda}[\hat{\mathbf{K}}]\, \hat{r}_i\, \hat{r}_j \int_0^{R} e^{i\mathbf{K}\cdot\mathbf{r}} \, dr \\
&=& \frac{e^{i \mathbf{K} \cdot \mathbf{r}_s}}{2\pi (KR)} \sum_{\ell=0}^{\infty} (i^{\ell}) (2\ell+1) \Big(\int_0^{KR} j_{\ell}(u)\, du \Big) \, \Big( \int d^2 \Omega_{\hat{r}}\, e_{ij}^{\gamma}[\hat{\mathbf{K}}]\, \hat{r}_i\, \hat{r}_{j}\, P_{\ell}(\cos\gamma) \Big)\,. \nonumber
\end{eqnarray}
In this way, integrations over the radial and spatial variables are decoupled. Here, $\hat{r} = (\sin\theta\cos\phi, \sin\theta\sin\phi, \cos\theta)$ and $\hat{\mathbf{K}} = (\sin\Theta_K\cos\Phi_K, \sin\Theta_K\sin\Phi_K, \cos\Theta_K)$ show the unit vectors in spherical coordinates, $d\Omega_{\hat{r}} = d(\cos\theta) d\phi$ and $\gamma$ represents the angle between $\hat{r}$ and $\hat{\mathbf{K}}$. Before proceeding, it is useful to obtain an asymptotic expression for $\mathcal{G}_{\mathbf{K}}^{\lambda}(R)$ in the case of a small detector, for which $KR \ll 1$. The radial part of Eq.~(\ref{c.13}) can be written as
\begin{eqnarray} \label{c.14}
&& \frac{\int_{0}^{KR} j_{\ell}(u) du}{(KR)}
\end{eqnarray}
In the small-detector approximation, we take $\epsilon = KR \ll 1$ and expand the spherical Bessel $j_{\ell}(u)$ around $u=0$, which gives rise to
\begin{eqnarray} \label{c.15}
j_{\ell}(u) \simeq j_{\ell}(0) + j'_{\ell}(0) (u-\epsilon) + j''_{\ell}(0) \, \frac{(u-\epsilon)^2}{2!} + \cdot\cdot\cdot\,.
\end{eqnarray}
Plugging the expansion Eq.~(\ref{c.15}) into Eq.~(\ref{c.14}) yields
\begin{eqnarray} \label{c.16}
&& \underset{\,KR\ll 1}{\text{lim}}\Big( \frac{\int_{0}^{KR} j_{\ell}(u) du}{(KR)} \Big) \simeq j_{\ell}(0) =
 \begin{cases}
   1 \quad,\quad \ell = 0 \\
   0 \quad,\quad \ell \neq 0 
 \end{cases}
\end{eqnarray}
and Eq.~(\ref{c.13}) reduce to 
\begin{eqnarray} \label{c.17}
\underset{\,KR\ll 1}{\text{lim}} \mathcal{G}_{\mathbf{K}}^{\lambda} &\simeq& \frac{\int d\Omega_{\hat{r}} \, e_{ij}[\hat{\mathbf{K}}] \, \hat{r}_i \, \hat{r_{j}}}{2\pi} \,.
\end{eqnarray}
which is a constant factor and independent of $KR$.

With the help of Eq.~(\ref{c.13}) we can write
\begin{eqnarray}\label{c.18}
\mathcal{G}_{\mathbf{K}}^{\lambda}(R_{1}) - \mathcal{G}_{\mathbf{K}}^{\lambda}(R_{2}) &=& \frac{e^{i \mathbf{K} \cdot \mathbf{r}_s}}{2\pi} \sum_{\ell=0}^{\infty} (i^{\ell}) (2\ell+1) \, \Big( \int d\Omega_{\hat{r}}\, e_{ij}^{\gamma}[\hat{\mathbf{K}}]\, \hat{r}_i\, \hat{r}_{j}\, P_{\ell}(\cos\gamma) \Big) \\
&\times& \Big( \frac{1}{(KR_{1})} \int_{0}^{KR_{1}} j_{\ell}(u)\, du - \frac{1}{(KR_{2})} \int_{0}^{KR_{2}} j_{\ell}(u)\, du \Big) \nonumber\,.
\end{eqnarray}
The radial part of Eq.~(\ref{3.18}) can be expanded as 
\begin{eqnarray}\label{c.19}
&&\frac{1}{(KR_{1})} \int_{0}^{KR_{1}} j_{\ell}(u)\, dxu- \frac{1}{(KR_{2})} \int_{0}^{KR_{2}} j_{\ell}(u)\, du \\
&& = \frac{1}{(KR_{1})} \int_0^{KR_{1}}j_{\ell}(u) du - \frac{1}{K(R_{2} - R_{1} + R_{1})} \int_{0}^{KR_{2}} j_{\ell}(u) du \nonumber\\
&& = \frac{1}{K} \bigg[ \frac{1}{R_{1}}\int_0^{KR_{1}} j_{\ell}(u) du - \frac{1}{R_{1}+(R_{2} - R_{1})} \int_{0}^{KR_{2}} j_{\ell}(u) du \bigg] \nonumber\\
&&= \frac{1}{K} \bigg[ \frac{1}{R_{1}}\int_0^{KR_{1}} j_{\ell}(u) du - \frac{1}{R_{1} \big( 1 + \frac{R_{2} - R_{1}}{R_{1}} \big)} \int_{0}^{KR_{2}} j_{\ell}(u) du \bigg] \nonumber\\
&&= \frac{1}{K} \bigg[ \frac{1}{R_{1}}\int_0^{KR_{2}} j_{\ell}(u) du + \frac{1}{R_{1}} \int_{KR_{2}}^{KR_{1}} j_{\ell}(u) du \nonumber\\
&& - \frac{1}{R_{1}} \int_{0}^{KR_{2}} j_{\ell}(u) du - \Big( \frac{R_{1} - R_{2}}{R_{1}} \Big) \, \frac{1}{R_{1}} \int_0^{KR_{2}} j_{\ell}(u) du \bigg] \nonumber\,.
\end{eqnarray}
In the last equality, the integral from $0$ to $KR_{1}$ is split into two sub-intervals, $0$ to $KR_{2}$ and $KR_{2}$ to $K R_{1}$. Moreover, we have used the fact that $(R_{2} -R_{1}) \ll R_{1}$. Hence, Eq.~(\ref{c.19}) can be written as
\begin{eqnarray}\label{c.20}
&& \frac{1}{(KR_{1})} \int_{0}^{KR_{1}} j_{\ell}(u)\, du - \frac{1}{(KR_{2})} \int_{0}^{KR_{2}} j_{\ell}(u)\, du \\
&& = \frac{1}{(KR_{1})} \int_{KR_{2}}^{KR_{1}} j_{\ell}(u) du - \frac{K \big( R_{1} - R_{2} \big) }{(K R_{1})^2} \, \Big( \int_0^{KR_{2}} j_{\ell}(u) du \Big) \nonumber \\ 
&& \equiv \mathcal{G}_{1}(R_{1},R_{2}) - \mathcal{G}_{2}(R_{1},R_{2}) \, , \nonumber
\end{eqnarray}
where in the last line we have defined
\begin{eqnarray}\label{c.21}
\mathcal{G}_{1}(R_{1},R_{2}) &\equiv& \frac{1}{(KR_{1})}  \Big( \int_{KR_{2}}^{KR_{1}} j_{\ell}(u) du \Big)\\
\mathcal{G}_{2}(R_{1},R_{2}) &\equiv& \frac{K\big( R_{1} - R_{2} \big) }{(K R_{1})^2} \, \Big( \int_0^{KR_{2}} j_{\ell}(u) du \Big) \nonumber\, .
\end{eqnarray}
The function $\mathcal{G}_{1}(R_{1},R_{2})$ can be written as
\begin{eqnarray}\label{c.22}
\mathcal{G}_{1}(R_{1},R_{2}) &=& \frac{1}{(KR_{1})} \int_{KR_{2}}^{K(R_{1}-R_{2}+R_{2})} j_{\ell}(u) du \\
&=& \frac{1}{(KR_{1})} \int_{KR_{2}}^{KR_{2} + K(R_{1} - R_{2})} j_{\ell}(u) du \nonumber\, .
\end{eqnarray}
Taking $y = KR_{2}$ and $\epsilon = K(R_{1} - R_{2})$ so that $\epsilon \ll y$, and using the Taylor series expansion of $j_{\ell}(u)$ around $y$, we obtain
\begin{eqnarray}\label{c.23}
j_{\ell}(u) &\simeq& j_{\ell}(y) + j'_{\ell}(y) (u-y) + \frac{j''_{\ell}(y)}{2!}(u-y)^2 +\cdot\cdot\cdot \, .
\end{eqnarray}
Inserting Eq.~(\ref{c.23}) into Eq.~(\ref{c.22}) yields
\begin{eqnarray}\label{c.24}
\mathcal{G}_{1}(R_{1},R_{2}) &=& \frac{1}{(KR_{1})} \int_{y}^{y+\epsilon} \Big( j_{\ell}(y) +  j'_{\ell}(y) (u-y) + \frac{j''_{\ell}(y)}{2!}(u-y)^2 +\cdot\cdot\cdot \Big) du \, \quad \\
&=& \frac{1}{(KR_{1})} \, j_{\ell}(y) \, \epsilon + \frac{1}{(KR_{1})} \, j'_{\ell}(y)\, \frac{\epsilon^2}{2} + \cdot\cdot\cdot \nonumber\\
&\simeq& \frac{ j_{\ell}(KR_{2})}{(KR_{1})} \big(K(R_{1} - R_{2}) \big) \, + \, \frac{j'_{\ell}(KR_{2})}{(KR_{1})} \, \frac{\big(K(R_{1} -R_{2}) \big)^2}{2} \nonumber\,.
\end{eqnarray}
The second term in Eq.~(\ref{c.24}) is proportional to $ \epsilon^2 = \big(K(R_{1}-R_{2})\big)^2$. From Eq.~(\ref{4.21}) one has $|\epsilon| \leq K \rho' w = K (\rho'/a) a (d/R) = K d \, (\theta/2) (\rho'/a) \leq K d\, \theta/2$, where $w=d/R$ is used. In a typical VLBI interferometer performing on the Earth, the detector separation $d$ can be as large as the Earth's radius, and the measured $\theta$ is of the order of $10^{-8}\,$rad. Thus, one may use $|\epsilon| \ll 1$ concerning the spectrum of PGWs. Therefore, we proceed by keeping terms up to $\mathcal{O} \big(K(R_{1}-R_{2})\big) = \epsilon$ and neglecting higher-order terms. Combining Eqs.~(\ref{c.20}, \ref{c.21}, \ref{c.24}) yields
\begin{eqnarray}\label{c.25}
&& \frac{1}{(KR_{1})} \int_{0}^{KR_{1}} j_{\ell}(u)\, du - \frac{1}{(KR_{2})} \int_{0}^{KR_{2}} j_{\ell}(u)\, du \\
&& = \bigg\{ \frac{j_{\ell}(KR_{2}) }{(KR_{1})} \, - \, \frac{\int_0^{KR_{2}} j_{\ell}(u) du }{(K R_{1})^2} \bigg\} \, \Big( K ( R_{1} - R_{2}) \Big) \nonumber\, .
\end{eqnarray}
Inserting Eq.~(\ref{c.25}) into Eq.~(\ref{c.18}) results in
\begin{eqnarray}\label{c.26}
\mathcal{G}_{\mathbf{K}}^{\lambda}(R_{1}) - \mathcal{G}_{\mathbf{K}}^{\lambda}(R_{2}) &=& \frac{e^{i \mathbf{K} \cdot \mathbf{r}_s}}{2\pi} \sum_{\ell=0}^{\infty} (i^{\ell}) (2\ell+1) \, \Big( \int d\Omega_{\hat{r}}\, e_{ij}^{\gamma}[\hat{\mathbf{K}}]\, \hat{r}_i\, \hat{r}_{j}\, P_{\ell}(\cos\gamma) \Big) \\
&\times& \bigg\{\frac{ j_{\ell}(KR_{2})}{(KR_{1})} \, - \, \frac{\int_0^{KR_{2}} j_{\ell}(u) du}{(K R_{1})^2} \, \bigg\} \, \Big( K(R_{1} - R_{2})\Big) \nonumber\,.
\end{eqnarray}
Using Eq.~(\ref{4.21}) for $R_{1} -R_{2}$, one obtains
\begin{eqnarray}\label{c.27}
\Big\vert \mathcal{G}_{\mathbf{K}}^{\lambda}(R_{1}) - \mathcal{G}_{\mathbf{K}}^{\lambda}(R_{2}) \Big\vert^2 &=&\Big\vert \sum_{\ell=0}^{\infty} (i^{\ell}) \frac{(2\ell+1)}{2\pi} \, \Big( \int d\Omega_{\hat{r}}\, e_{ij}^{\gamma}[\hat{\mathbf{K}}]\, \hat{r}_i\, \hat{r}_{j}\, P_{\ell}(\cos\gamma) \Big) \\
&\times& \Big( \frac{j_{\ell}(KR)}{(KR)} \, - \, \frac{\int_0^{KR} j_{\ell}(u) du}{(K R)^2} \, \Big) \Big\vert^2 \Big( K\, \rho'\, w\, \cos(\phi'-\psi) \Big)^2 \nonumber\,.
\end{eqnarray}
and Eq.~(\ref{4.22}) for $\mathcal{D}^{\text{ts}}(R_{1},t;R_{2},t)$ outcomes.


\subsection{Expression of the time of flight $t$ and distance $R$ versus redshift $z$}\label{app:C.3}

Typical quasars detected by VLBI means are located at cosmological distances. Hence, it seems necessary to express physical quantities in terms of the redshift of the source, $z$, rather than the time of flight $t$. These objects are located at the redshift range $0.46 \leq z \leq 2.73$, so during their journey to the Earth, they have been affected not only by PGWs background but also by the expansion of the Universe. One may incorporate the expansion of the universe by expressing time, distance, and frequency in terms of the redshift. In particular, one may replace the redshifted EM frequency according to $\omega_k \rightarrow \omega_k/(1+z)$. The light detected today was emitted at some earlier time. The time of flight $t(z)$, i.e., the time interval between emission at time $t_{z}$, corresponding to the redshift $z$, and detection at present $t_{z_H}$, is determined by \cite{mukhanov2005physical},
\begin{eqnarray}\label{c.28}
t(z) = t_{z_H} - t_{z} = \int_{z_{H}}^{z} \frac{dz'}{(1+z')H(z')},
\end{eqnarray}
\noindent where $H(z')$ is the Hubble parameter at redshift $z$ and $z_{H}=0$ is the redshift at the present. We proceed assuming a flat universe comprising only cold dark matter and a cosmological constant, so that $\Omega_\Lambda + \Omega_{m} = 1$. The Hubble parameter versus redshift is thus determined by
\begin{eqnarray}\label{c.29}
H(z) = H_0 \sqrt{(1-\Omega_m)+ \Omega_m(1+z)^3}.
\end{eqnarray}
\noindent By plugging Eq.~(\ref{c.25}) into Eq.~(\ref{c.24}) and setting $ z_H = 0$, the time of flight $t(z)$ turns out
\begin{eqnarray}\label{c.30}
t (z) = H_0^{-1} \int_{0}^{z} \frac{dz'}{\sqrt{(1-\Omega_m)(1+z')^2 + \Omega_m (1+z')^5}},
\end{eqnarray}
Moreover, the angular diameter distance $R(z)$ of an object located at redshift $z$ is determined by \cite{mukhanov2005physical}
\begin{eqnarray}\label{c.31}
R(z) = \frac{c}{H_0 (1+z)}\bigg( \int_{0}^z \frac{dz'}{\sqrt{\Omega_m (1+z')^3 + (1-\Omega_m)}} \bigg) \,.
\end{eqnarray}


\section{Derivation of the visibility $\mathcal{V}_{\text{ts}}(\mathbf{r}_C,t)$} \label{app:D}

The integration over the polar coordinate $\phi'$ in Eq.~(\ref{4.27}) for the visibility $\mathcal{V}(\mathbf{r}_C,t)$ can be written in the following compact form:
\begin{eqnarray}\label{d.1}
I_{\phi'} &\equiv& \int_{0}^{2\pi} e^{-A^2\, \cos^2\phi'} \, e^{-ik\xi_{\text{ts}}A\cos\phi'} d\phi' \\
&=& \int_{0}^{2\pi} e^{-\frac{A^2}{2} - \frac{A^2}{2} \cos2\phi'} \, e^{-ik\xi_{\text{ts}}A\cos\phi'} d\phi' \nonumber \,,
\end{eqnarray}
where we have defined $A\equiv \xi_{\text{ts}}^{-1} \rho' w$ and replaced $\phi'-\psi \rightarrow \phi'$. With the help of the Jacobi-Anger identity
\begin{eqnarray}\label{d.2}
e^{iz\cos\phi'} = \sum_{m=-\infty}^{\infty} i^m \, J_{m}(z) e^{im\phi'}\, ,
\end{eqnarray}
Eq.~(\ref{d.1}) can be written as
\begin{eqnarray}\label{d.3}
I_{\phi'} &=& e^{-\frac{A^2}{2}} \sum_{m=-\infty} ^{\infty} \int_{0}^{2\pi} e^{- \frac{A^2}{2} \cos2\phi'} J_{m} \big( -k\xi_{\text{ts}}A \big) \, e^{im\phi'} d\phi'\, .
\end{eqnarray}
Substituting the expression
\begin{eqnarray}\label{d.4}
e^{- \frac{A^2}{2} \cos2\phi'} = e^{i \big( i\frac{A^2}{2} \cos2\phi' \big)} = \sum_{n=-\infty}^{\infty} i^n J_{n}\big( i \frac{A^2}{2} \big) e^{i2n\phi'}\, ,
\end{eqnarray}
into Eq.~(\ref{d.3}) and using the properties $J_{m}(-x) = (-1)^m J_{m}(x)$ and $J_n(ix) = i^n I_n(x)$ where $I_n(x)$ stands for the modified Bessel function of the first kind, one ends up with
 \begin{eqnarray}\label{d.5}
I_{\phi'} &=& (2\pi)\, e^{-\frac{A^2}{2}} \sum_{m=-\infty} ^{\infty} J_{2m} \big( k\xi A \big) \, I_{m} \big( \frac{A^2}{2} \big)\, . 
\end{eqnarray}
Inserting Eq.~(\ref{d.5}) into Eq.~(\ref{4.27}) yields
\begin{eqnarray}\label{d.6}
\mathcal{V}_{\text{ts}}(\mathbf{r}_C,t) &=& \Big\vert \frac{2\pi}{\pi a^2} \sum_{m=-\infty}^{\infty} \int_0^a \rho'\, e^{- \frac{w^2}{2\xi^2_{\text{ts}} } \rho'^2 } \, J_{2m} \big( k w \rho' \big) \, I_{m} \big( \frac{w^2}{2 \xi^2_{\text{ts}}} \rho'^2 \big) \, d\rho' \Big\vert \\
&=& \Big\vert \frac{2}{a^2} \sum_{m=-\infty}^{\infty} (-i)^n \int_0^a \rho'\, e^{- \frac{w^2}{2\xi^2_{\text{ts}} } \rho'^2 } \, J_{2m} \big( k w \rho' \big) \, J_{m} \big( i \frac{w^2}{2 \xi^2_{\text{ts}}} \rho'^2 \big) \, d\rho' \Big\vert \nonumber \\
&=& \frac{2}{a^2} \Big\vert \int_0^a \rho' \, e^{- \frac{w^2}{2\xi^2_{\text{ts}} } \rho'^2 } \Big( J_{0} \big( k w \rho' \big) \, J_{0} \big( i \frac{w^2}{2 \xi^2_{\text{ts}}} \rho'^2 \big) + 2 \sum_{m=1}^{\infty} (-i)^m J_{2m} \big( k w \rho' \big) \, J_{m} \big( i \frac{w^2}{2 \xi^2_{\text{ts}}} \rho'^2 \big) \Big) \, d\rho' \Big\vert \,. \nonumber
\end{eqnarray}
In the last line, we have used the symmetry properties of the Bessel functions. Changing the variable $\rho' \rightarrow \rho'/a$ results in Eq.~(\ref{4.28}).


\bibliography{main}
\bibliographystyle{unsrt}

\end{document}